\documentclass[fleqn,usenatbib]{mnras}

\usepackage{newtxtext,newtxmath}

\usepackage[T1]{fontenc}
\usepackage[flushleft]{threeparttable}
\usepackage[font=normalsize]{caption}
\usepackage[english]{babel}
\usepackage{multicol}
\usepackage{relsize}
\usepackage{bm}
\usepackage[none]{hyphenat}
\usepackage{multirow}
\usepackage{fix-cm}

\newcommand{\ptorus}{p_{\text{torus}}}
\newcommand{\qtorus}{q_{\text{torus}}}
\newcommand{\Rin}{R_{\text{in}}}
\newcommand{\Rout}{R_{\text{out}}}
\newcommand{\OAtorus}{\text{OA}_{\text{torus}}}
\newcommand{\Rcl}{R_{\text{cl}}}
\newcommand{\Ncl}{N_{\text{cl}}}
\newcommand{\fcl}{f_{\text{cl}}}
\newcommand{\ffill}{f_{\text{fill}}}
\newcommand{\opttau}{\tau^{9.7}_{\text{torus}}}
\newcommand{\convtau}{\tau_{\text{0.55}}}
\newcommand{\convtauz}{\tau_{\text{0.55},z}}
\newcommand{\res}{\text{R}_{\lambda}}
\newcommand{\globalres}{\overline{\text{R}}}
\newcommand{\chisquare}{\chi^{2}_{N}}
\newcommand{\chisquaremin}{\chi^{2}_{\text{min},N}}
\newcommand{\alphanir}{\rm{\alpha_{5.5-7.5 \ \mu m}}}
\newcommand{\alphamir}{\rm{\alpha_{7.5-14 \ \mu m}}}
\newcommand{\alphafir}{\rm{\alpha_{25-30 \ \mu m}}}
\newcommand{\strength}{\rm{S_{10 \mu m}}}
\newcommand{\peak}{\rm{C_{10 \mu m}}}
\newcommand{\ewidth}{\rm{EW_{10 \mu m}}}

\DeclareRobustCommand{\VAN}[3]{#2}
\let\VANthebibliography\thebibliography
\def\thebibliography{\DeclareRobustCommand{\VAN}[3]{##3}\VANthebibliography}

\usepackage{graphicx, url}	
\usepackage{subcaption}
\usepackage{amsmath}	

\title[AGN dust models]{Towards an observationally motivated AGN dusty torus model – II. The roles of density distribution and chemical composition of the dust}

\author[Reyes-Amador et al.]{%
Omar Ulises Reyes-Amador,$^{1}$\thanks{E-mail: o.reyes@irya.unam.mx }
Omaira González-Martín,$^{1}$
Jacopo Fritz,$^{1}$
Maarten Baes,$^{2}$
\newauthor
Sundar Srinivasan,$^{1}$
Ismael García-Bernete,$^{3}$
Donaji Esparza-Arredondo,$^{1}$ \& 
Marko Stalevski$^{4}$ \\
$^{1}$Instituto de Radioastronomía y Astrofísica (IRyA), Universidad Nacional Autónoma de México (UNAM), Antigua Carretera a Pátzcuaro, 8701, \\ 
Ex-Hda. San José de la Huerta, Morelia, Michoacán, 58089, México\\
$^{2}$Sterrenkundig Observatorium, Universiteit Gent, Krijgslaan 281 S9, 9000 Gent, Belgium \\
$^{3}$Centro de Astrobiología (CAB), CSIC-INTA, Camino Bajo del 497 Castillo s/n, E-28692 Villanueva de la Cañada, Madrid, Spain \\
$^{4}$Astronomical Observatory, Volgina 7, 11060 Belgrade, Serbia\\\\
}

\date{Accepted XXX. Received YYY; in original form ZZZ}

\pubyear{2024}

\begin{document}
\label{firstpage}
\pagerange{\pageref{firstpage}--\pageref{lastpage}}
\maketitle

\begin{abstract}
Several models of nuclear dust in active galactic nuclei (AGN) have been presented in the literature to determine its physical and geometrical properties, usually assuming the dust density distribution as the main aspect producing differences in the mid-infrared (MIR) emission of AGNs. We present a study of the MIR emission of nearby AGNs by exploring the effects of dust  distribution and chemical composition on the spectral energy distributions (SEDs) using radiative transfer simulations. Our model grid includes smooth, clumpy, and two-phase dust distributions, combined with two dust compositions: the interstellar medium (ISM) dust composition including large grains (up to $\rm{10 \ \mu m}$), and the oxide/silicate-based composition from \citet{Reyes-Amador2024}. A synthetic SED library was generated and analysed both on a model-to-model basis and with observed MIR spectra from 68 AGNs. We found that dust density distribution and dust composition significantly influence the spectral shapes and silicate features at $10$ and $\rm{18 \ \mu m}$, especially at edge-on orientations. The smooth distribution produces stronger and broader silicate absorption features, while the clumpy distribution generates stronger features in emission. The two-phase distributions exhibit intermediate characteristics depending on the clumpiness fraction ($\fcl$) and filling factor ($\ffill$). The ISM dust composition with large grains is more suited to reproduce the observed features and a higher fraction of good fits, particularly with Type-2 SEDs, independently of dust density distributions. The \citet{Reyes-Amador2024} composition provides a larger number of good fits with Type-1 SEDs for $\fcl \leq 0.5$, and with Type-2 SEDs for $\fcl \geq 0.9$. This work shows that no single dust distribution or composition reproduces all observations.
\end{abstract}




\section{Introduction} \label{sec:introduction}

It is well known that the infrared (IR) emission from active galactic nuclei (AGNs) originates from a dusty structure known as the ``dusty torus'' that surrounds the central supermassive black hole (SMBH) and accretion disc \citep{Antonucci1993}. Polarimetric observations \citep{Antonucci1985} revealed that broad line region is present even in obscured (type 2) AGN, and must be hidden by a geometrically and optically thick structure, which absorbs and scatters the photons coming from the accretion disc \citep[see][]{Netzer2015,Ramos-Almeida2017}. The presence of this structure allows us to consistently explain the differences in the spectral characteristics observed in AGNs that have driven the established Type-1/Type-2 dichotomy and the proposition of the ``unification scheme'' \citep{Antonucci1993,Urry1995}. 

Due to its compact size and distance, direct imaging of the torus is limited to a few nearby AGNs. However, IR interferometry has successfully resolved this structure in galaxies such as NGC 1068 and Circinus  \citep[e.g.,][]{Jaffe2004,Tristram2007,Tristram2014,Pfuhl2020,GamezRosas2022,Isbell2022,Isbell2025}, and observations with the ALMA interferometer have have probed AGN dust and molecular gas structures in various sources \citep[e.g.,][]{Combes2019,García-Burillo2021}. Yet, the most common approach for studying the torus has been through comparison of radiative transfer (RT) models with mid-infrared (MIR) observations, where torus emission dominates and contamination is minimal \citep{Barvainis1987}. 

Dust geometry and distribution have historically shaped torus model classifications. ``Smooth'' models assume a homogeneous dust distribution \citep{Pier1992, GranatoDanese1994,Efstathiou1995,vanBemmel2003,Fritz2006,Efstathiou2013,Efstathiou2022}, offering computational simplicity. ``Clumpy'' models later emerged to explain early MIR spectra that lacked silicate emission features \citep{Nenkova2002,Nenkova2008a,Nenkova2008b,Honig2006,Honig&Kishimoto2010}, and quickly gained popularity for their improved fit to early observations. However, limitations in clumpy models, such as their ability to reproduce the current observations, have been identified  \citep{Dullemond2005,Sturm2005,Siebenmorgen2005,Gonzalez-Martin2019II,Gonzalez-Martin2023,Garcia-Bernete2024}. Motivated by the 3D hydrodynamical simulations of the medium surrounding AGNs proposed by \citet{Wada2009} \citep[see also][]{Wada2012,Wada2016}, ``two-phase'' models emerged, incorporating clumps within a low-density medium \citep{Stalevski2012,Siebenmorgen2015,Stalevski2016,Stalevski2019,Gonzalez-Martin2023}. These have shown that some torus and dust grain parameters, like maximum grain size, play a dominant role in shaping the MIR spectral energy distributions (SEDs). For example, \citet{Gonzalez-Martin2023} demonstrated that allowing grain sizes up to $\rm{10 \ \mu m}$ improved model accuracy in reproducing observed MIR spectra.

Diverse studies have also attempted to determine which dust density distribution best reproduces observed MIR spectra. \citet{Gonzalez-Martin2019II} found that the clumpy models by \citet{Nenkova2008b} successfully reproduce the $\sim 30\%$ of the SEDs in their sample. \citet{Esparza-Arredondo2021} found that 80\% of their sample of MIR spectra were best fit by a combination of smooth and clumpy dust and gas distributions. \citet{Martinez-Paredes2020} found that although none of the models they studied could perfectly replicate the silicate features, clumpy models provided a better match than smooth models in quasars. In contrast, \citet{Efstathiou2022} and \citet{Varnava2025} found that their smooth torus model with a tapered disc geometry is more effective for ultra-luminous infrared galaxies (ULIRGs).

In order to compare the characteristics between the different dust density distributions, \citet{Stalevski2012} constructed a grid of two-phase models and, for each case, generated corresponding clumpy and smooth models with identical global physical parameters. They found that two-phase models with small clumps resemble smooth models in Type-1 views but differ more in Type-2 SEDs, and that the largest discrepancies between the two-phase and clumpy models arise in the near-infrared (NIR) range, particularly for face-on orientations. \cite{Feltre2012} found that smooth models by \citet{Fritz2006} and clumpy models by \citet{Nenkova2008b} yield different SEDs even under matched parameters that could be due to differences in the dust chemical composition and in the properties of the primary radiation source. 


One of the least-explored aspects of AGN dusty torus models is dust chemical composition. The SED libraries built by all the models assume that the AGN dust composition is the Galactic interstellar medium (ISM) composition \citep[e.g.,][]{GranatoDanese1994,Fritz2006,Nenkova2008b,Honig&Kishimoto2010,Honig&Kishimoto2017,Stalevski2012,Stalevski2016,Gonzalez-Martin2023}. 
Several studies have investigated this through analytical modelling and SED fitting techniques applied to AGN observations \citep{Markwick-Kemper2007,Srinivasan2017,Tsuchikawa2021,Tsuchikawa2022,Reyes-Amador2024}. In our previous work \citep{Reyes-Amador2024}, we explored models with dust species such as oxides, amorphous silicates, crystalline silicates, and the Galactic ISM silicates. We found that a dust composition of oxides and amorphous silicates better reproduced the MIR spectra of the sample than the Galactic ISM silicates. \citet{Tsuchikawa2022} modelled the MIR spectra of a sample of heavily obscured AGN using 1D RT calculations with four different dust species. They simplified the calculation by assuming the dust is distributed within a sphere instead of a torus, finding that 97\% of the sample prefer a porous silicate dust model without micron-sized large grains. This is consistent with the SED fitting analysis provided by \citet{Gonzalez-Martin2023}, where Spitzer spectra are best fitted to torus models with large grain sizes. 

Many of the studies mentioned above have attempted to determine the dust distribution that best represents AGNs. However, they differ in several key aspects that could bias their results. For example, the compared models do not come from the same RT code, so their parameters cannot be matched directly. Therefore, a comprehensive approach is needed that simultaneously accounts for all relevant factors. Motivated by this challenge, we present self-consistently calculated models with the three different dust density distributions: smooth, clumpy, and two-phase, treating the dust mass fraction locked in clumps as one of the explored parameters entering the calculation and analysis of a comprehensive SED library using 3D RT simulations (created with the same RT code), and hence compared with observations. This is the first SED library that includes an observationally motivated dust chemical composition, different from the ones assumed in previous models, which we also compare with the Galactic ISM dust composition. We will leave the exploration and analysis of the effects of a polar dust distribution in a future paper (Reyes-Amador et al. in prep.). 

This paper is structured as follows: Section~\ref{sec:the_model} describes the RT code, the characteristics of the model, and the parameter space. Section~\ref{sec:data_sample} presents the data sample we used in the models-observations comparison. Section~\ref{sec:analysis_models} outlines the methodology used to analyse the grid of models. The results are presented in Section~\ref{sec:results}, discussed in Section~\ref{sec:discussion}, and summarised in Section~\ref{sec:conclussions}.

\section{The model} \label{sec:the_model}

In this section, we briefly describe SKIRT, the RT code used to create the grid of models for the dusty torus. We present the characteristics of the assumed SED for the primary source of emission, as well as the detailed characteristics of the dusty torus model, i.e., the chemical and physical properties of the dust grains and of the torus geometry itself. Finally, we discuss the characteristics of the spatial grid within the RT code used to compute the models and the parameter space explored in the grid of models to simulate the different dust density distributions.

\subsection{The radiative transfer code}
\label{sec:RT_code} 

We use the version 9.0 of SKIRT \citep{Baes2011,Camps2015,Camps2020}, a generic 3D RT code based on the Monte Carlo method. In its dust RT mode, it emulates absorption, scattering, emission and polarisation in dusty astrophysical systems such as galaxies, galactic nuclei, and star forming regions. It is fully 3D and equipped with a suite of advanced grids \citep{Camps2013,Saftly2013,Saftly2014,Lauwers2024}, which makes it ideally suited to generate synthetic observations of hydrodynamic simulations \citep{Camps2018,Kapoor2021,Jaquez2023,Gebek2024,Baes2024,Baes2025}. SKIRT is highly customisable in terms of geometry, SEDs, dust models, and physical processes involved.

SKIRT is on a continuous development track, and the developers are explicitly willing to take into account possible requests for modifications and additions to tackle unforeseen issues. In addition, it has been benchmarked against other Monte Carlo and RT codes \citep{Gordon2017,VanderMeulen2023} which aim to define and provide a suite of benchmark problems covering all the various numerical problems arising in dust RT, and assess their validity and reliability. The benchmarks can be used to further develop and improve the existing codes as well as to test newly developed codes containing dust RT calculations. 

Finally, SKIRT has already been used by various authors to successfully produce dusty torus models and calculate their MIR \citep{Stalevski2012,Stalevski2016,Stalevski2019,Victoria-Ceballos2022,Gonzalez-Martin2023} and X-ray \citep{VanderMeulen2023} emission.

\subsection{The accretion disc SED}
\label{sec:the_accretion_disc_SED}

The accretion disc is the primary source of emission, which we model as a central point source with isotropic emission that produces a SED described by a composition of power laws as proposed by \citet{Schartmann2005} and used by \citet{Stalevski2012,Stalevski2016,Gonzalez-Martin2023}. This SED is motivated by observations and theoretical modelling of accretion disc spectra \citep[e.g.][]{Schartmann2005}. The values for the spectral indices and spectral ranges in terms of $L_{\lambda}$ are:
\begin{equation} \label{eq:accretion_disc_SED}
    L_{\lambda} =
    \begin{cases}
       \lambda^3, &\quad 0.001 < \lambda \leq 0.05 \ [\mu m]\\
       \lambda, &\quad 0.05 < \lambda \leq 0.1 \ [\mu m]\\
       \lambda^{-0.2}, &\quad 0.1 < \lambda \leq 5 \ [\mu m]\\
       \lambda^{-1.54}, &\quad 5 < \lambda \leq 1000 \ [\mu m]\\ 
     \end{cases}
\end{equation}

For the accretion disc bolometric luminosity, we used  $L_{\text{bol}} = 10^{11}$$\rm{L_{\odot}}\approx \rm{3.8\times10^{44}\ erg \ s^{-1}}$, which is the typical one used for the accretion disc in dusty torus models \citep[e.g.,][]{Schartmann2005,Stalevski2012,Stalevski2016,Gonzalez-Martin2023}. However, the choice of specific value for the bolometric luminosity is of no importance due to the scaling of the dust emission, size and mass \citep[see][]{Ivezic1997,Fritz2006,Honig&Kishimoto2010}.

\subsection{The geometry of the torus}
\label{sec:the_torus_geometry}

The geometrical distribution of the AGN dust is modelled as a flared disc (\textsc{TorusGeometry} within SKIRT), with a dust density distribution described by a power law that allows density gradients in the radial ($r$) and polar ($\theta$) directions. This is the same geometry used in the models from \citet{GranatoDanese1994,Fritz2006,Stalevski2012,Stalevski2016,Gonzalez-Martin2023}. The equation used to describe the dust density distribution within this geometry is calculated according to the following: 
\begin{equation} \label{eq:torus_geometry}
    \rho (r) \propto r^{-\ptorus} e^{-\qtorus|\cos\theta|},
\end{equation}
for $\Rin < r < \Rout$ and $\rm{90^{\circ}-OA_{torus} < \theta < 90^{\circ}+OA_{torus}}.$ The five free parameters describing this geometry are: the inner ($\Rin$) and outer ($\Rout$) radii of the torus, the radial ($\ptorus$) and polar ($\qtorus$) indices, and the opening angle of the torus ($\rm{OA_{torus}}$). The $\Rin$ is fixed to be the dust sublimation radius ($R_{\text{sub}}$), which is described by \citet{Barvainis1987} and used in the models cited above, as follows:
\begin{equation} \label{eq:sublimation_radius}
    R_{\text{sub}} = 1.3(L_{\text{bol}}/10^{46})^{1/2}(T_{\text{sub}}/1500)^{-2.8} \ \text{pc},
\end{equation}
where the $L_{\text{bol}}$ is the accretion disc bolometric luminosity (see Sec.~\ref{sec:the_accretion_disc_SED}) and $T_{\text{sub}}$ is the sublimation temperature (in Kelvin) of the dust. We used $T_{\text{sub}} = 1250$ K, which is an average between $\rm{1000 \ K}$ and $\rm{1500 \ K}$. These are the sublimation temperatures typically assumed for silicates  and graphites, respectively  \citep[e.g.,][]{GranatoDanese1994,Fritz2006,Gonzalez-Martin2023}. Introducing the values for sublimation temperature and bolometric luminosity into the Eq.~(\ref{eq:sublimation_radius}), we obtained $R_{\text{in}} = R_{\text{sub}} = 0.42$ pc. As $R_{\text{in}}$ is always fixed (although rescalable with $L_{\text{bol}}$, see Sect.~\ref{sec:adequacy_models}), it is more convenient to use the ratio between the outer and the inner radius of the torus, $Y=\Rout/\Rin$, to parametrise the dusty torus size.

\subsection{The dust density distribution of the torus}
\label{sec:the_torus_dust_density_distribution}

\begin{figure*}
    \centering
    \includegraphics[trim=0cm 0 0 0, clip,scale=0.45]{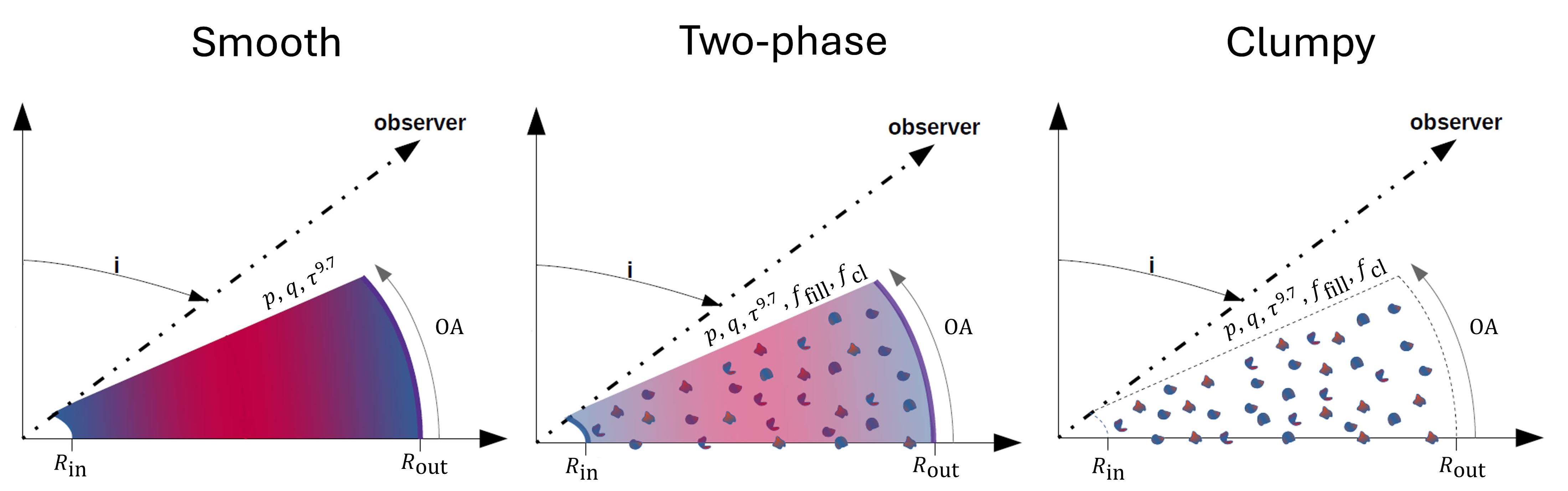}
    \caption{Scheme illustrating the three different dust density distributions within the torus models studied in this work and their corresponding parameters. Note that this figure is for illustrative purposes only, it is not a direct representation of the models. Figure adapted from \citet{Garcia-Bernete2022c}.}
    \label{fig:scheme_dust_distributions}
\end{figure*}

We use the dust density distribution as a free parameter, i.e., we create models with the three distributions: smooth, clumpy, and two-phase (see Figure~\ref{fig:scheme_dust_distributions}). The creation of a clumpy or two-phase dust density distribution is achieved with the use of the so-called ``decorators'' \citep{Baes2015}, which design patterns that allow the modification of the characteristics of basic functions. For this specific case, we used the \textsc{ClumpyGeometryDecorator}, which modifies a continuous density distribution (smooth), creating high-density spheres of given characteristics, randomly located within any defined geometry. It assigns a fraction of the mass of the original geometry to compact clumps, which are distributed within the same geometry. The characteristics that describe the clumpiness are the fraction of the mass locked in clumps $\fcl$, the total number of clumps $\Ncl$, the scale radius of a single clump $\Rcl$, and the kernel that describes the mass distribution of a single clump. For the latter, we used a \textsc{CubicSplineSmoothingKernel}\footnote{For details on the cubic spline density function, visit \url{https://skirt.ugent.be/skirt9/class_cubic_spline_smoothing_kernel.html}}. 

The parameter that distinguishes between the clumpy and the two-phase density distributions is $\fcl$: for the clumpy models, $\fcl=1$, while for the two-phase models $0 < \fcl< 1$. Although in practice, the smooth dust density distribution is achieved in SKIRT without the need to use \textsc{ClumpyGeometryDecorator} and its associated parameters, we can obtain smooth models using $\fcl=0$.

The scale radius of the clumps, $\Rcl$, is assumed to be the $5\%$ of the radial extent of the torus, so it depends on the size of the system, which is usually done in other works \citep[e.g.,][]{Stalevski2012,Gonzalez-Martin2023}. Hence, the prescription for $\Rcl$ is 

\begin{equation} \label{eq:size_cloud}
    \Rcl = 0.05(\Rout - \Rin).
\end{equation}

The total number of clouds, $\Ncl$, is calculated from the filling factor ($\ffill$), defined as the ratio between the volume occupied by clumps ($V_{\text{cl}}$) and the volume of the torus ($V_{\text{torus}}$). Since the clumps have a spherical geometry, $\ffill$ is given by the following equation:

\begin{equation} \label{eq:filling_factor}
    \begin{split}
        \ffill = \frac{V_{\text{cl}}}{V_{\text{torus}}} = \frac{\frac{4 \pi}{3} \Ncl \Rcl^{3}}{\frac{4 \pi}{3} (\Rout^{3} - \Rin^{3}) \cos(90^{\circ}-\OAtorus)} \\
    = \frac{\Ncl(\Rcl/\Rin)^3}{(Y^3-1) \cos(90^{\circ}-\OAtorus)}.
    \end{split}
\end{equation}

\subsection{The dust chemical composition}
\label{sec:the_dust_chemical_composition}

\begin{table*}
\begin{threeparttable}
\begin{center}
\caption{Dust chemical compositions included in our model grid.}
\begin{tabular}{ccccc}\hline \hline
Dust composition & Components & Size distribution & Grain size ($\rm{ \ \mu m}$) & Optical properties \\
\hline
\multirow{4}{*}{\citet{Reyes-Amador2024}} & $\sim26.3\%$ porous alumina & CDE & $0.1 $ & \citet{Begemann1997} \\
& $\sim 10.3\%$ periclase & CDE & 0.1 & \citet{Hofmeister2003} \\ 
& $\sim16.1\%$ olivine (small) & - & 0.1 & \citet{Dorschner1995} \\
& $\sim0.3\%$ olivine (large) & - & 3 & \citet{Dorschner1995} \\
& 47\% graphites & MRN & 
0.005 - 0.25 & \citet{LaorDraine1993} \\
\hline
\multirow{2}{*}{ISM with large grains} & $49\%$ graphites & MRN & 0.005 - 10 & \citet{LaorDraine1993} \\
& $51\%$ silicates & MRN & 0.005 - 10 & \citet{LaorDraine1993} \\
\hline\hline
\end{tabular}
\label{tab:dust_compositions}
\begin{tablenotes}
\item \noindent \textbf{Notes:} CDE: continuous distribution of ellipsoids \citep{Bohren1998}. MRN: Mathis, Rumpl \& Nordsieck \citep{Mathis1977}. Small and large olivines from the \citet{Reyes-Amador2024} dust composition do not follow a size distribution, since they are spherical grains with a fixed size.
\end{tablenotes}
\end{center}
\end{threeparttable}
\end{table*}

To investigate the impact of dust chemical composition on RT dusty torus models, we explore two different dust compositions. We generated a set of models using a modified version of the Galactic ISM dust composition, consisting of $49\%$ of graphite and $51\%$ of silicate (instead of the standard 47\% and 53\%, respectively). Motivated by the success of the models from \citet{Gonzalez-Martin2023} in reproducing observed MIR spectra when including larger dust grains, we adopted a grain size range from 0.005 to $\rm{10 \ \mu m}$ (instead of the standard $\rm{0.005 - 0.25 \ \mu m}$) with the MRN grain size distribution ($dn(a)\propto a^{-3.5}$). Throughout this paper, we refer to this dust composition as "ISM dust composition with large grains" or simply "ISM dust composition". In addition to the commonly used Galactic ISM silicates (also known as "astronomical silicates"), we tested an alternative oxide/silicate-based composition obtained in \citet{Reyes-Amador2024}, which has been demonstrated to work better than the astronomical silicates when reproducing the MIR spectra of AGNs that exhibit silicate features in emission. The details of both dust compositions are summarised in Table~\ref{tab:dust_compositions}.

\subsection{The parameter space and grids of models} \label{sec:paramter_space}

Each set of models corresponding to different chemical compositions and dust distributions, as mentioned in Sections~\ref{sec:the_torus_dust_density_distribution} and~\ref{sec:the_dust_chemical_composition}, is characterised by a set of parameters: 
\begin{itemize}
    \item The optical depth of the torus at $\rm{9.7 \ \mu m}$ ($\opttau$), which is defined in the full x-axis (i.e., along the diameter of the torus).
    \item The exponents of the spatial dust density dependency in the radial ($\ptorus$) and polar ($\qtorus$) direction.
    \item The half-opening angle of the torus ($\OAtorus$), measured from the equatorial plane to its edge. 
    \item The ratio of the outer to inner radius ($Y = \Rout/\Rin$).
    \item The viewing angle ($i$).
\end{itemize}
For clumpy and two-phase models, two additional parameters are included:
\begin{itemize}
    \item The fraction of dust mass locked in clumps ($\fcl$).
    \item The filling  factor ($\ffill$), which provides the number of clumps ($\Ncl$) through Eq.~\ref{eq:filling_factor}.
\end{itemize}
Although the size of the clumps ($\Rcl$) is an important parameter for these models, it is not a free parameter because we define it as being dependent on $Y$ (see Equation~\ref{eq:size_cloud}).

\begin{table}
\begin{threeparttable}
\begin{center}
\caption{Parameter space explored in the grid of models with smooth, clumpy, and two-phase dust density distributions for each of the two dust compositions.}
\begin{tabular}{c@{\hspace{2mm}}c@{\hspace{3mm}}c@{\hspace{2mm}}c}\hline \hline
\textbf{Parameter} &  \multicolumn{3}{c}{\textbf{Values}}  \\
 & \textbf{Smooth} & \textbf{Clumpy} & \textbf{Two-phase}  \\ \hline
$\opttau$ & 3, 5, 7, 9, 11, 13 & 3, 5, 7, 9, 11, 13 & 3, 5, 7, 9, 11, 13 \\
$\ptorus$ & 0, 0.75, 1.50 & 0, 0.75, 1.50 & 0, 0.75, 1.50 \\
$\qtorus$ & 0, 0.75, 1.50 & 0, 0.75, 1.50 & 0, 0.75, 1.50  \\
$\OAtorus$ & $10^{\circ}$, $45^{\circ}$, $80^{\circ}$ & $10^{\circ}$, $45^{\circ}$, $80^{\circ}$ & $10^{\circ}$, $45^{\circ}$, $80^{\circ}$ \\
$Y$ & $2^*$,10, 20, 40 & $2^*$,10, 20, 40 & $2^*$,10, 20, 40\\
$\ffill$ & 0 & 0.10, 0.15, 0.25 & 0.10, 0.15, 0.25\\
$\fcl$ & 0 & 1 & 0.25, 0.50, 0.90, 0.97 \\ \hline
\textbf{Simulations} & 486 & 1\,458 & 5\,832 \\
\textbf{SEDs$^{**}$} & 9\,234 & 27\,702 & 110\,808 \\
\hline\hline
\end{tabular}
\label{tab:model_parameters}
\begin{tablenotes}
\item \noindent \textbf{Notes:} $^*$Models with $Y=2$ were excluded from the further analysis, so the numbers of simulations and SEDs do not include this parameter value. $^{**}$ Each simulation generates SEDs at arbitrarily inclinations, so we computed SEDs at 19 values from $i=0^{\circ}$ to $i=90^{\circ}$ in steps of $5^{\circ}$.
\end{tablenotes}
\end{center}
\end{threeparttable}
\end{table}

The selection of most parameter values explored in this work is based on the one studied in \citet{Gonzalez-Martin2023}. For $\opttau$ we selected the same six values, while for $Y$, $\OAtorus$, $\ptorus$, and $\qtorus$, we selected only their minimum, median and maximum values. Table~\ref{tab:model_parameters} summarises the parameter values used in this work. It is important to mention that we also explored $Y=2$. Such models were eventually not taken into account for further analysis because SEDs with this particular torus size have spectral shapes that are unsuitable to reproduce any of the MIR spectra used in this work (see Appendix~\ref{sec:models_Y2}). For the viewing angle, we used values from $i=0^{\circ}$ to $i=90^{\circ}$ in steps of $5^{\circ}$, resulting in a total of 19 values. It is worth mentioning that the viewing angle is not really a model parameter like the others, in the sense that a single RT model can generate SEDs at arbitrarily many inclinations.

In the case of clumpy and two-phase models, we adopted the $\ffill$ values explored in \citet{Stalevski2012} (i.e., $\ffill=0.15$ and $0.25$), with the addition of one more value ($\ffill=0.1$). This led to different values for $\Ncl$ according to Eq.~\ref{eq:filling_factor}. Among the values explored for $\fcl$, we included the ones used in \citet{Gonzalez-Martin2023} and \citet{Stalevski2012}, $\fcl=0.25$ and~$0.97$, respectively. The explored parameter space is summarised in Table~\ref{tab:model_parameters}.

Using the parameter space described in this section, we constructed a grid of models for both dust compositions outlined in Section~\ref{sec:the_dust_chemical_composition}. We created $15\,552$ RT simulations, $7\,776$ for each of the two dust compositions. Specifically, we performed a total of $972$, $2\,916$, and $11\,664$ RT simulations for smooth, clumpy, and two-phase models, respectively. Each RT simulation produces one synthetic SED for each viewing angle, resulting in $18\,468$, $55\,404$, and $221\,616$ synthetic SEDs for smooth, clumpy, and two-phase models, respectively. In total, we generated $295\,488$ synthetic SEDs in the wavelength range of 0.09-1000 $\rm{\mu m}$, with all parameters evaluated at least at three grid values. 

The wavelength grid at which the SEDs are calculated consists of 182 logarithmically spaced points, 101 of which are embedded within the 1-50 $\rm{\mu m}$ range. However, since dust emission starts to be significant at wavelength larger than 3 $\rm{ \ \mu m}$, the analysis of all the synthetic SEDs was performed only in the wavelength range $\rm{\lambda > 3 \ \mu m }$.

According to the Unified Model, in Type-1 AGNs, the line of sight does not intercept a significant amount of dust within the torus, whereas in Type-2 AGNs, at least for a smooth torus, it does. Based on this, we have classified our synthetic SEDs into Type 1 and Type 2 depending on the $\OAtorus$ and $i$ parameter values, as follows:
\begin{equation} \label{eq:type1_type2_SEDs}
\begin{split}
    \text{If} \quad i < 90^{\circ} - \OAtorus, \quad \text{Type 1}, \\
    \text{if} \quad i \geq 90^{\circ} - \OAtorus, \quad \text{Type 2}.
\end{split}
\end{equation}

\noindent We obtained $~48\%$ of Type-1 and $~52\%$ of Type-2 synthetic SEDs. 

\section{The sample} \label{sec:data_sample}

\begin{table*}
\caption{Identifier, redshift and classification for the AGNs in our sample, followed by quantities computed in this paper for the observed spectra. Respectively, they are the location of the peak of $\rm{10 \ \mu m}$ silicate features, the feature strength, the equivalent width, and spectral slopes for various wavelength ranges.}
\scriptsize 
\renewcommand{\arraystretch}{0.8806}
\begin{center}
\begin{tabular}{lcccccccc}\hline \hline
Object name & z & Class & $\rm{C_{10\mu m}}$ ($\mu$m) & $\rm{S_{10\mu m}}$ & $\rm{EW_{10\mu m}}$ ($\mu$m)&  $\rm{\alpha_{5.5-7.5\mu m}}$ &  $\rm{\alpha_{7.5-14\mu m}}$ &  $\rm{\alpha_{25-30\mu m}}$ \\
(1) & (2) & (3) & (4) & (5) & (6) & (7) & (8) & (9)\\ \hline
2MASXJ11454045-1827149   &  0.033 & S1 & 10.390 & -0.244 & -0.901 & -0.610 & -1.000 &  0.352 \\
3C120                    &  0.033 & S1 & 10.710 & -0.299 & -0.977 & -0.788 & -1.240 & -0.679 \\
Ark120                   &  0.032 & S1 & 10.680 & -0.275 & -0.908 & -0.309 & -0.498 &  0.231 \\
ESO141-G055              &  0.037 & S1 & 10.400 & -0.211 & -0.487 & -0.437 & -0.715 &  0.067 \\
ESO511-G030              &  0.015 & S1 & 10.590 & -0.405 & -1.347 & -0.107 & -0.503 &  0.692 \\
ESO548-G081              &  0.014 & S1 & 10.640 & -0.296 & -1.033 & -0.252 & -0.397 &  0.561 \\
Fairall51                &  0.011 & S1 & 11.100 & -0.225 & -0.697 & -1.156 & -1.131 & -0.863 \\
FAIRALL9                 &  0.047 & S1 & 10.660 & -0.181 & -0.595 & -0.543 & -0.799 &  0.372 \\
IC4329A                  &  0.016 & S1 & 11.100 & -0.100 & -0.294 & -0.767 & -1.077 &  0.109 \\
IISZ010                  &  0.034 & S1 & 10.820 & -0.220 & -0.696 & -1.568 & -1.379 &  0.687 \\
IIZw136                  &  0.078 & S1 & 11.190 & -0.058 & -0.153 & -0.980 & -0.982 & -0.366 \\
IZw1                     &  0.059 & S1 &  9.890 & -0.296 & -1.002 & -0.844 & -1.301 & -0.955 \\
M106                     &  0.002 & S1 & 11.280 & -0.277 & -0.914 & -0.948 & -1.285 & -1.153 \\
MCG-01-13-025            &  0.016 & S1 & 10.780 & -0.614 & -2.226 &  0.320 & -0.619 &  0.223 \\
MCG+04-22-042            &  0.032 & S1 & 10.480 & -0.201 & -0.640 & -0.912 & -1.092 &  0.584 \\
MCG-06-30-015            &  0.008 & S1 & 10.770 & -0.155 & -0.403 & -1.038 & -1.137 & -0.782 \\
Mrk1018                  &  0.042 & S1 & 10.500 & -0.274 & -0.891 & -0.483 & -0.521 &  0.615 \\
Mrk110                   &  0.033 & S1 & 10.520 & -0.274 & -0.843 & -0.502 & -0.791 &  0.862 \\
Mrk1210                  &  0.013 & S1 & 11.140 & -0.178 & -0.466 & -2.353 & -2.054 & -0.329 \\
Mrk1392                  &  0.036 & S1 & 10.820 & -0.140 & -0.437 & -1.674 & -1.718 & -0.482 \\
Mrk1393                  &  0.054 & S1 & 10.740 & -0.102 & -0.310 & -0.444 & -1.759 & -0.284 \\
Mrk279                   &  0.030 & S1 & 11.110 & -0.093 & -0.259 & -1.094 & -1.209 & -0.570 \\
Mrk290                   &  0.030 & S1 & 10.900 & -0.251 & -0.869 & -0.993 & -1.270 & -0.185 \\
Mrk590                   &  0.021 & S1 & 10.510 & -0.190 & -0.530 & -1.299 & -2.131 & -0.210 \\
Mrk705                   &  0.029 & S1 & 11.300 & -0.136 & -0.354 & -1.603 & -1.384 & -0.817 \\
Mrk841                   &  0.036 & S1 & 10.910 & -0.096 & -0.286 & -1.356 & -1.656 & -0.373 \\
NGC1052                  &  0.005 & S1 & 10.910 & -0.260 & -0.813 & -1.173 & -1.696 & -0.541 \\
NGC3783                  &  0.011 & S1 & 10.810 & -0.164 & -0.452 & -0.847 & -1.562 & -0.389 \\
NGC4151                  &  0.002 & S1 & 10.530 & -0.124 & -0.335 & -1.070 & -1.589 & -0.323 \\
NGC526A                  &  0.019 & S1 & 11.180 & -0.194 & -0.643 & -0.920 & -1.369 &  1.064 \\
NGC5548                  &  0.025 & S1 & 11.170 & -0.194 & -0.558 & -1.902 & -1.650 & -0.554 \\
NGC6814                  &  0.003 & S1 & 11.440 & -0.122 & -0.255 & -0.576 & -1.097 & -0.894 \\
NGC7213                  &  0.005 & S1 & 10.710 & -0.708 & -2.805 & -0.347 & -1.402 & -0.026 \\
PG0804+761               &  0.100 & S1 & 10.110 & -0.344 & -0.969 & -0.236 & -0.586 &  0.272 \\
PG1211+143               &  0.090 & S1 & 10.410 & -0.289 & -0.944 & -0.890 & -0.910 &  0.488 \\
PG1351+640               &  0.088 & S1 &  9.930 & -0.752 & -3.205 & -0.870 & -1.879 & -0.603 \\
PG1448+273               &  0.065 & S1 & 11.170 & -0.151 & -0.476 & -1.068 & -1.288 & -0.323 \\
PG2304+042               &  0.042 & S1 & 10.530 & -0.588 & -2.129 & -0.192 & -1.150 &  0.515 \\
PICTORA                  &  0.035 & S1 & 10.720 & -0.539 & -1.957 & -0.976 & -1.430 &  0.443 \\
UGC6728                  &  0.007 & S1 & 10.190 & -0.211 & -0.523 & -0.599 & -1.054 &  0.268 \\
UM614                    &  0.033 & S1 & 10.670 & -0.262 & -0.913 & -0.998 & -0.994 &  0.546 \\
2MASXJ05580206-3820043   &  0.034 & S1 &  9.680 &  0.317 &  0.698 & -0.294 & -0.594 &  0.590 \\
MCG-03-34-064            &  0.020 & S1 &  9.200 &  0.276 &  0.467 & -2.391 & -1.812 & -1.063 \\
Ark347                   &  0.022 & S2 & 10.490 & -0.084 & -0.163 & -1.212 & -1.254 & -0.714 \\
CGCG420-015              &  0.029 & S2 & 11.160 & -0.117 & -0.349 & -1.346 & -1.198 & -0.262 \\
ESO138-G001              &  0.009 & S2 & 10.740 & -0.228 & -0.715 & -1.769 & -1.395 & -0.694 \\
ESO374-G044              &  0.028 & S2 & 11.710 & -0.070 & -0.091 & -2.652 & -2.497 & -0.441 \\
Mrk417                   &  0.033 & S2 & 11.630 & -0.076 & -0.131 & -1.258 & -1.750 &  0.550 \\
NGC1275                  &  0.016 & S2 & 10.500 & -0.308 & -1.005 & -2.329 & -2.525 & -1.113 \\
NGC4507                  &  0.012 & S2 & 11.130 & -0.140 & -0.347 & -1.097 & -1.152 & -1.300 \\
NGC4939                  &  0.009 & S2 & 10.160 & -0.259 & -0.877 & -1.868 & -2.263 & -0.745 \\
2MASSXJ10594361+6504063  &  0.084 & S2 &  9.870 &  0.949 &  2.037 & -1.083 & -1.468 & -0.560 \\
2MASXJ05054575-2351139   &  0.035 & S2 &  9.520 &  0.194 &  0.305 & -1.814 & -1.241 & -0.052 \\
ESO103-G35               &  0.013 & S2 &  9.700 &  0.919 &  1.499 & -2.185 & -1.919 & -0.506 \\
ESO439-G009              &  0.025 & S2 &  9.620 &  0.458 &  0.970 & -3.700 & -2.463 & -0.824 \\
IC4518W                  &  0.016 & S2 &  9.840 &  1.418 &  1.984 & -2.113 & -1.315 & -2.260 \\
IC5063                   &  0.009 & S2 &  9.710 &  0.343 &  0.698 & -2.631 & -2.005 & -0.837 \\
MCG-05-23-016            &  0.008 & S2 &  9.750 &  0.311 &  0.574 & -1.730 & -1.647 & -0.418 \\
MCG+07-41-03             &  0.056 & S2 &  9.680 &  0.706 &  1.252 & -3.159 & -2.800 & -1.794 \\
Mrk348                   &  0.005 & S2 &  9.670 &  0.232 &  0.336 & -1.056 & -1.288 & -0.185 \\
Mrk3                     &  0.014 & S2 &  9.850 &  0.143 &  0.083 & -2.500 & -2.364 & -0.604 \\
Mrk78                    &  0.037 & S2 &  9.690 &  0.545 &  0.957 & -2.112 & -1.820 & -1.140 \\
NGC3081                  &  0.006 & S2 &  9.350 &  0.161 &  0.220 & -2.370 & -2.106 & -1.116 \\
NGC4388                  &  0.005 & S2 &  9.820 &  0.962 &  1.612 & -2.369 & -1.630 & -1.310 \\
NGC7314                  &  0.004 & S2 &  9.650 &  0.598 &  1.152 & -2.150 & -1.860 & -1.699 \\
NGC788                   &  0.014 & S2 &  9.710 &  0.152 &  0.273 & -1.934 & -1.916 & -0.571 \\
PKS2356-61               &  0.096 & S2 &  9.630 &  0.458 &  0.533 & -1.223 & -1.968 & -0.509 \\
ESO426-G002              &  0.022 & S2 &  9.400 &  0.225 &  0.514 & -1.593 & -1.545 & -0.737 \\
\hline\hline
\end{tabular}
\end{center}
\label{tab:sample}
\end{table*}

We performed a comprehensive comparison of all synthetic SEDs with observational data. For this purpose, we used the Spitzer/IRS spectra from the sample of 68  local ($z \leq 0.1$) AGNs used by \citet{Gonzalez-Martin2023}. According to NASA/IPAC Extragalactic Database (NED\footnote{\url{https://ned.ipac.caltech.edu/}}), 43 AGNs in this sample are classified as Type 1 and 25 as Type 2. Additionally, in agreement with \citet{Gonzalez-Martin2023}, we determined that 49 AGNs exhibit a $\rm{10 \ \mu m}$ silicate feature in emission, while 19 show it in absorption. The 49 spectra of AGNs showing $\rm{10 \ \mu m}$ silicate features in emission were previously analysed in \citet{Reyes-Amador2024} to obtain the dust chemical composition we adopt in this work (see Section~\ref{sec:the_dust_chemical_composition}). 

Section~3.1 of that work explains the technique used to remove the emission lines emitted by gas within the wavelength range of the spectra, which was used in this work as well to remove them from the spectra of the 19 AGN that exhibit a $\rm{10 \ \mu m}$ silicate feature in absorption. Using spectra without emission lines, both the continuum and the silicate features can be properly measured and analysed. In Section~\ref{sec:spectra_properties_measurement} we perform this analysis. Additionally, it is important to mention that the spectra of each source on this sample is dominated ($> 80\%$ of the MIR flux) by the AGN emission. Table~\ref{tab:sample} presents relevant information on the sample and contains the results of our analysis of the silicate features.

\section{Model grid analysis methods} \label{sec:analysis_models}

In this section, we describe the techniques that we used to analyse the grid of models obtained from the RT simulations. We performed a model-to-model comparison focused on the effects of $\ffill$ and $\fcl$ from the dust density distribution and the dust chemical composition (see Section~\ref{sec:models_vs_models}). Additionally, we compared the model SEDs with observed spectra through chi-square analysis and through the parametrisation of the spectral shape and silicate features (see Section~\ref{sec:model_vs_observations}). 

\subsection{The model-to-model comparison} \label{sec:models_vs_models}

For the first time, we are in a position that allows us to properly perform a direct comparison between models constructed with both different dust density distributions and dust chemical compositions while keeping all other parameters identical. This approach allows us to unambiguously determine how different dust density distributions and dust chemical compositions affect the resulting SEDs and to identify, characterize, and quantify these differences. We selected each synthetic SED from the library and compared it with its counterparts that share the same parameter values, only differing in the parameter to explore the differences ($\ffill$, $\fcl$, or dust composition). In this work, we explored:
\begin{enumerate}
    \item Differences produced by the dusty density distribution, which is divided into two parts:
    \begin{enumerate}
    \item Smooth models are taken as a reference, comparing them with their clumpy and two-phase counterparts. 
    \item Clumpy models are taken as a reference, comparing them with their smooth and two-phase counterparts.
    \end{enumerate}
    \item Differences produced by the dust chemical composition. 
\end{enumerate}

\noindent The differences between models were quantified globally using a mean absolute relative error formula, which we call averaged-global residual ($\globalres$), defined as: 

\begin{equation}
        \globalres = \frac{1}{N} \mathlarger{\mathlarger{\sum}} \left| \frac{S^{\lambda}_{\text{reference}}-S^{\lambda}_{\text{counterpart}}}{S^{\lambda}_{\text{reference}}} \right|, \label{eq:ave_glob_res} \\
\end{equation}

where $S^{\lambda}_{\text{reference}}$ are the referenced SEDs from smooth, clumpy, and \citet{Reyes-Amador2024} dust composition in the comparisons (ia), (ib), and (ii), respectively. $S^{\lambda}_{\text{counterpart}}$ represents their clumpy and two-phase counterparts, smooth and two-phase counterparts, and ISM dust composition with large-grain counterparts, respectively. Finally, $N$ is the number of wavelength points in the SEDs. $\globalres$ was computed within two wavelength ranges: 3-1000\,$\mu$m, which allows us to perform a comparison of the overall average SED; and 5-32\,$\mu$m, which is the same wavelength range for the MIR observed spectra.

To investigate differences at specific wavelengths, we calculated residuals again on a one-to-one model basis analysing the points (i,a), (i.b), and (ii), as well. The equation for the residuals ($\res$) is defined as:
\begin{equation}
        \res = \frac{S^{\lambda}_{\text{reference}}- S^{\lambda}_{\text{counterpart}}}{S^{\lambda}_{\text{reference}}}, \label{eq:residuals} \\
\end{equation}



The analysis of $\res$ was performed within the 5-32\,$\mu$m range. In the comparison (i,a), Equations~\ref{eq:ave_glob_res} and~\ref{eq:residuals} were applied to every smooth-model SED and its corresponding clumpy ($\fcl=1$) and two-phase ($\fcl=0.25,0.5,0.9,0.97$) counterparts for each value of the filling factor ($\ffill$), only for the SEDs with the \citet{Reyes-Amador2024} dust composition. Since we explored three values of $\ffill$ (see Table~\ref{tab:model_parameters}), each smooth model SED has 15 counterparts. This resulted in 15 sets of $\globalres$ and $\res$, each corresponding to a specific combination of $\fcl$ and $\ffill$. This comparison should be undertaken for each value of the line of sight angle under which the torus is viewed. To keep the analysis simple, we focussed on differences for model SEDs viewed only at $i=0^{\circ}$ (face-on) and $90^{\circ}$ (edge-on), as these orientations are expected to show the most significant differences and are also representative of both Type-1 and Type-2 AGNs, respectively. Consequently, we analysed 30 subsets of $\globalres$ and $\res$ when smooth model SEDs are the reference (i,a). 

The same was done in the comparison (i,b), where clumpy model SEDs are used as the reference. In this case, each clumpy model SED with a given $\ffill$ value has 13 counterparts (one smooth and 12 two-phase with different combinations of $\fcl$ and $\ffill$). Therefore, for the three $\ffill$ values of clumpy model SEDs and the two viewing angles (face-on and edge-on), we have 78 subsets of $\globalres$ and $\res$ when clumpy model SEDs are the reference (i,b). In the comparison (ii), each SED with the \citet{Reyes-Amador2024} dust composition has one counterpart with the ISM dust composition. As we analysed their differences as a function of the dust density distribution, we obtained 32 subsets of $\globalres$ and $\res$, one smooth and 15 clumpy/two-phase (three values of $\ffill$ per each $\fcl$ value), and one per each viewing angle (face-on and edge-on).  
\begin{figure*}
    \centering
    \begin{subfigure}{0.45\textwidth}
        \centering
        \includegraphics[trim=0.9cm 0 0 0, clip,scale=0.45]{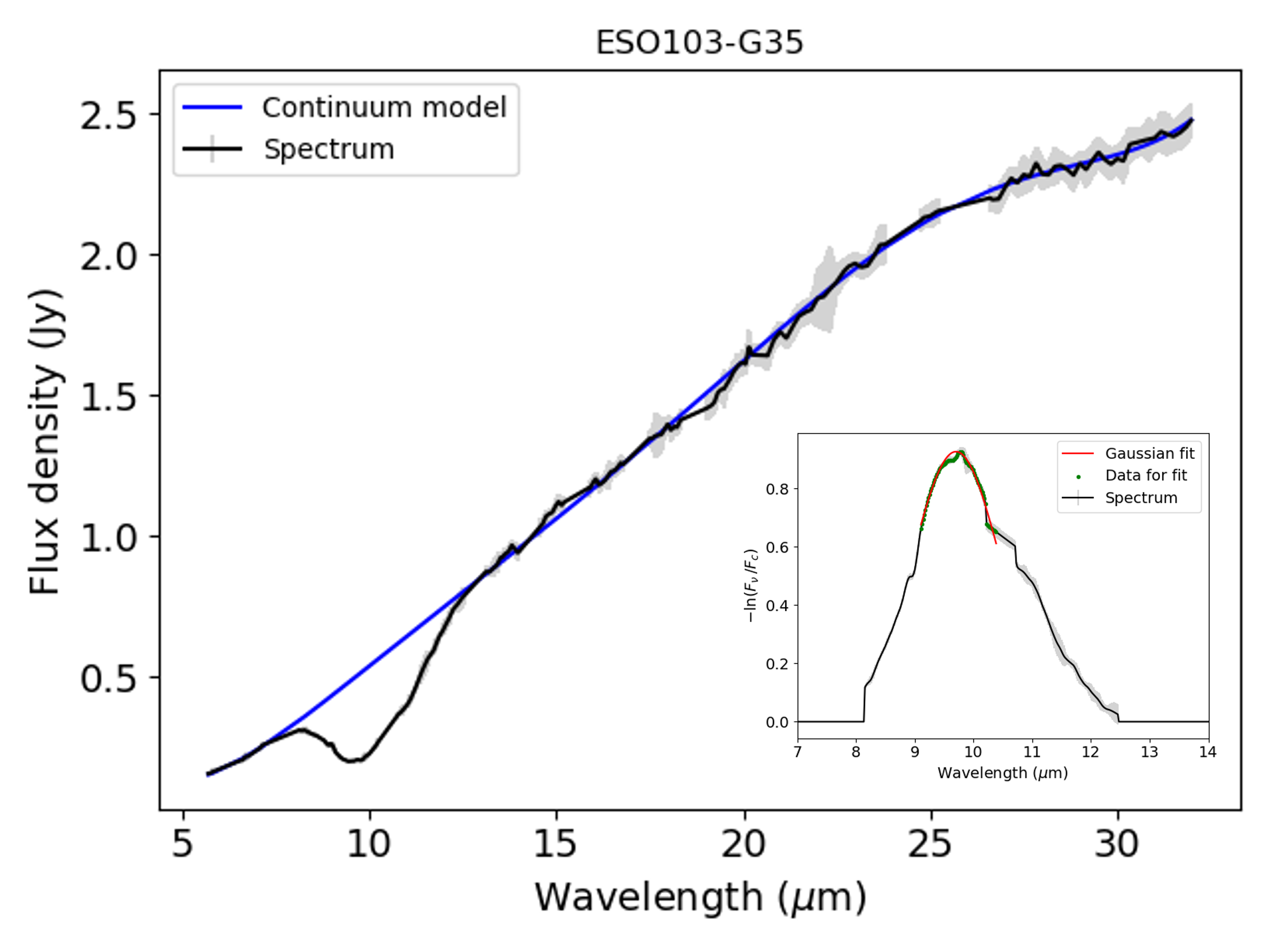}
    \end{subfigure}
    \begin{subfigure}{0.45\textwidth}
        \centering
        \includegraphics[trim=0.9cm 0 0 0, clip, scale=0.45]{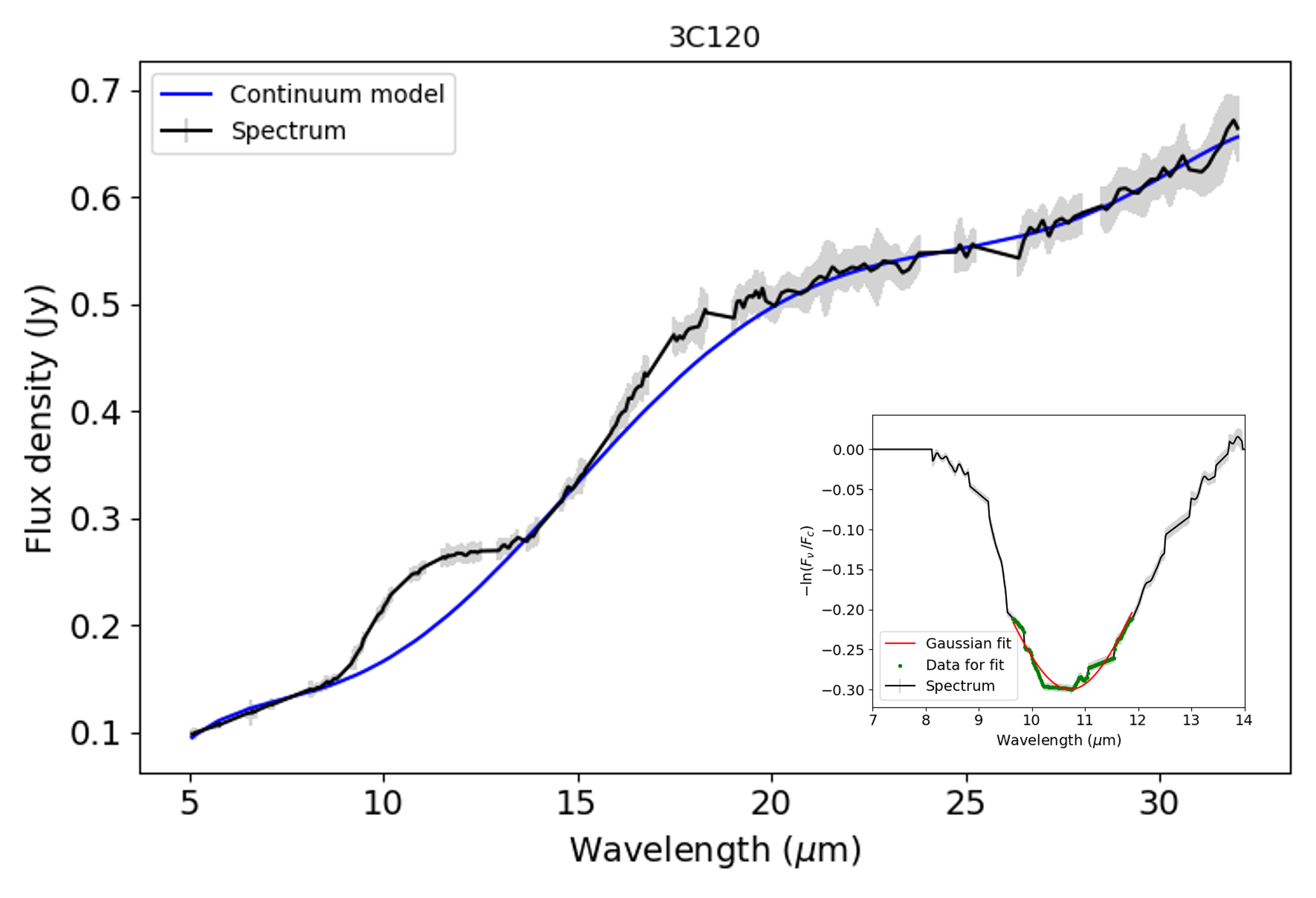}    \end{subfigure}
    \caption{Examples of the fitting procedure of the underlying continuum (blue line) used to fit the spectra (black line) and measure the spectral slopes and $\rm{10 \ \mu m}$ silicate feature properties. The left panel illustrates the case for the spectrum of ESO~103-G35, which shows a $\rm{10 \ \mu m}$ silicate feature in absorption. The right panel presents the case for the spectrum of 3C~120, which shows a $\rm{10 \ \mu m}$ silicate feature in emission. The inset figures in both panels show the negative of the natural logarithm of the continuum-divided spectra (black line) zoomed in the $\rm{10 \ \mu m}$ silicate feature. The green points are the data to which the Gaussian was fitted and then used to determine $\peak$ and $\strength$. \textbf{Note:} In the inset figures, the absorption (emission) feature appears as emission (absorption) because the definition of $\strength$ (see Eq.~\ref{eq:strength_silicate_feature}) is plotted in the y-axis.}
    \label{fig:fitting_underlying_continuum}
\end{figure*}

\subsection{Model SEDs versus observed spectra} \label{sec:model_vs_observations}

We analysed all the synthetic SEDs by performing a chi-square analysis on the spectra from the sample observed with Spitzer. The objective of this analysis is to identify synthetic SEDs with shapes that significantly differ from the observed spectra, hence identifying regions of the parameter space that are not worth including in the model grid. For this purpose, we focused on the wavelength range of the SEDs that overlaps with the observed spectra, namely $\rm{5-32\ \mu m}$. The chi-square per data point ($\chisquare$) was calculated using the following equation:

\begin{equation} \label{eq:reduced_chi_square}
    \chisquare = \frac{1}{N} \mathlarger{\mathlarger{\sum}} \left(\frac{S^{\lambda}_{\text{data}}- C\cdot S^{\lambda}_{\text{model}}}{\sigma^{\lambda}_{\text{data}}}\right)^2,
\end{equation}
where $S^{\lambda}_{\text{data}}$ is the observed spectrum, $\sigma^{\lambda}_{\text{data}}$ are their flux errors, $N$ is the number of wavelength points in the spectrum, $S^{\lambda}_{\text{model}}$ is the synthetic SED interpolated to match the wavelength of the observed spectrum, also ensuring the same number of data points, and $C$ is a scale factor that multiplies $S^{\lambda}_{\text{model}}$ to minimize the sum of squared differences and align the flux of the model with the observed spectrum as closely as possible. This scale factor was calculated using the following equation:

\begin{equation} \label{eq:scale_factor}
    C = \frac{\mathlarger{\sum} \left(S^{\lambda}_{\text{data}} \cdot S^{\lambda}_{\text{model}}\right)}{\mathlarger{\sum} \left(S^{\lambda}_{\text{model}}\right)^2}.
\end{equation}

\noindent Equations~\ref{eq:reduced_chi_square} and~\ref{eq:scale_factor} were computed for each model-observation pair, resulting in 68 $\chisquare$ values per synthetic SED. Among these, the minimum value of $\chisquare$ ($\chisquaremin$) was identified as the best fit and evaluated whether it represents a good fit. After dividing the global distribution into Type-1 and Type-2 SEDs, we investigated the similarity of the $\chisquaremin$ distribution of each of these subsets with the combined (Type-1 and Type-2) distribution. We quantified the similarity of the distributions using the Kullback-Leibler Divergence \footnote{To calculate this quantity, we used the \texttt{scipy.stats.entropy} function \url{https://docs.scipy.org/doc/scipy/reference/generated/scipy.stats.entropy.html}.} \citep{Kullback-Leibler}. We did not employ the Kolmogorov-Smirnov or Anderson-Darling tests, as these are only valid for uncorrelated samples. We also applied KL to distributions of $\chisquaremin$ obtained from models focusing on each value of the parameter space to examine the suitability of each parameter for reproducing the observations. Finally, for model SEDs with $\chisquaremin<2$ (considered as good fits), we analysed how the various dust density distributions and the two dust chemical compositions are able to produce good fits, also discerning Type-1 and Type-2 SEDs. 

\subsection{Spectral properties measurement} \label{sec:spectra_properties_measurement}

The MIR spectrum of an AGN can be characterized by quantities that provide insights into both its overall shape (through its slope) and spectral characteristics determined by dust properties, such as the two silicate absorption/emission features at 9.7 and 18\,$\mu$m. To describe the spectral shape, we calculated three spectral slopes that are defined by two wavelengths, $\lambda_1$ and $\lambda_2$, were $\lambda_1 < \lambda_2$, and the corresponding flux densities $F_{\nu}(\lambda_1)$ and $F_{\nu}(\lambda_2)$. To compute these quantities, we applied the methodology outlined in \citet{Gonzalez-Martin2023} to both the observed spectra and synthetic SEDs within the $\rm{5 - 32 \, \mu m}$, enabling a direct comparison and allowing us to check whether the models cover a spectral parameter space similar to that of the observations. We use the following definition for spectral slopes:
\begin{equation} \label{eq:spectral_slopes}
    \alpha_{\lambda_1-\lambda_2} = -\frac{\log(F_{\nu}(\lambda_1)/F_{\nu}(\lambda_2))}{\log(\lambda_1/\lambda_2)}.
\end{equation}

\noindent We selected the same slopes as in \citet{Gonzalez-Martin2019I,Gonzalez-Martin2023}: $\alphanir$, $\alphamir$, and $\alphafir$, calculated consistently for both observed spectra and synthetic SEDs. It is important to note that, using the prescription of Eq.~\ref{eq:spectral_slopes}, negative (positive) values are found when the flux increases (decreases) with wavelength.  

Additionally, we characterized the $\rm{10 \ \mu m}$ silicate feature (both in emission and absorption) by determining its peak wavelength, strength, and equivalent width. In order to do this, we estimated the continuum under this feature as follows. For the observed spectra, we used a sixth-order polynomial fitted to visually selected wavelength regions, excluding the silicate features. For the synthetic SEDs, which exhibit diverse spectral shapes, we fitted an $n$-th order polynomial, where $2\leq n \leq 8$. This broader range of polynomial orders was necessary because the synthetic SEDs could not be uniformly fitted with a sixth-order polynomial as the observed spectra were. Due to the large number of synthetic SEDs (344~736 in total), it was not feasible to manually select the fitting regions for each SED. Instead, we defined specific wavelength ranges: $\rm{\lambda < 7 \ \mu m}$, $\rm{13 < \lambda < 14 \ \mu m}$, and $\rm{\lambda > 24 \ \mu m}$. The best-fitting polynomial order for each SED was identified using an iterative F-test\footnote{In the absence of random errors, the application of the F-test in this case does not have a true statistical significance; however, it is a convenient way to choose the lowest-order polynomial that gives a good fit.}. After dividing the observed spectra (or synthetic SEDs) by these continuum estimates, we determined the flux density ($F_{\nu}$) and wavelength ($\peak$) of the maximum (minimum) of the emission (absorption) feature, and the flux density of the continuum ($F_{\nu,\text{cont}}$) at that wavelength by fitting a Gaussian profile to the core of the feature. Then, we computed the feature strength, $\strength$, as follows:

\begin{equation} \label{eq:strength_silicate_feature}
    \text{S}_{10 \mu m} = -\ln\left(\frac{F_{\nu}}{F_{\nu,\text{cont}}}\right) ,       
\end{equation}

\noindent Figure~\ref{fig:fitting_underlying_continuum} illustrates the fitting procedure for two objects in the sample: ESO~103-G35 and 3C~120, which exhibit the $\rm{10 \ \mu m}$ silicate feature in absorption and emission, respectively. The equivalent width, $\ewidth$, was determined using:

\begin{equation} \label{eq:EW_silicate_feature}
    \text{EW}_{10 \mu m} = \int_{\lambda_1}^{\lambda_2} \frac{F_{\nu,\text{cont}}-F_{\nu}}{F_{\nu,\text{cont}}}\,d\lambda .    
\end{equation}

where the integration range $[\lambda_1,\lambda_2]$ was defined at $\rm{[7,14] \ \mu m}$ for the calculations of both synthetic SEDs and observed spectra, following previous works \citep[e.g.][]{Nenkova2008b,Gonzalez-Martin2019II,Gonzalez-Martin2023}. It is important to highlight that the sign convention in Equations~\ref{eq:strength_silicate_feature} and~\ref{eq:EW_silicate_feature} assigns positive values to absorption features and negative values to emission features for $\strength$ and $\ewidth$ and that all calculations described in this section were systematically performed in $F_{\nu}$ units (Jansky) for both the observed spectra and synthetic SEDs.

\section{Results} \label{sec:results}

\begin{figure*}
    \centering
    \begin{subfigure}{\textwidth}
        \centering
        \includegraphics[trim=0cm 0 0 0, clip,scale=0.45]{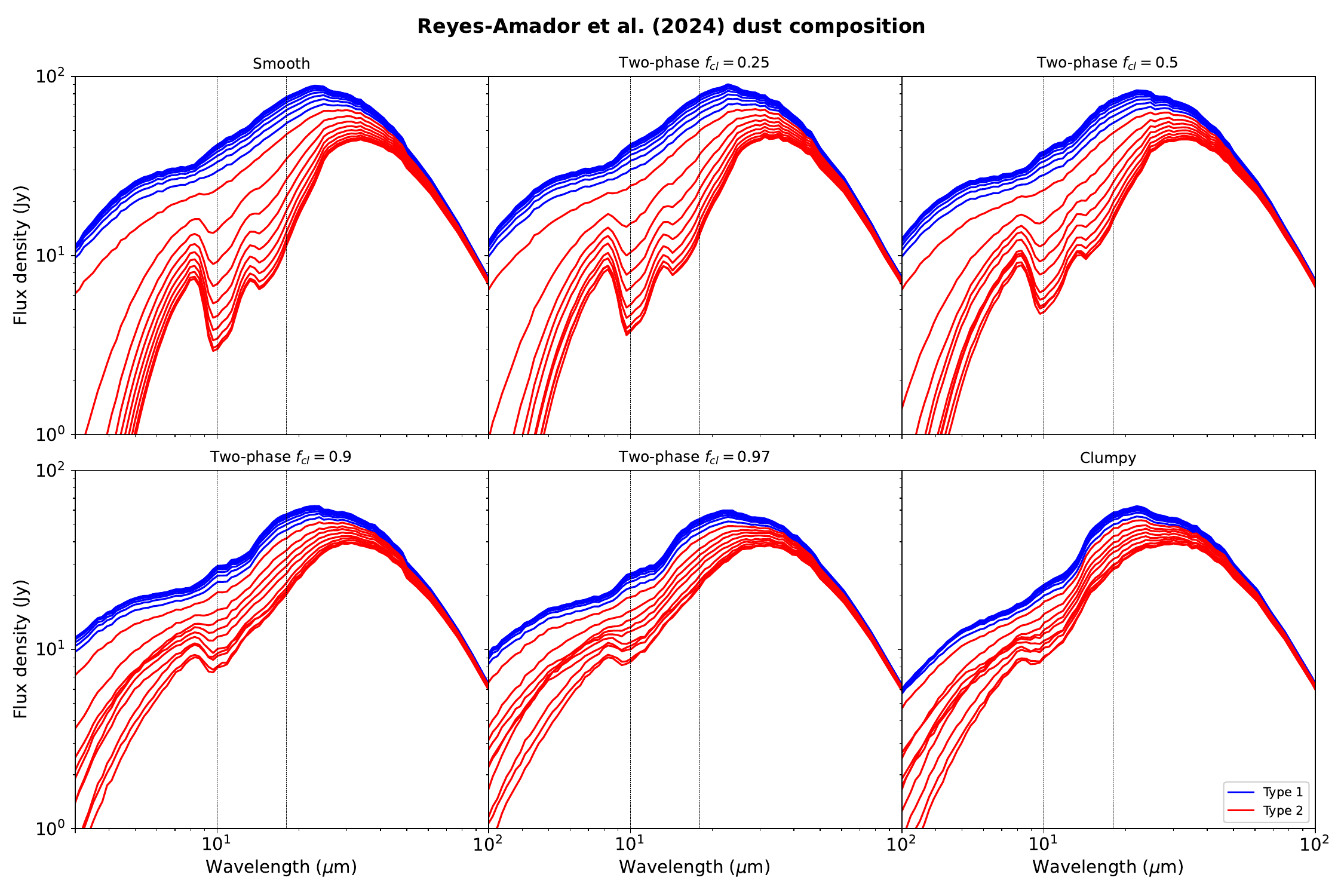}
    \end{subfigure}
    \begin{subfigure}{\textwidth}
        \centering
        \includegraphics[trim=0cm 0 0 0, clip,scale=0.45]{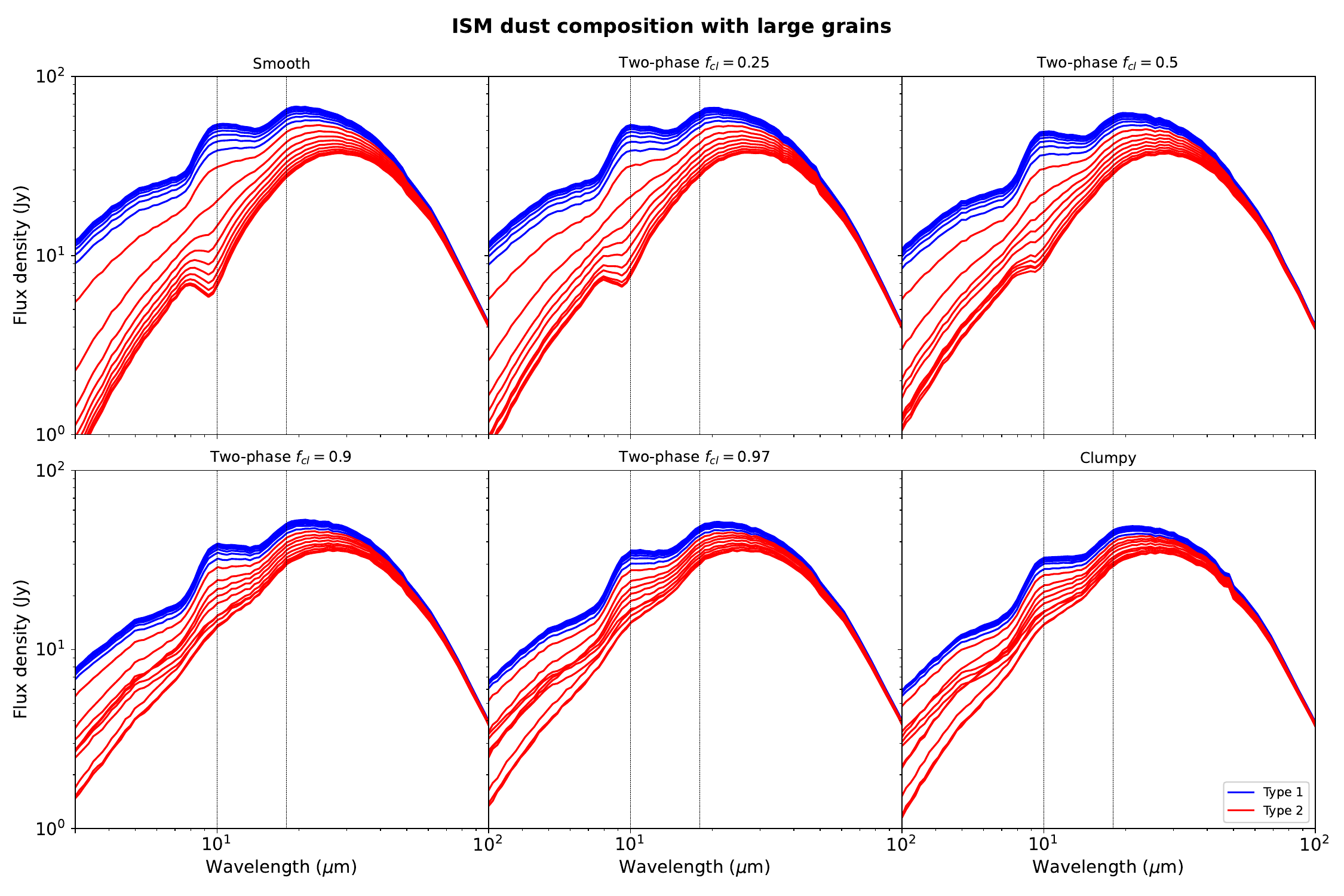}   
    \end{subfigure}
    \caption{Synthetic SEDs in the 3-100\,$\mu$m range, obtained from the models with comparable parameter values and the six clumpiness values using the \citet{Reyes-Amador2024} dust composition (top panels) and the ISM dust composition (bottom panels). For all the models, the parameter values are $\ptorus=0.75$, $\qtorus=0.75$, $\OAtorus=45^\circ$, $Y=20$, $\opttau=9$. For the two-phase and clumpy models, $\ffill=0.15$. Each panel shows the Type-1 (blue) and Type-2 (red) SEDs, corresponding to $i<45^{\circ}$ and $i\geq45^{\circ}$, respectively. The vertical dotted lines indicate $\rm{\lambda = 10 \ \mu m}$ and $\rm{18 \ \mu m}$.}
    \label{fig:example_SEDs}
\end{figure*}

\begin{figure*}
    \centering
    \begin{subfigure}{\textwidth}
        \centering
        \includegraphics[trim=0cm 0 0 0, clip,scale=0.45]{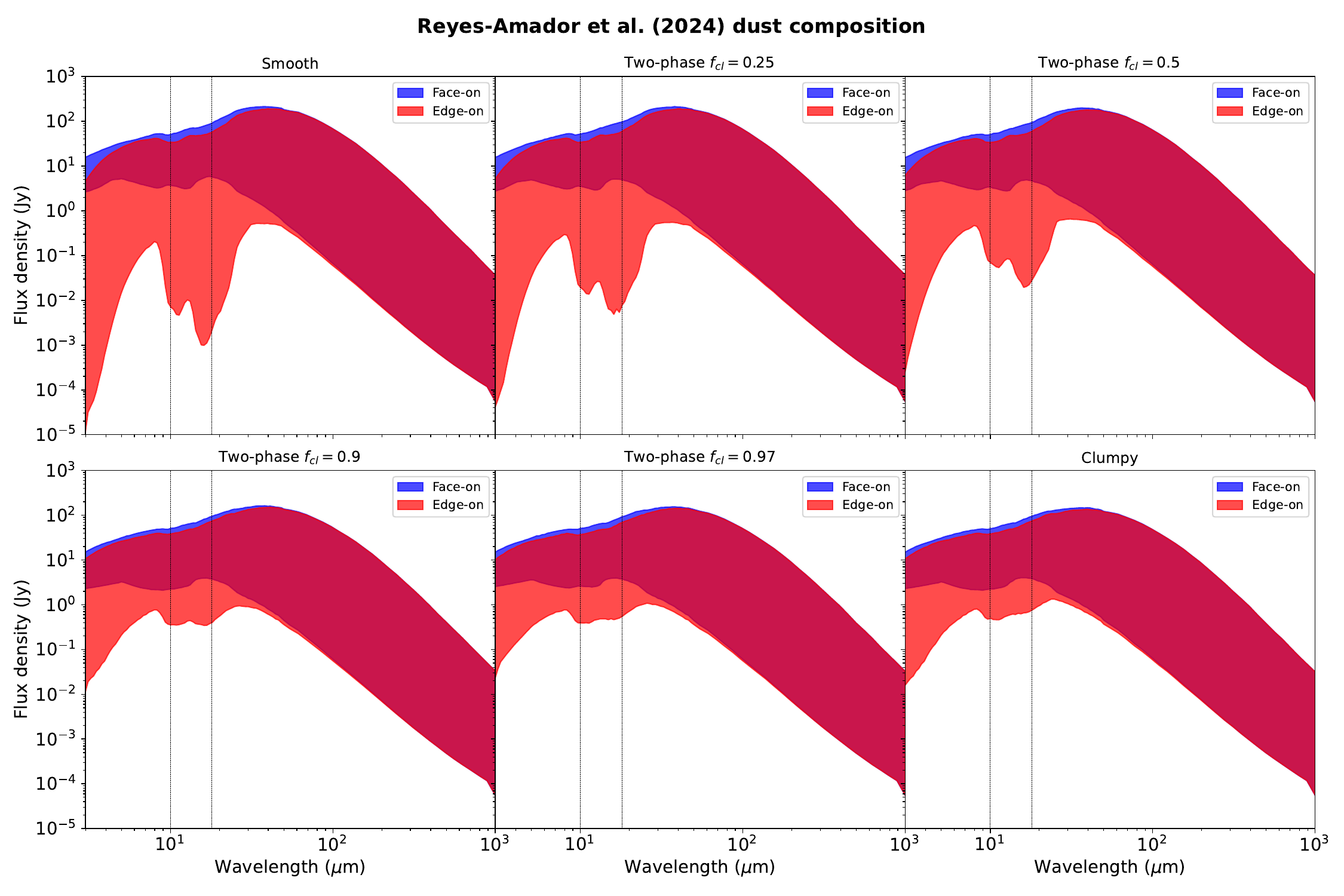}
    \end{subfigure}
    \begin{subfigure}{\textwidth}
        \centering
        \includegraphics[trim=0cm 0 0 0, clip,scale=0.45]{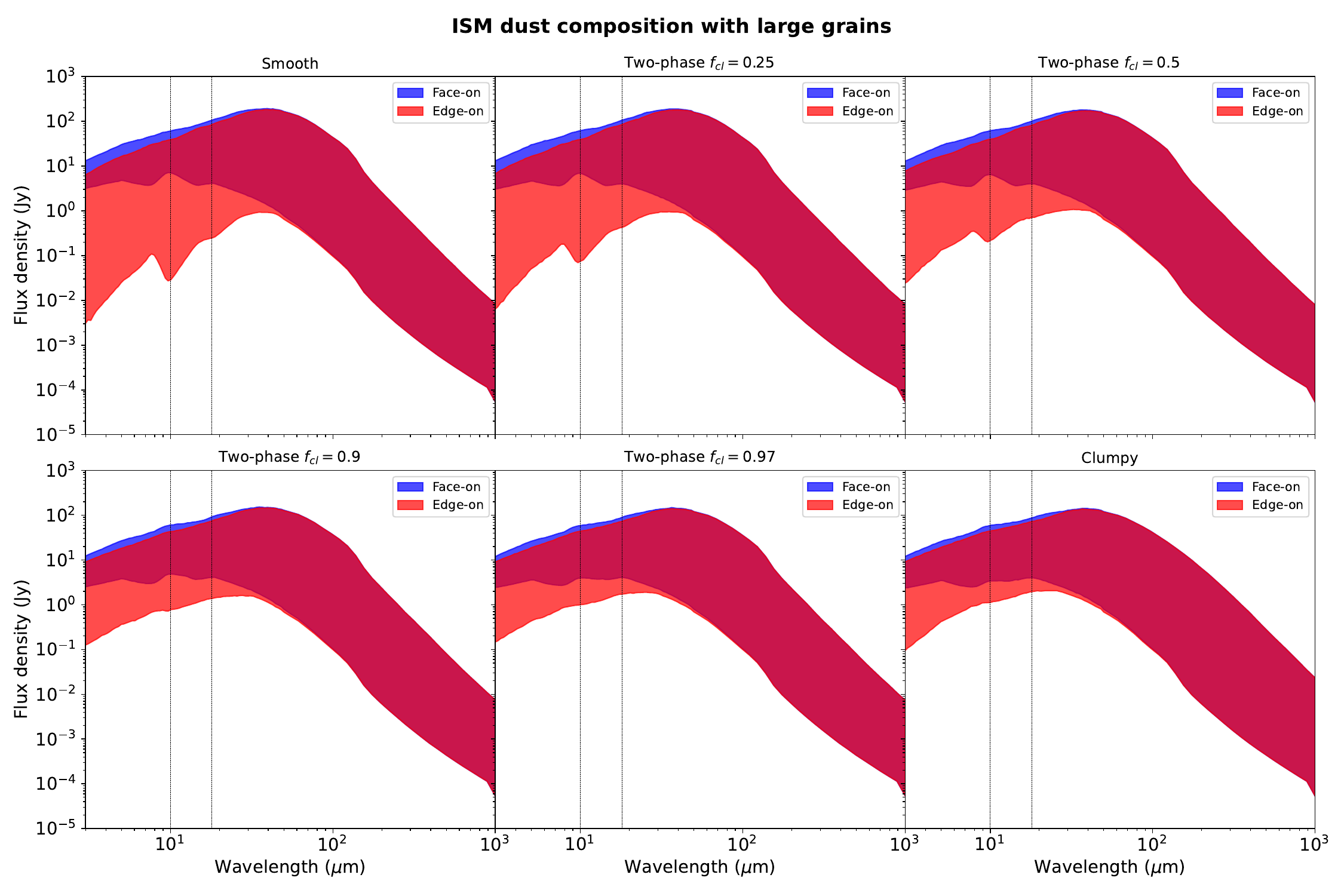}   
    \end{subfigure}
    \caption{The range of SEDs covered by the models with the six clumpiness values using the \citet{Reyes-Amador2024} dust composition (top panels) and the ISM dust composition (bottom panels). The SED coverage is shown for face-on (blue) and edge-on (red) views. The vertical dotted lines indicate $\rm{\lambda = 10 \ \mu m}$ and $\rm{18 \ \mu m}$.}
    \label{fig:coverage_SEDs}
\end{figure*}

\subsection{The effects of the dust density distribution} \label{sec:result_density_distribution}

Figure~\ref{fig:example_SEDs} shows the synthetic SEDs in the 3-100\,$\mu$m range (where differences are observed) for an example model with given parameters. The synthetic SEDs are presented at the 19 values of $i$ ($\rm{\Delta i = 5^{\circ}}$), obtained from the six clumpiness values and two dust compositions. For the \citet{Reyes-Amador2024} dust composition (six top panels of that figure), we can see that, at wavelengths $\rm{\lambda < 7 \ \mu m}$, for Type-2 SEDs (those with $i\geq45^{\circ}$ for this specific model, see Eq.~\ref{eq:type1_type2_SEDs}), the flux is lower for smooth models compared to clumpy and it increases getting a flatter slope as the $\fcl$ increases, while for Type-1 SEDs (those with $i<45^{\circ}$ for this specific model), the flux slightly decreases only for clumpy models and two-phase with $\fcl=0.97$. At wavelengths around the $\rm{10 \ \mu m}$ and $\rm{18 \ \mu m}$ silicate features, no significant differences are visible for Type-1 SEDs when comparing among the clumpiness values, showing weak emission silicate features, except for the clumpy models, where the $\rm{18 \ \mu m}$ silicate feature shows a stronger emission. In contrast, for Type-2 SEDs, both silicate features are shown in absorption for smooth and two-phase models with $\fcl=0.25$ and~$\fcl=0.5$, becoming stronger with the largest values of $i$ and the lowest values of $\fcl$. For clumpy and two-phase models with $\fcl=0.9$ and~$\fcl=0.97$, the $\rm{18 \ \mu m}$ silicate feature disappears completely, and the $\rm{10 \ \mu m}$ silicate feature is even weaker than in the models with lower $\fcl$. Additionally, the Type-1 SEDs show a higher flux in the IR bump at $\rm{\lambda \sim 25 \ \mu m}$ for lower $\fcl$. 


It is important to emphasize that the characteristics described in the previous paragraphs about the Figure~\ref{fig:example_SEDs} are for a particular model with specific parameter values and their corresponding counterparts. 
Figure~\ref{fig:coverage_SEDs} shows the SED coverage (i.e., the ranges of flux covered by the complete library of SEDs at each wavelength) for the models with the six clumpiness values using both dust compositions and for face-on and edge-on views\footnote{Note that ``face on/edge on'' terminology is different to ``Type 1/Type 2'', since face-on and edge-on SEDs correspond only to $i=0^{\circ}$ and $i=90^{\circ}$, respectively, while Type-1 and Type-2 SEDs include inclinations given by Eq.~\ref{eq:type1_type2_SEDs}.} to maximise the contrast between the SED shapes. In the SED coverages using the \citet{Reyes-Amador2024} dust composition (six top panels of that figure), we note that, in the lower limit of smooth edge-on SEDs, both silicate features are observed to be strongly in absorption and they peak at wavelengths which are respectively shorter and longer as compared to their nominal values (namely 9.7 and 18\,$\mu$m). The intensity of this absorption on the silicate features decreases as the $\fcl$ increases. In the case of the lower limit of face-on SEDs for the models with the six clumpiness values, both silicate features can be observed in emission, the $\rm{18 \ \mu m}$ feature more intense than the $\rm{10 \ \mu m}$ and peaking to shorter wavelength than $\rm{18 \ \mu m}$. In the case of the upper limit, the silicate features are not visible in face-on SED coverage, while in edge-on SED coverage, they are visible in absorption only for smooth and two-phase models with $\fcl=0.25$ and $0.5$. Similar trends are found in the SED coverages of the ISM dust composition for the different clumpiness values. See Section~\ref{sec:result_dust_composition} for the details in the comparison between the dust compositions.

\begin{figure*}
    \centering
    \includegraphics[trim=1.5cm 3cm 0 4cm, clip,scale=0.7]{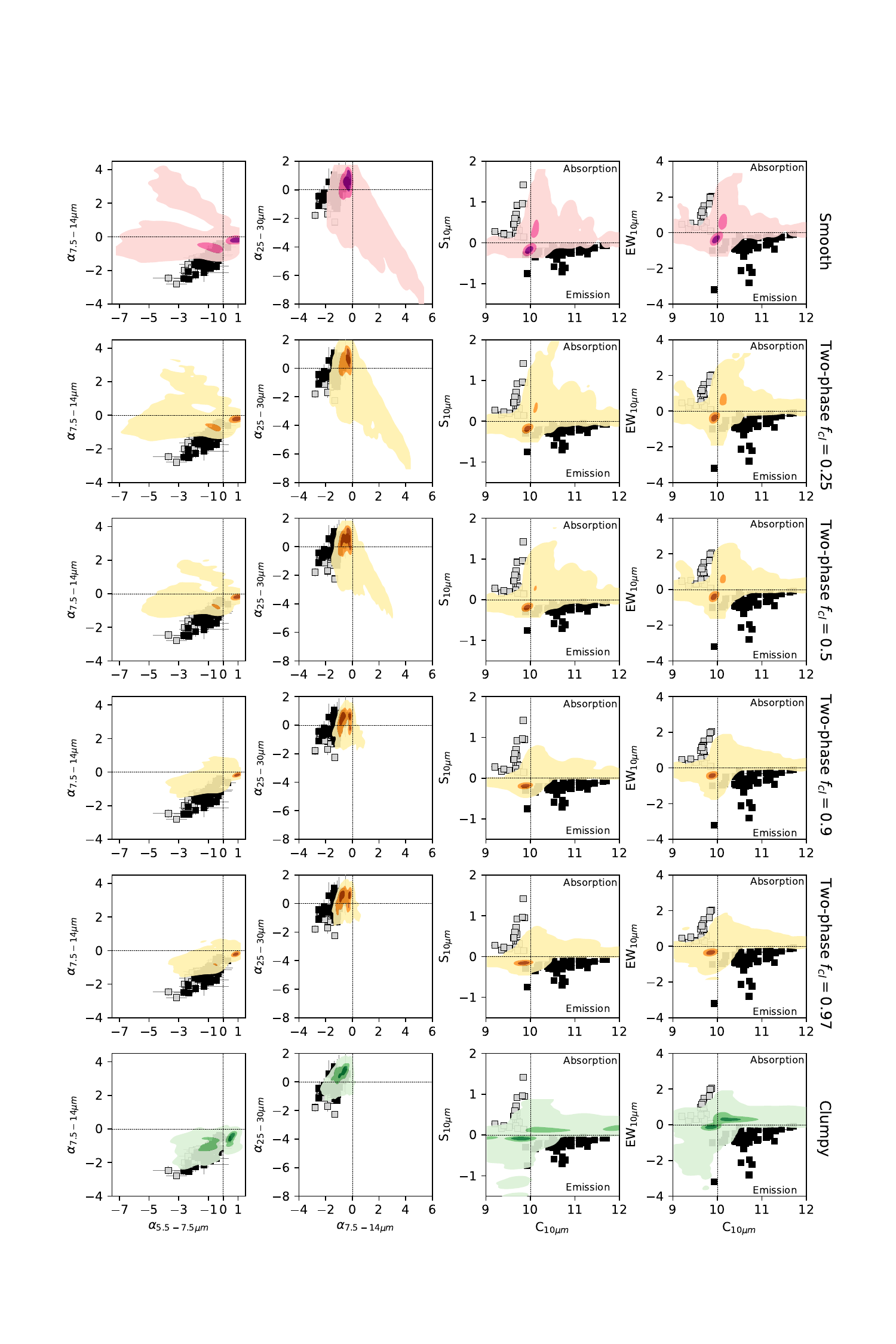}
    \caption{Density contours of the spectral measurements of the model SEDs (in colours) using the \citet{Reyes-Amador2024} dust composition and the observed spectra (black squares for AGN with silicate feature in emission and gray squares for AGN with silicate feature in absorption). \textbf{From left to right:} mid-infrared vs near-infrared slopes, far-infrared vs mid-infrared slopes, the $\rm{10 \ \mu m}$ feature silicate strength vs its peak wavelength, and the $\rm{10 \ \mu m}$ feature silicate equivalent width vs its peak wavelength. \textbf{From top to bottom:} smooth (pink), two-phase (yellow), and clumpy (green) models.}
    \label{fig:silicate_properties_slopes_all_models}
\end{figure*}

\begin{figure*}
    \centering
    \includegraphics[trim=1.5cm 3cm 0 4cm, clip,scale=0.7]{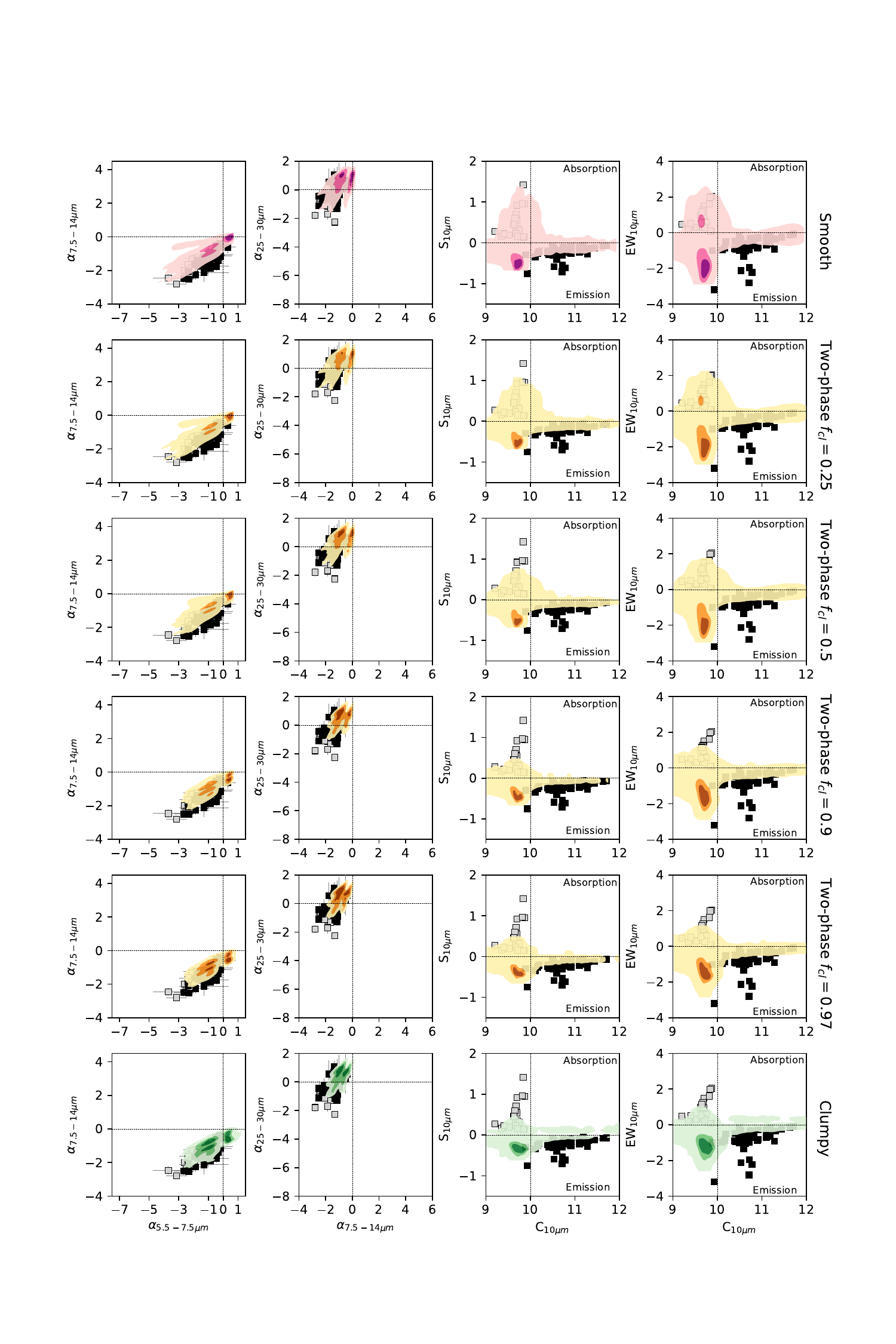}
    \caption{Same as Fig.~\ref{fig:silicate_properties_slopes_all_models} but for the models with the ISM dust composition.}
    \label{fig:silicate_properties_slopes_all_models_omaira}
\end{figure*}

Figures~\ref{fig:silicate_properties_slopes_all_models} and ~\ref{fig:silicate_properties_slopes_all_models_omaira} present the results of the silicate features and spectral shape characterization for both the synthetic SEDs and the observed spectra. Columns\,1 and 2 of those figures reveal that for both dust compositions, the range of values for the three spectral slopes ($\alphanir$, $\alphamir$, and $\alphafir$) from the synthetic SEDs becomes narrower as $\fcl$ increases, but their ranges are always broader than the ones of the observed spectra, particularly for models with $\fcl \leq 0.5$ (smooth and two-phase with $\fcl=0.25$ and~$\fcl=0.5$). With the set of parameter values here explored, the clumpy dust density distribution reproduces the spectral shape measurements of the observed spectra more accurately, $\sim 93\%$ and $\sim87 \%$ of the objects in the $\alphanir$ versus $\alphamir$ and $\alphamir$ versus $\alphafir$, respectively. As $\fcl$ decreases, $\alphanir$ and $\alphafir$ become increasingly negative (indicating a steeper upward slope), while $\alphamir$ shows a different trend with $\fcl$ according to the dust composition. For the \citet{Reyes-Amador2024} dust composition, $\alphamir$ becomes increasingly positive (indicating a steeper downward slope), creating ``tails'' in the distributions shown in the top-left panels of Figure~\ref{fig:silicate_properties_slopes_all_models}, while in the ISM dust composition, it does not show significant changes with $\fcl$. 

In examining the $\rm{10 \ \mu m}$ silicate feature properties in Fig.~\ref{fig:silicate_properties_slopes_all_models} (cols. 3 and 4), we found that models with $\fcl \leq 0.5$ tend to produce stronger and broader absorption features, but they produce weaker and narrower emission features. Another notable effect of the dust density distribution is observed in synthetic SEDs with the ISM dust composition (Fig.~\ref{fig:silicate_properties_slopes_all_models_omaira}, cols. 3 and 4). In this case, models with $\fcl \leq 0.5$ reproduce a larger amount ($\gtrsim 65\%$) of observed silicate properties, both in absorption and emission, compared to models with $\fcl \geq 0.9$. Conversely, synthetic SEDs using the \citet{Reyes-Amador2024} dust composition do not show significant differences in their ability to reproduce the observed silicate properties (most of them produce $\sim 30\%$) based on the dust density distribution, except by models with $\fcl = 0.9$ and $\fcl = 0.97$, which produce the lowest percentage ($\sim 15\%$) of observed silicate properties.

\begin{figure*}
    \centering
    \begin{subfigure}{0.45\textwidth}
        \centering
        \includegraphics[trim=1.3cm 0 0 0, clip,scale=0.35]{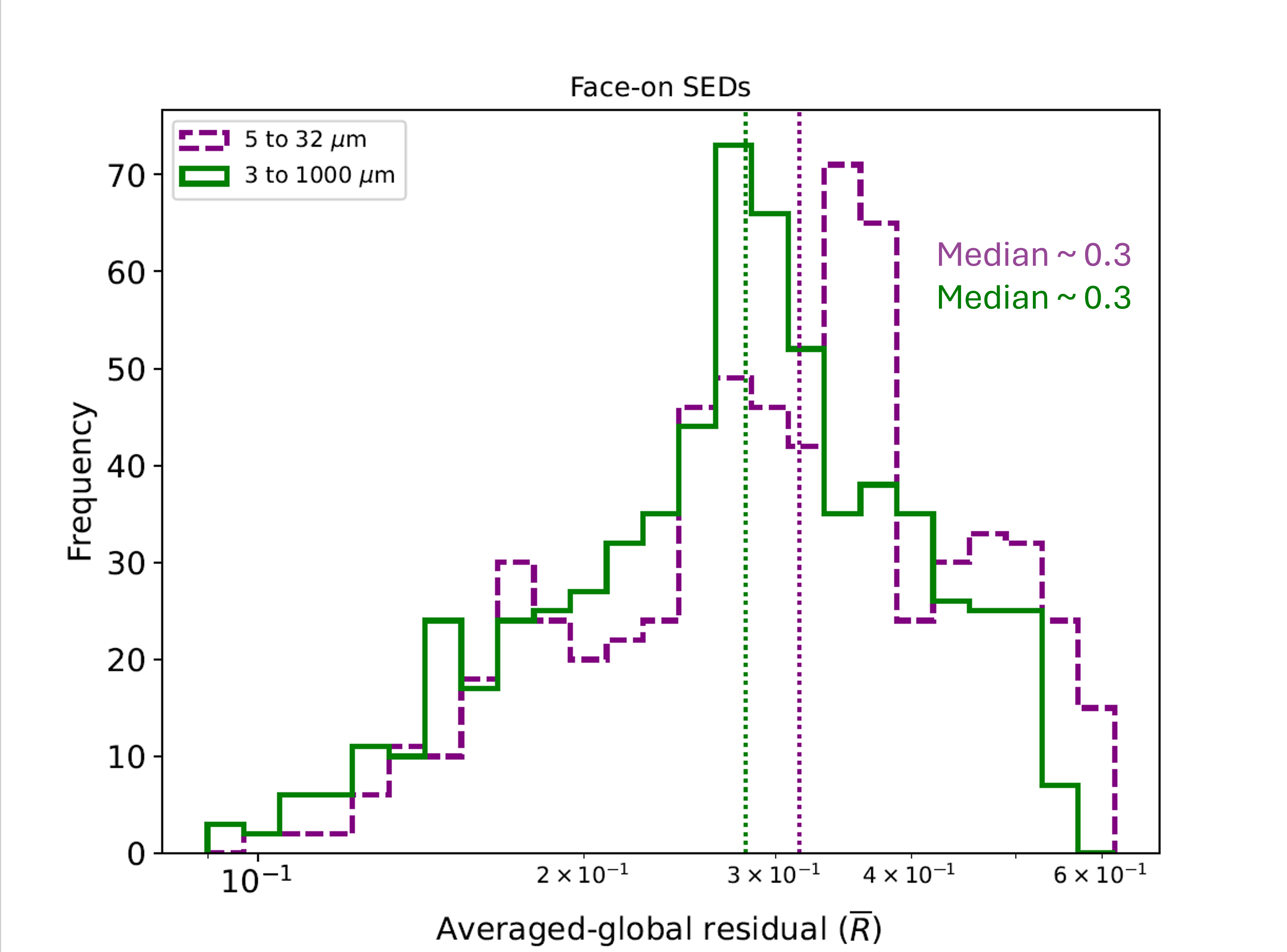}
    \end{subfigure}
    \begin{subfigure}{0.45\textwidth}
        \centering
        \includegraphics[trim=0.8cm 0 0 0, clip, scale=0.35]{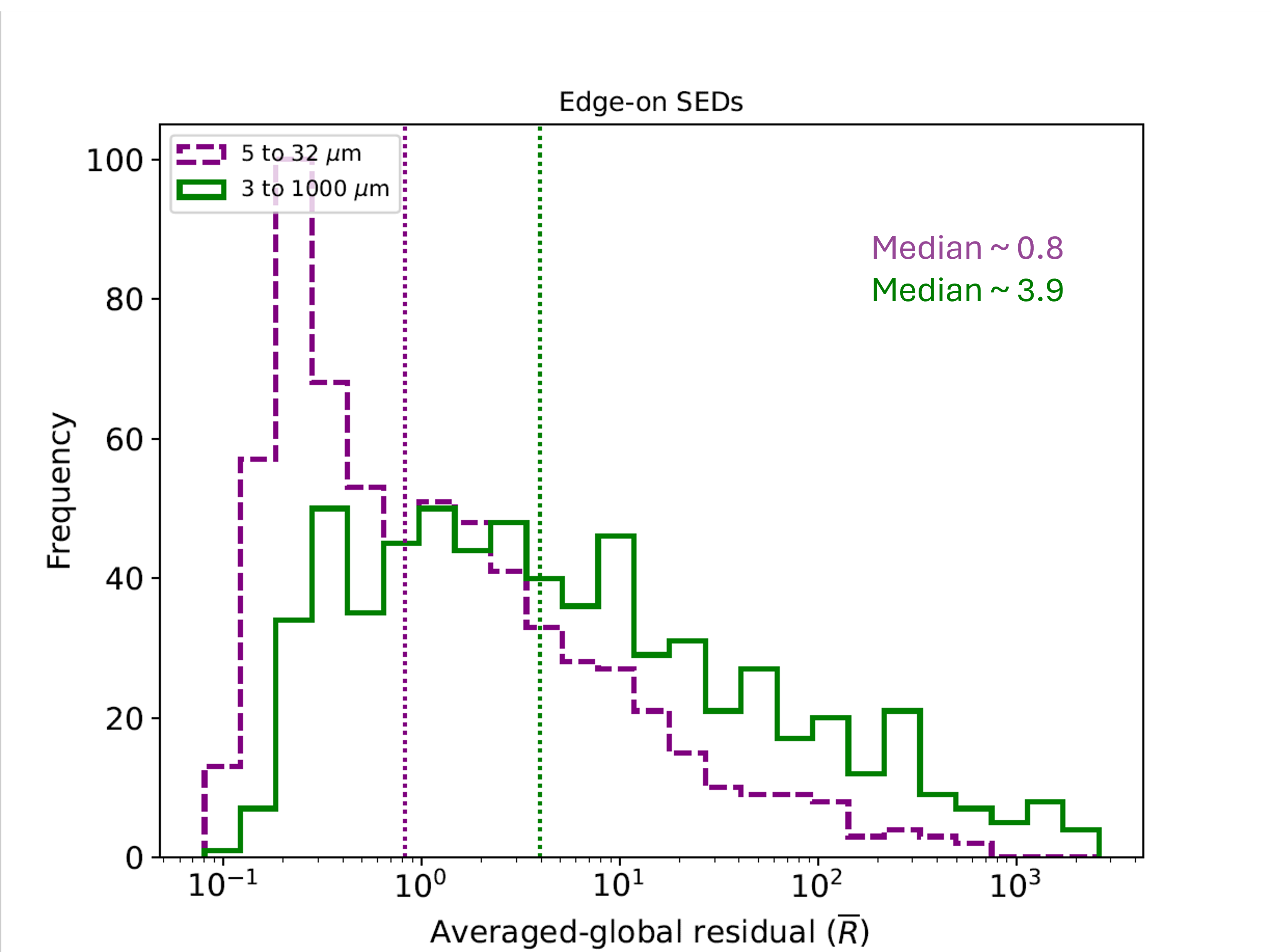}    
    \end{subfigure}
    \caption{Distributions of $\globalres$ obtained by comparing all the smooth model SEDs with their clumpy ($\fcl=1$) counterparts with $\ffill=0.1$. The dashed purple line histogram corresponds to the calculations of $\globalres$ considering a wavelength range in the SEDs of 5 to $\rm{32 \ \mu m}$, while the green line histogram considers 3 to $\rm{1000 \ \mu m}$. The vertical dotted lines correspond to the medians of each distribution, whose values are shown in the upper right corner. Left panel shows the values of $\globalres$ for SEDs with $i=0^{\circ}$ (face-on), while right panel those for $i=90^{\circ}$ (edge-on). }
    \label{fig:distributions_ave_glob_res}
\end{figure*}

\begin{figure*}
    \centering
    \includegraphics[trim=0cm 0 0 0, clip,scale=0.45]{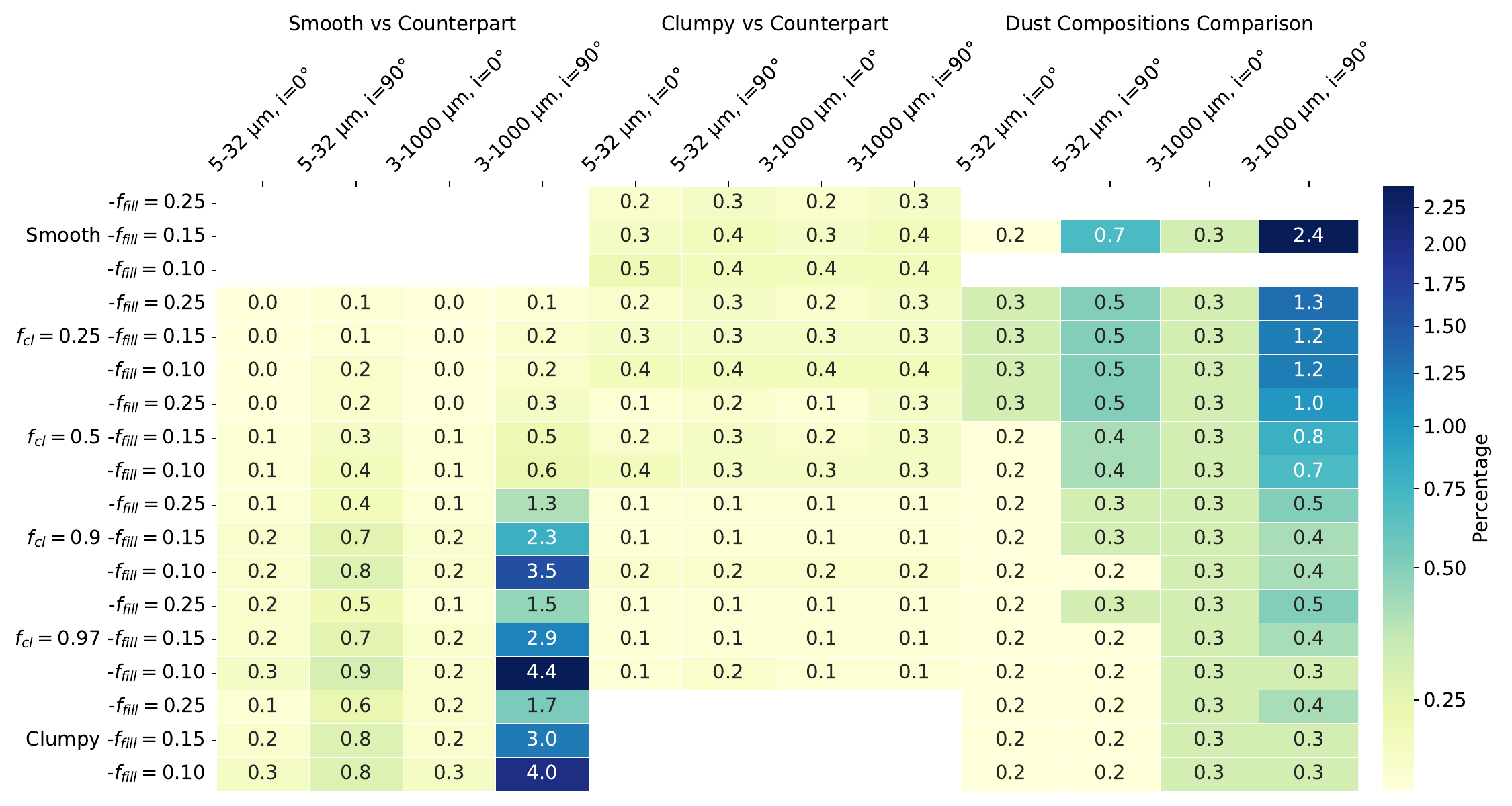}
    \caption{Colour map of the medians of $\globalres$ in the model-to-model comparison of smooth versus their counterparts, clumpy versus their counterparts, and \citet{Reyes-Amador2024} dust composition versus ISM dust composition.}
    \label{fig:distributions_medians_glob_res}
\end{figure*}

\begin{figure*}
    \centering
    \begin{subfigure}{0.49\textwidth}
        \centering
        \includegraphics[trim=0 0 55 0, clip,scale=0.48]{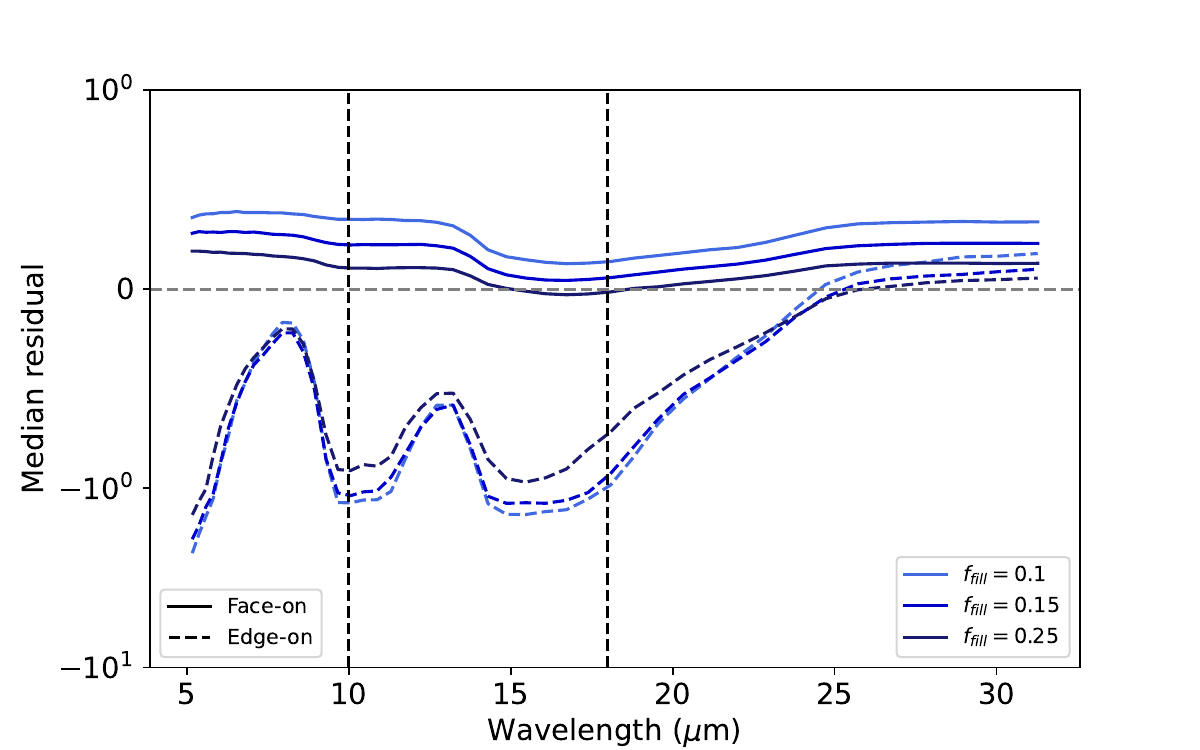}
    \end{subfigure}
    \begin{subfigure}{0.49\textwidth}
        \centering
        \includegraphics[trim=65 0 0 0, clip, scale=0.48]{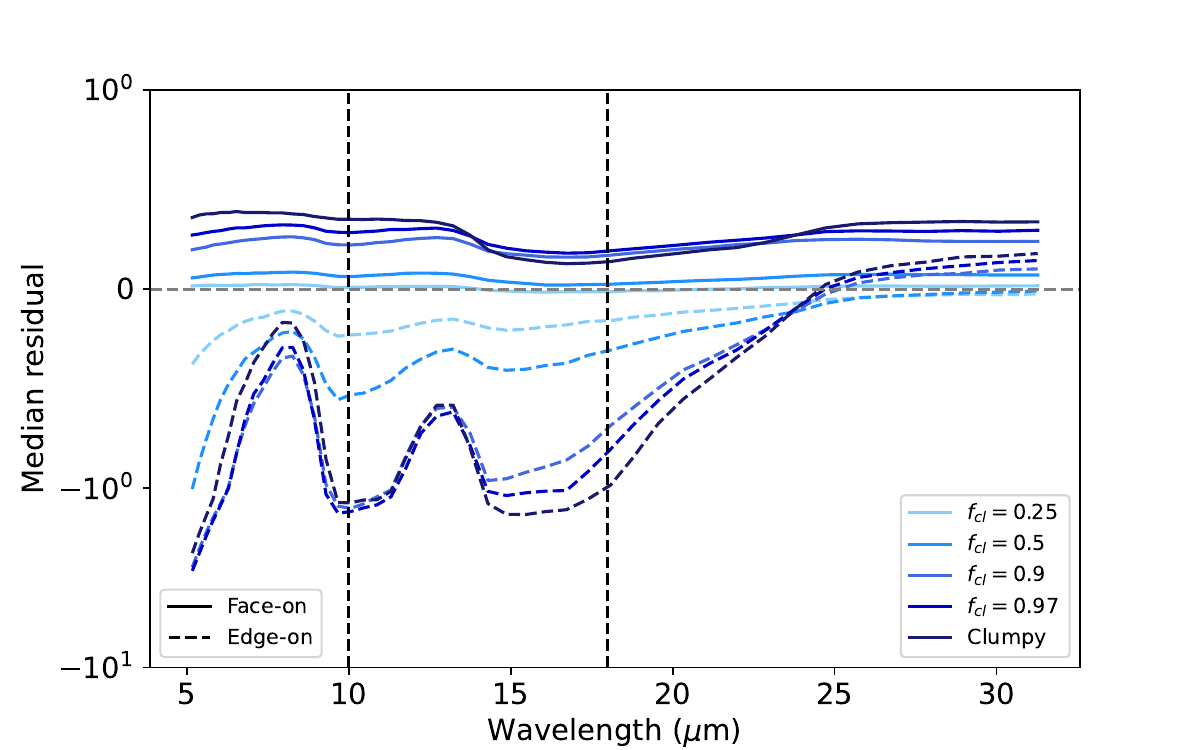}
    \end{subfigure}
    \caption{Median residuals obtained from comparing smooth model SEDs with their counterparts. \textbf{Left panel:} The counterparts are clumpy ($\fcl=1$) with different values of $\ffill$. \textbf{Right panel:} The counterparts have different values of $\fcl$, but fixed $\ffill=0.1$. For both panels, solid and dashed lines correspond to face-on and edge-on, respectively.}
    \label{fig:median_residuals_per_wavelength}
\end{figure*}

\begin{figure*}
    \centering
    \begin{subfigure}{0.49\textwidth}
        \centering
        \includegraphics[trim=0 0 55 0, clip,scale=0.48]{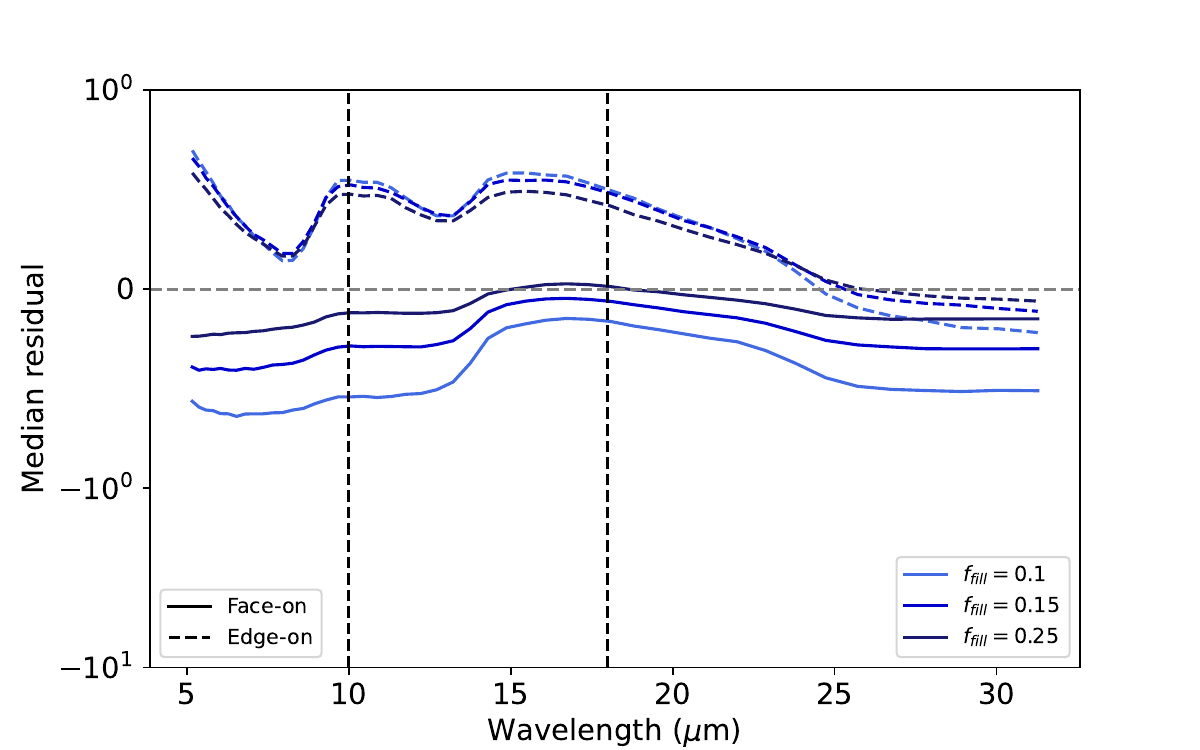}
    \end{subfigure}
    \begin{subfigure}{0.49\textwidth}
        \centering
        \includegraphics[trim=65 0 0 0, clip, scale=0.48]{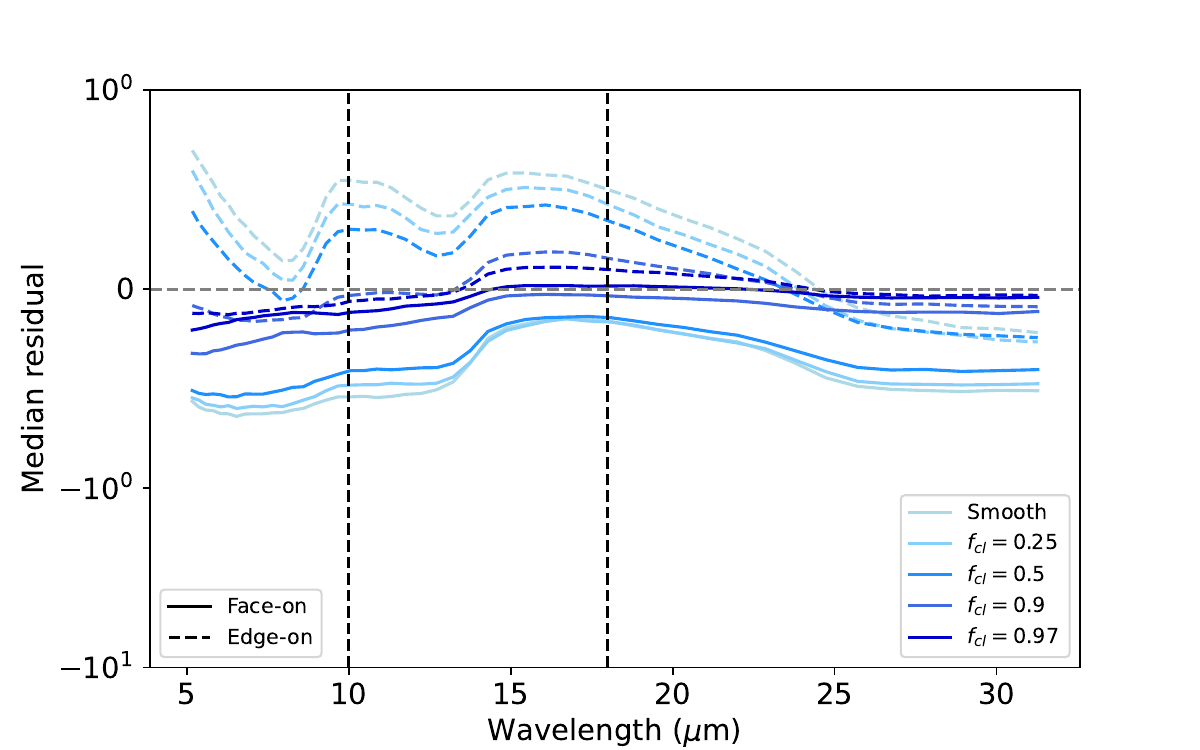}
    \end{subfigure}
    \caption{Median residuals obtained from the comparison between clumpy model SEDs and their smooth counterparts. \textbf{Left panel:} The clumpy model SEDs have different values of $\ffill$. \textbf{Right panel:} The clumpy model SEDs have different values of $\fcl$, but fixed $\ffill=0.1$. For both panels, solid and dashed lines correspond to face-on and edge-on, respectively.}
\label{fig:median_residuals_per_wavelength_clumpy_vs_counter}
\end{figure*}

\subsubsection{Smooth versus clumpy/two-phase counterparts} \label{sec:smooth_vs_counterpart_results}

As an example, and to illustrate the origin of the median residuals shown in subsequent figures, Figure~\ref{fig:distributions_ave_glob_res} shows the distributions of the averaged-global residuals, $\globalres$, obtained by comparing all the face-on and edge-on smooth model SEDs with their clumpy counterparts with $\ffill=0.1$. We can observe that, at a face-on viewing angle (left panel), the distribution of residuals for both wavelength ranges is narrow with medians around $\sim 0.3$. In contrast, at the edge-on viewing angle (right panel), the medians of the distributions of both wavelength ranges are located at higher values, being $\sim 13.9$ times larger in the 3-1000\,$\mu$m range and $\sim 2.6$ times larger in the 5-32\,$\mu$m range when compared to face-on. We calculated the medians of the $\globalres$ distributions obtained from comparing the face-on and edge-on smooth model SEDs and all their respective counterparts, which are presented in Figure~\ref{fig:distributions_medians_glob_res}. In this figure, we can see that the medians of $\globalres$ for edge-on viewing angles obtained when considering the 3-1000\,$\mu$m range (see Col.\,4) are larger than those for face-on (Col.\,3). The same result is exhibited for the 5-32\,$\mu$m range (see Cols.\,1 and 2). The difference between medians of $\globalres$ for edge-on and face-on orientations becomes larger when comparing smooth to their counterparts with $\fcl\geq0.5$ in the two wavelength ranges.

Figure~\ref{fig:median_residuals_per_wavelength} illustrates the median $\res$ at each wavelength for different values of $\ffill$ (left panel) and $\fcl$ (right panel). For all values of $\ffill$ or $\fcl$, edge-on SEDs produce larger median residuals than face-on SEDs, especially in the wavelength ranges $\rm{\lambda \lesssim 7 \ \mu m}$, $\rm{8 \lesssim \lambda \lesssim 12 \ \mu m}$, and $\rm{13 \lesssim \lambda \lesssim 25 \ \mu m}$. Furthermore, the median residuals for edge-on SEDs in these ranges are negative, while the opposite is obtained for face-on SEDs. From the left panel of Figure~\ref{fig:median_residuals_per_wavelength}, we observe that the median residuals for face-on SEDs approach zero across all wavelengths as $\ffill$ increases. However, this trend is not systematically observed for edge-on SEDs, where noticeable changes across different $\ffill$ values occur mainly in the $\rm{8 \lesssim \lambda \lesssim 25 \ \mu m}$ range. The right panel of Figure~\ref{fig:median_residuals_per_wavelength} shows that median residuals for both face-on and edge-on SEDs move closer to zero at most wavelengths ($\rm{\lambda \lesssim 25 \ \mu m}$ for edge-on) as $\fcl$ decreases. Finally, in both panels, it is notable that, at $\rm{\lambda \gtrsim 25 \ \mu m}$, the median residuals for edge-on SEDs are positive and become increasingly so as the counterpart SED becomes more clumpy (i.e., lower $\ffill$ and larger $\fcl$).

\subsubsection{Clumpy versus smooth/two-phase counterparts} \label{sec:clumpy_vs_counterpart_results}

In the previous section, we described the differences in the SEDs produced by the dust density distributions with respect to the smooth distribution. In this section, we present the differences with respect to the clumpy distribution to see how the two-phase distribution compares with the clumpy in terms of the resulting SEDs. 
 
In Figure~\ref{fig:distributions_medians_glob_res}, we also present the medians of the distributions of $\globalres$ from clumpy versus smooth/two-phase comparisons, also separated into face-on and edge-on viewing angles. We found that, in the 3-1000\,$\mu$m range, the medians of $\globalres$ are slightly higher for edge-on SEDs as compared to face-on for $\fcl \leq 0.5$, while for $\fcl>0.5$, the medians of $\globalres$ are very similar for both viewing angles. In the 5-32\,$\mu$m range, the medians of $\globalres$ are higher for edge-on SEDs in most cases compared to face-on. The only exception occurs in the comparisons between clumpy models with $\ffill = 0.1$ and smooth, where the medians of $\globalres$ are slightly higher for face-on SEDs. However, the medians of $\globalres$ in the clumpy versus smooth/two-phase comparisons do not reach such high values as in the smooth versus clumpy/two-phase comparisons. 

Figure~\ref{fig:median_residuals_per_wavelength_clumpy_vs_counter} shows the results from the $\res$ analysis of clumpy versus smooth/two-phase. The left panel of that figure shows the median residuals at each wavelength for different values of $\ffill$. We observe that edge-on SEDs produce median residuals that remain positive and nearly constant with $\ffill$ at most wavelengths ($\rm{\lambda \lesssim 25 \ \mu m}$). The largest median residuals occur within the wavelength ranges $\rm{\lambda \lesssim 7 \ \mu m}$, $\rm{8 \lesssim \lambda \lesssim 12 \ \mu m}$, and $\rm{13 \lesssim \lambda \lesssim 25 \ \mu m}$. In contrast, face-on SEDs produce negative median residuals across most wavelengths, with a systematic decrease as $\ffill$ increases. Additionally, we note that median residuals are closer to zero in the $\rm{13 \lesssim \lambda \lesssim 25 \ \mu m}$ range. The right panel of Figure~\ref{fig:median_residuals_per_wavelength_clumpy_vs_counter} shows the median residuals as a function of $\fcl$, revealing a similar behaviour for both viewing angles in smooth and two-phase with $\fcl=0.25$ and $\fcl=0.5$, comparable to the trend observed as a function of $\ffill$. However, for clumpy model SEDs with $\fcl=0.9$ and~$0.97$, the median residuals remain close to zero and exhibit minimal variation across most wavelengths, except in $\rm{13 \lesssim \lambda \lesssim 25 \ \mu m}$ range, where deviations are more noticeable. Finally, at $\rm{\lambda > 25 \ \mu m}$, both face- and edge-on SEDs present negative residuals.

\subsection{The effects of the dust chemical composition} \label{sec:result_dust_composition}

In the example displayed in Figure~\ref{fig:example_SEDs}, we see that Type-2 SEDs built with the \citet{Reyes-Amador2024} dust composition present a much more prominent decrease in flux in the
$\rm{\lambda < 7 \ \mu m}$ range, as compared to the Type-2 SEDs with the ISM dust composition. In the $\rm{\lambda > 30 \ \mu m}$ range, the behaviour for both dust compositions is similar for both Type-1 and Type-2 SEDs. In the $\rm{7 <\lambda < 30 \ \mu m}$ range, there are notable differences between the two dust compositions. The SEDs with the ISM dust composition produce a stronger $\rm{10 \ \mu m}$ silicate feature when it is in emission but weaker when it is in absorption, and they do not produce $\rm{18 \ \mu m}$ silicate feature in absorption at any clumpiness. In contrast, the SEDs with the dust \citet{Reyes-Amador2024} produce weaker $\rm{10 \ \mu m}$ silicate feature when it is in emission but stronger when it is in absorption, and very strong $\rm{18 \ \mu m}$ silicate features in absorption at smooth and two-phase models with $\fcl=0.25$ and~$\fcl=0.5$. These differences in the SEDs between the two dust compositions are observed not only for the model with those specific parameters but are also valid in general for all other parameters (see Figs.~\ref{fig:coverage_SEDs}, ~\ref{fig:silicate_properties_slopes_all_models}, and~\ref{fig:silicate_properties_slopes_all_models_omaira}). 

When looking at the extreme characteristics of the models as presented in Figure~\ref{fig:coverage_SEDs}, we observe that, compared to the SED coverage of models using the \citet{Reyes-Amador2024} dust composition, the lower limits of edge-on SEDs built with the ISM dust composition exhibit weaker absorption silicate features, their peaks are not shifted, and their intensity also decreases as the $\fcl$ increases, disappearing in the SED coverage of models with $\fcl \geq 0.9$. In the upper limits of edge-on SEDs, the only difference is that the \citet{Reyes-Amador2024} dust composition shows slightly stronger absorption silicate features, mainly in the models with $\fcl \leq 0.5$. On the other hand, face-on SEDs with the ISM dust composition present, in the lower limits, that the $\rm{10 \ \mu m}$ emission silicate feature is stronger than the $\rm{18 \ \mu m}$ feature, and that, in the upper limits, the silicate features are weak in emission. These characteristics contradict the observed cases with the \citet{Reyes-Amador2024} dust composition. Finally, we observe that in the six dust configurations, the \citet{Reyes-Amador2024} dust composition displays a wider flux range at $\lambda>\rm{100 \ \mu m}$ as compared to the ISM dust composition.

\begin{figure*}
    \centering
    \begin{subfigure}{0.49\textwidth}
        \centering
        \includegraphics[trim=0 0 55 0, clip,scale=0.48]{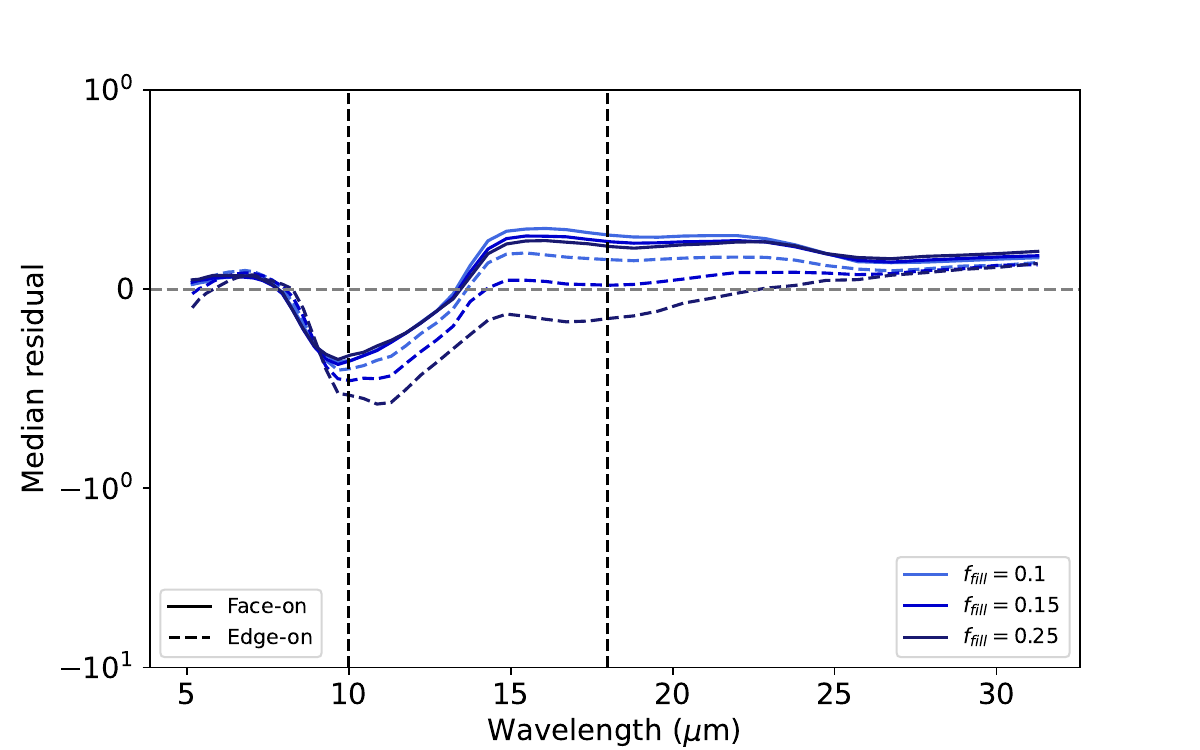}
    \end{subfigure}
    \begin{subfigure}{0.49\textwidth}
        \centering
        \includegraphics[trim=65 0 0 0, clip, scale=0.48]{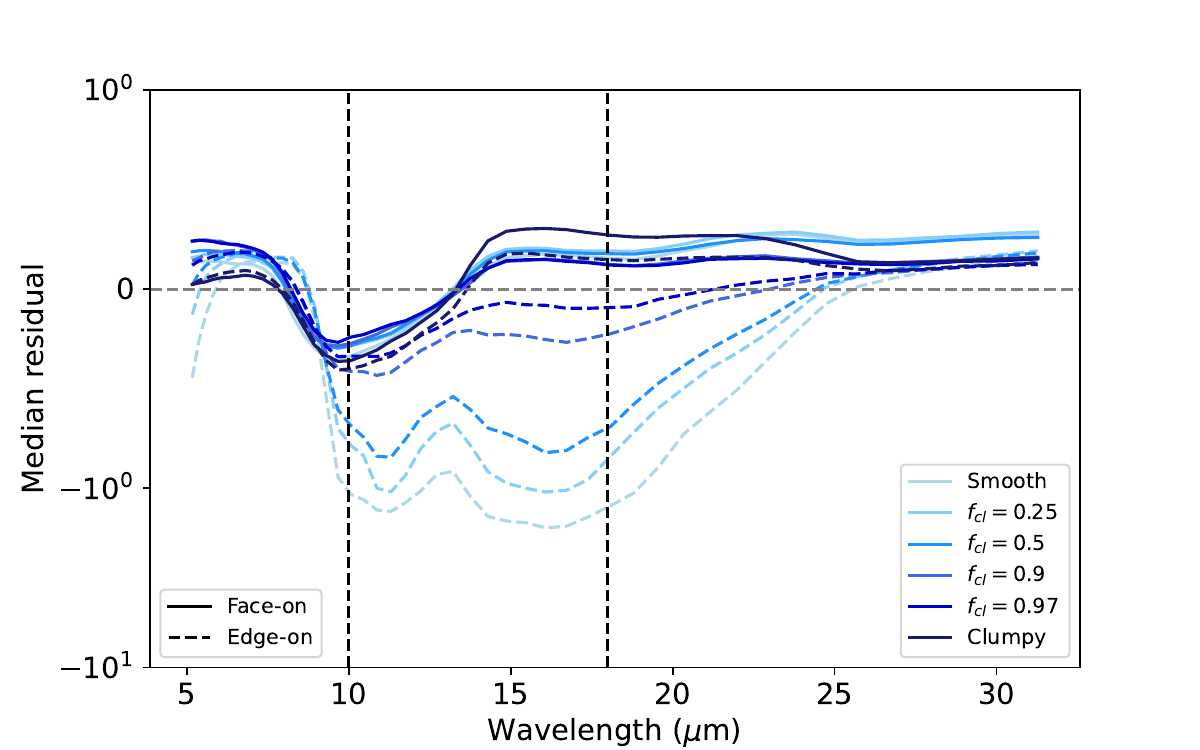}
    \end{subfigure}
    \caption{Median residuals obtained from the comparison between model SEDs with the \citet{Reyes-Amador2024} dust composition and their counterparts with the ISM dust composition. \textbf{Left panel:} The comparison is between clumpy ($\fcl=1$) model SEDs for different values of $\ffill$. \textbf{Right panel:} The comparison is between model SEDs with $\ffill=0.1$ (except by the smooth) for different values of $\fcl$. For both panels, solid and dashed lines correspond to face-on and edge-on, respectively.}
\label{fig:median_residuals_per_wavelength_my_comp_vs_omaira_comp}
\end{figure*}

Our results clearly show that the characteristics (shape, intensity, wavelength peak) of the silicate features vary for the two chemical compositions. In Figure~\ref{fig:silicate_properties_slopes_all_models_omaira}, we observe that synthetic SEDs using the ISM dust composition exhibit narrower ranges for the three spectral slopes, and their variations are less sensitive to $\fcl$, with $\alphamir$ showing no significant dependence on $\fcl$, as compared to those using the \citet{Reyes-Amador2024} dust composition. Most of the values for the three spectral slopes are negative for the two dust compositions, except for the $\alphamir$ obtained with the \citet{Reyes-Amador2024} dust composition, which exhibits large positive values. The results obtained using the ISM dust composition show that a greater number of synthetic SEDs exhibit silicate features in emission rather than absorption, and most silicate features, whether in absorption or emission, peak around $\peak=9.7$, with a significant fraction of emission features peaking in the range $10 < \peak < 12$. However, when comparing with the observed silicate features, Figure~\ref{fig:silicate_properties_slopes_all_models_omaira} shows this dust composition reproduces a larger fraction ($\sim 80\%$ in average) of observed absorption silicate features rather than the observed emission silicate features ($\sim50\%$ in average). The \citet{Reyes-Amador2024} dust composition produces a large number of synthetic SEDs with silicate features within $10 < \peak < 12$, predominantly in absorption rather than emission. Furthermore, most synthetic SEDs with silicate features that peak $\peak < 10$ exhibit them in emission rather than absorption. When comparing with the observed silicate features, we find that this dust composition works better to reproduce the observed absorption silicate features ($\sim 33\%$ in average) rather than the observed emission silicate features ($\sim 21\%$ in average, see Figure~\ref{fig:silicate_properties_slopes_all_models}).

The medians of $\globalres$ from the comparison between the model SEDs with the two dust compositions are presented in Figure~\ref{fig:distributions_medians_glob_res}. We found that, at both wavelength ranges and for face-on SEDs, the medians of $\globalres$ do not change significantly with the dust density distribution, showing medians of $\globalres$ always around $\sim 0.2$ and $\sim0.3$ for the 5-32\,$\mu$m and 3-1000\,$\mu$m ranges, respectively. Conversely, for edge-on SEDs, the medians of $\globalres$ diminish as the $\fcl$ increases, being always higher than for face-on SEDs for $\fcl\leq0.9$. For the smooth dust density distribution, we found the largest medians of $\globalres$ for edge-on SEDs, being much larger in the 3-1000\,$\mu$m range than in the 5-32\,$\mu$m range.

Examining the resulting $\res$ within the 5-32\,$\mu$m wavelength range, we observe in the left panel of Figure~\ref{fig:median_residuals_per_wavelength_my_comp_vs_omaira_comp} that when taking clumpy models into account, differences in the SED within this range are mainly driven by chemical composition over $\ffill$. Such differences are, for both face-on and edge-on SEDs, within $\rm{8 \lesssim \lambda \lesssim 13 \ \mu m}$, showing prominent negative median residuals, but getting positive and less prominent within $\rm{14 \lesssim \lambda \lesssim 25 \ \mu m}$, except for median residuals of edge-on SEDs with $\ffill=0.25$, which remain negative. From the same figure, we also note that for face-on SEDs, the median residuals show no significant variation with $\ffill$ across most wavelengths. However, for edge-on SEDs, the median residuals within $\rm{10 \lesssim \lambda \lesssim 25 \ \mu m}$ decrease with $\ffill$, switching from negative to positive. 

On the right panel of Figure~\ref{fig:median_residuals_per_wavelength_my_comp_vs_omaira_comp}, we focus on the differences given by the two chemical compositions for different $\fcl$ values. First of all, we notice that SEDs corresponding to a face-on view display the least difference in this spectral range, with no significant trends dependent on $\fcl$. The difference in dust composition produces an enhancement of the differences in the SEDs within $\rm{8 \lesssim \lambda \lesssim 25 \ \mu m}$, which is a function of $\fcl$, being largest for a smooth dust distribution. These differences are --as somehow expected-- the most dramatic.

\subsection{Goodness of the models at reproducing observations} \label{sec:goodness_of_models}

\begin{figure*}
    \centering
    \begin{subfigure}{\textwidth}
        \centering
        \includegraphics[trim=0cm 0 0 0, clip,scale=0.45]{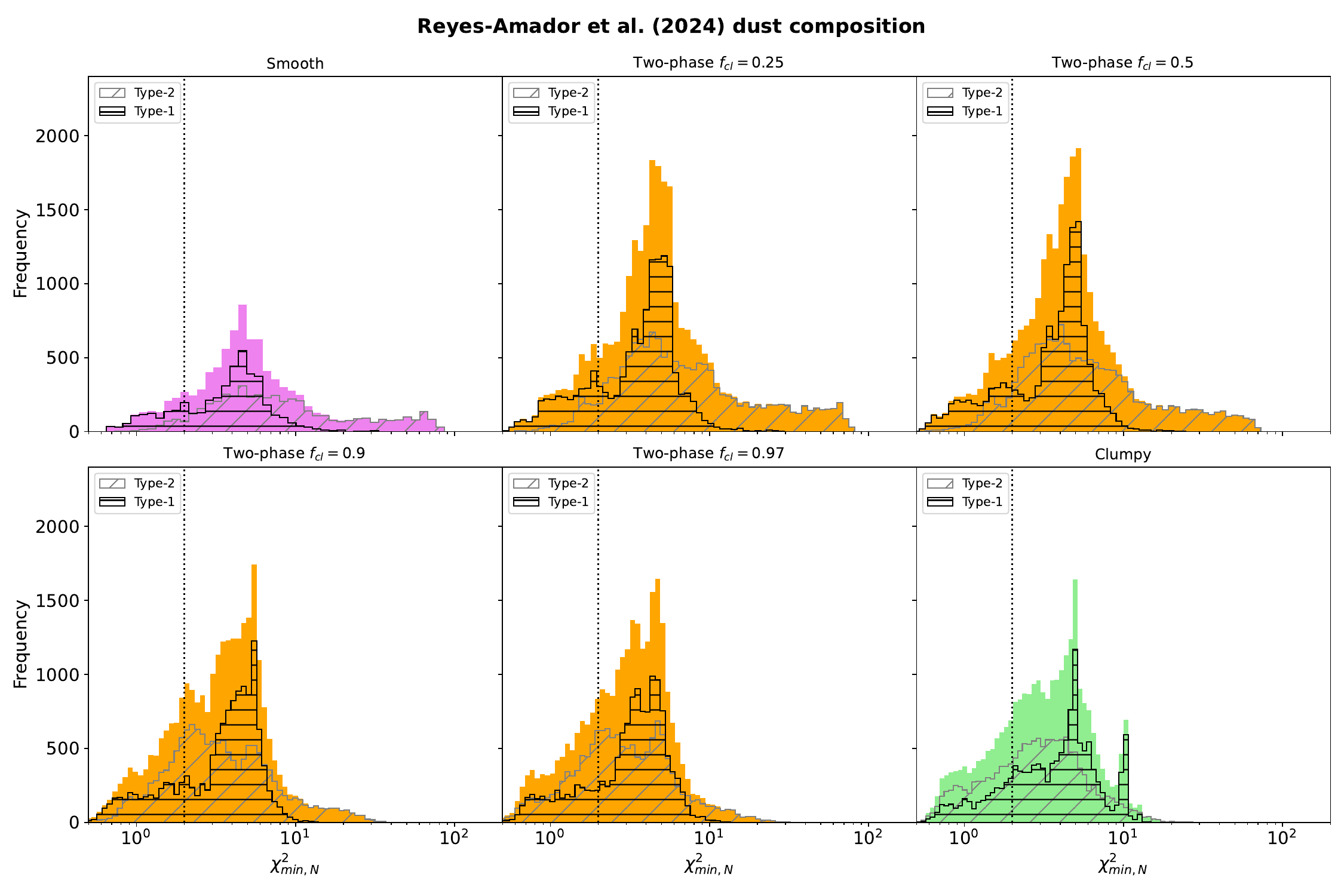}
    \end{subfigure}
    \begin{subfigure}{\textwidth}
        \centering
        \includegraphics[trim=0cm 0 0 0, clip,scale=0.45]{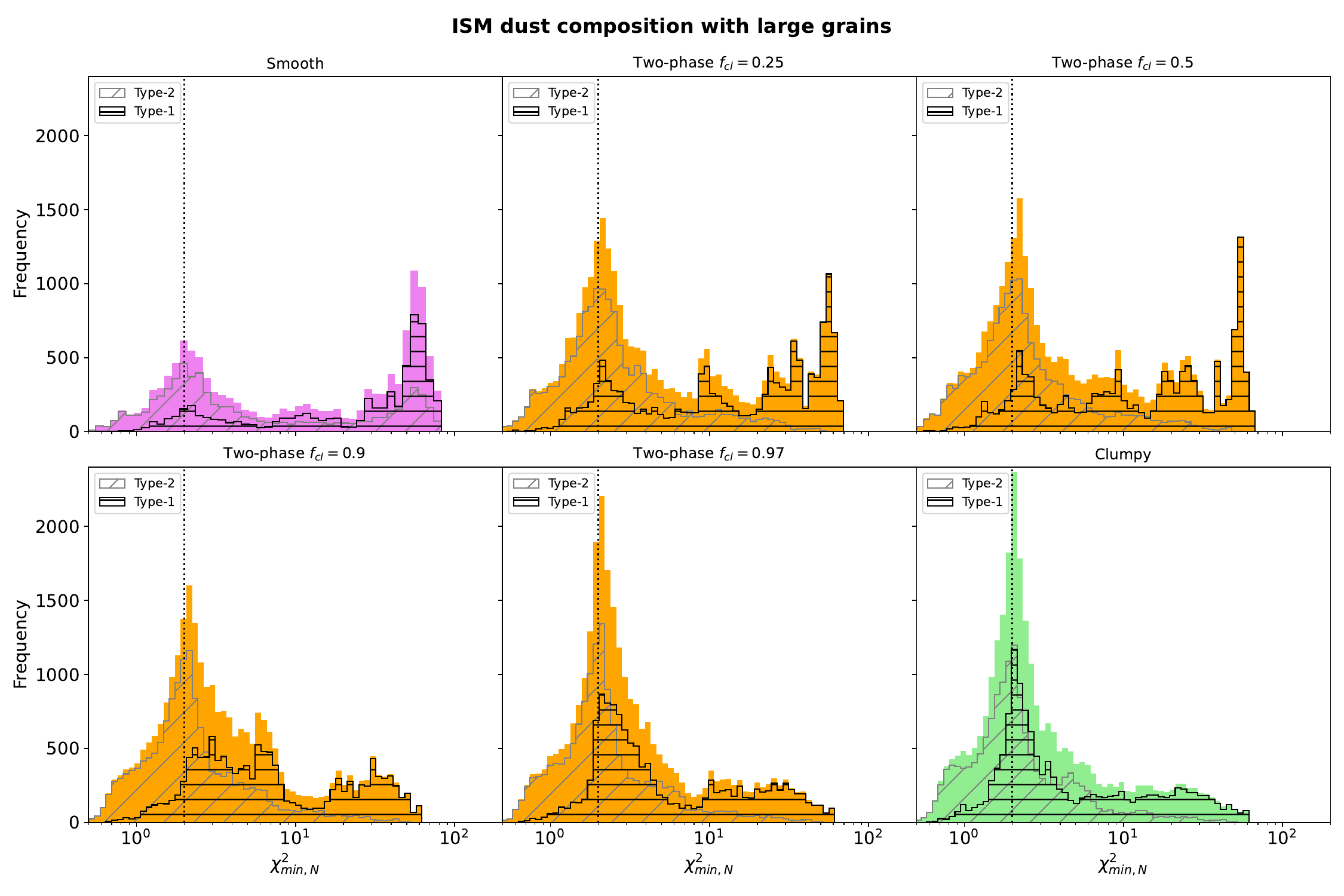}   
    \end{subfigure}
     \caption{Distributions of $\chisquaremin$ for the smooth (pink), two-phase (orange) and clumpy (light-green) model SEDs using the \citet{Reyes-Amador2024} dust composition (top panels) and the ISM dust composition (bottom panels). Histograms with black (horizontal lines) and gray (inclined lines) edge colours show the Type-1 and Type-2 SEDs, respectively. The vertical dotted line corresponds to $\chisquaremin=2$.} \label{fig:chi2_histograms_models_vs_obs}
\end{figure*}

Figures~\ref{fig:chi2_histograms_models_vs_obs} shows the resulting $\chisquare$ statistics when fitting the observed spectra to our model SED library (see Sec.~\ref{sec:model_vs_observations}). It shows the distributions of the $\chisquaremin$ for the models with the six clumpiness values using the two dust compositions analysed in this work. It is worth remembering to the reader here that the two-phase and clumpy models are three times more numerous than the smooth models due to the three $\ffill$ values. 

\begin{figure*}
    \centering
    \begin{subfigure}{0.45\textwidth}
        \centering
        \includegraphics[trim=0 0 0 0, clip,scale=0.5]{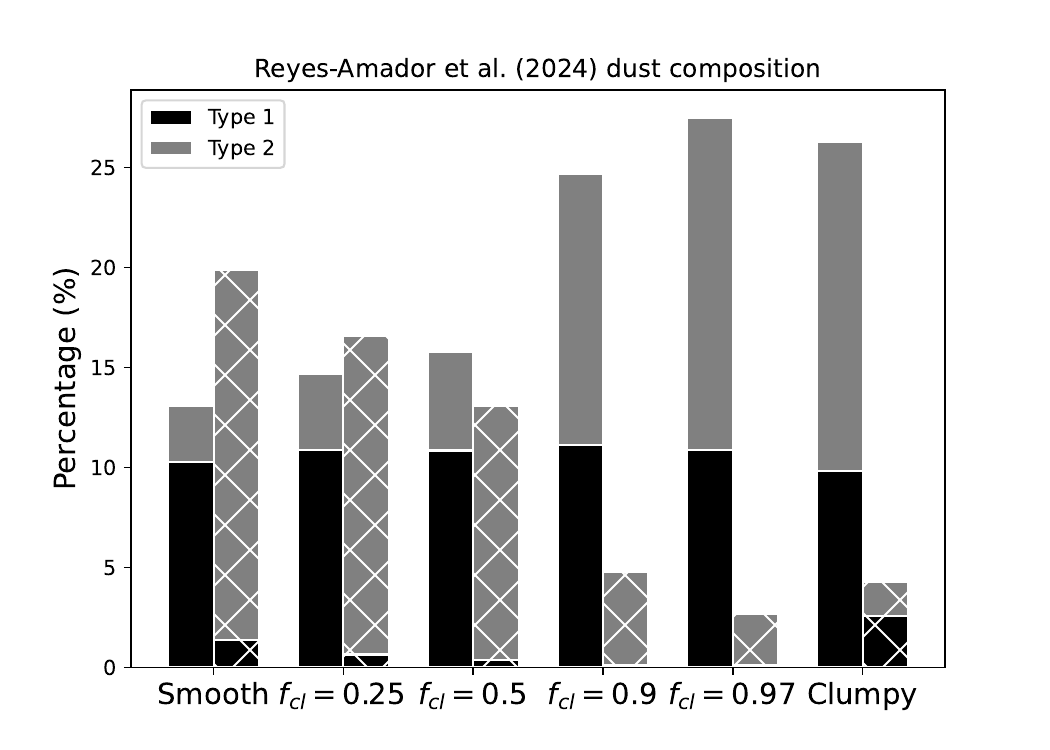}
    \end{subfigure}
    \begin{subfigure}{0.45\textwidth}
        \centering
        \includegraphics[trim=0 0 0 0, clip, scale=0.5]{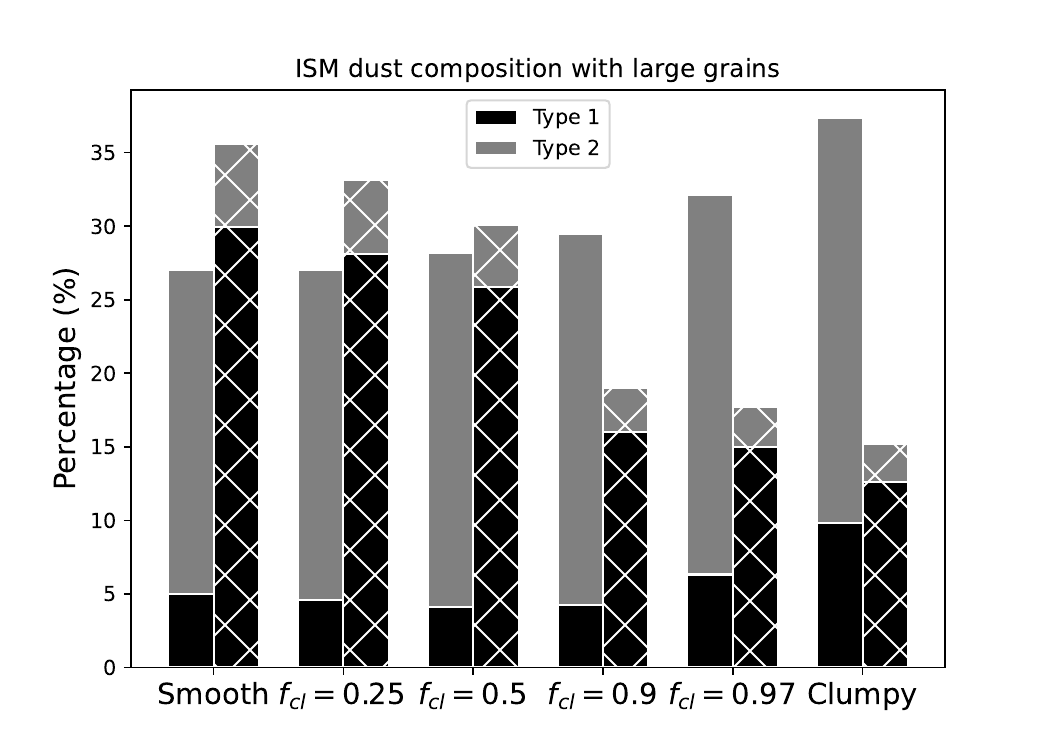}
    \end{subfigure}
    \caption{Percentages of SEDs with $\chisquaremin<2$ (solid bars) and $\chisquare>10$ (bars with pattern) for each clumpiness. Type-1 and Type-2 SEDs correspond to the black and gray bars, respectively. Left: for the SEDs using the \citet{Reyes-Amador2024} dust composition. Right: for the SEDs using the ISM dust composition.}
\label{fig:bar_plot_percentages_chi2_type1_type2}
\end{figure*}

The six top panels in Fig.\,\ref{fig:chi2_histograms_models_vs_obs} show the distributions of the $\chisquaremin$ for models using the \citet{Reyes-Amador2024} dust composition. These distributions of the models with the six clumpiness values present a peak around $\chisquaremin \approx 5$, which is dominated by Type-1 SEDs. All the models display a wing towards high $\chisquaremin$ values ($\chisquaremin>10$), with the number of such models decreasing as $\fcl$ increases, almost disappearing for purely clumpy models. The number of models providing an acceptable fit to the data (which we take to be those with $\chisquaremin<2$, see Fig.~\ref{fig:fitting_model_vs_obs}) varies depending on the $\fcl$ (see Figure~\ref{fig:bar_plot_percentages_chi2_type1_type2} for the quantification of this). A similar tendency is also observed when dividing between Type-1 and Type-2 model SEDs, with the latter being increasingly less dominating the high $\chisquaremin$ wing for higher $\fcl$ values. Except for the two-phase models with $\fcl=0.9$ and $\fcl=0.97$, which exhibit a bimodal distribution with one peak at $\chisquaremin \approx2$ dominated by Type-2 SEDs and the other at $\chisquaremin \approx5$ dominated by Type-1 SEDs.

The six bottom panels of Figure~\ref{fig:chi2_histograms_models_vs_obs} show the same for the ISM dust composition. Smooth models exhibit a bimodal distribution of $\chisquaremin$, with one peak centred at $\chisquaremin \approx2$, primarily associated with Type-2 synthetic SEDs, and another peak at $\chisquaremin \approx 60$, dominated by Type-1 SEDs. However, this bimodality disappears for two-phase and clumpy models. For two-phase models with $\fcl = 0.25$ and~$0.5$, the distributions display a main peak at $\chisquaremin \approx 2$, dominated by Type-2 SEDs, and secondary peaks at $\chisquaremin \geq 10$, which are primarily associated with Type-1 SEDs. In contrast, for clumpy and two-phase models with $\fcl = 0.9$ and~$0.97$, the secondary peaks vanish, but they keep exhibiting an important fraction of SEDs extending to $\chisquaremin \geq 10$, dominated by Type-1 SEDs, while the main peak remains dominated by Type-2 SEDs.

\begin{table*}
\begin{center}
\caption{Results from the Kullback-Leibler Divergence statistic for the distributions shown in Figure~\ref{fig:chi2_histograms_models_vs_obs_Y2} and~\ref{fig:chi2_histograms_models_vs_obs}.}
\begin{tabular}{cccccc}\hline \hline
 & & \multicolumn{2}{c}{\textbf{\citet{Reyes-Amador2024} dust composition}} & \multicolumn{2}{c}{\textbf{ISM dust composition large grains}} \\
\textbf{Parent} &  \textbf{Child} & \textbf{D-value} & \textbf{D-value} & \textbf{D-value} & \textbf{D-value}\\
\textbf{distribution} & \textbf{distribution} & \textbf{(vs Parent)} & \textbf{(Type 1 vs Type 2)} & \textbf{(vs Parent)} & \textbf{(Type 1 vs Type 2)}\\ \hline
\multirow{3}{*}{Smooth} & Type 1 & 0.001 & & 0.003 & \\ 
 & Type 2 & 0.002 & 0.010 & 0.0 & 0.005\\
 & Models $Y=2$ & 0.037 & & - & \\ \hline
\multirow{3}{*}{$\fcl=0.25$} & Type 1 & 0.001 &  & 0.003 & \\
& Type 2 & 0.002 & 0.008 & 0.0 & 0.009 \\
 & Models $Y=2$ & 0.041 & & - & \\ \hline
\multirow{3}{*}{$\fcl=0.5$} & Type 1 & 0.001 & & 0.004 & \\
& Type 2 & 0.002 & 0.007 & 0.0 & 0.010 \\
 & Models $Y=2$ & 0.042 & & - & \\ \hline
\multirow{3}{*}{$\fcl=0.9$} & Type 1 & 0.001 & & 0.004 & \\
& Type 2 & 0.001 & 0.003 & 0.001 & 0.009 \\
 & Models $Y=2$ & 0.048 & & - & \\ \hline
\multirow{3}{*}{$\fcl=0.97$} & Type 1 & 0.0 & & 0.003 & \\
& Type 2 & 0.0 & 0.002 & 0.001 & 0.007\\
 & Models $Y=2$ & 0.052 & & - & \\ \hline
 \multirow{3}{*}{Clumpy} & Type 1 & 0.0 & & 0.002 & \\
& Type 2 & 0.0 & 0.001 & 0.001 & 0.005 \\
 & Models $Y=2$ & 0.050 & & - &  \\
\hline\hline
\end{tabular}
\label{tab:KL_statistic}
\end{center}
\end{table*}

In Table~\ref{tab:KL_statistic} we summarize the results from the Kullback-Leibler Divergence statistic applied to quantify the similarity between Type-1 and Type-2 distributions. In columns 5 and 9, we observe that, although not systematically, the D-value decreases as $\fcl$ increases for both dust compositions, meaning that Type-1 distribution is more similar to Type-2 distribution in models with $\fcl \geq 0.9$ rather than in models with $\fcl \leq 0.5$.

We also analysed the distributions of $\chisquaremin$ obtained from SEDs, focusing on each of the parameter space values. We found that all the SEDs produced by models having a dusty torus size of $Y=2$ have $\chisquaremin > 10$ (see Figure~\ref{fig:chi2_histograms_models_vs_obs_Y2}) and very large D-values (see column 3 in Table~\ref{tab:KL_statistic}). On the other hand, the rest of the explored parameter values show a uniform-like distribution of $\chisquaremin$, covering the entire domain of $\chisquaremin$ from their respective global distribution of model SEDs, giving low D-values in the KL statistic.

From the distributions shown in Figure~\ref{fig:chi2_histograms_models_vs_obs}, we have calculated the percentages of SEDs with $\chisquaremin<2$ and $\chisquaremin>10$ for each clumpiness, discerning Type-1 SEDs from Type-2 SEDs, and we present them in the bar plot of Figure~\ref{fig:bar_plot_percentages_chi2_type1_type2}. This figure shows that models with the largest percentage of SEDs reproducing the observed spectra well ($\chisquaremin<2$) are the clumpy and the two-phase with $\fcl=0.97$ and that the ones reproducing them badly ($\chisquaremin>10$) are the smooth models.
In the left panel of this figure, we can see that, for the SEDs using the \citet{Reyes-Amador2024} dust composition, the percentage of Type-2 SEDs with $\chisquaremin<2$ ($\chisquaremin>10$) increases (decreases) as the $\fcl$ increases, whereas the percentage of Type-1 SEDs with $\chisquaremin<2$ is constant at all $\fcl$. The percentage of Type-1 SEDs yielding $\chisquaremin>10$ decreases as the $\fcl$ increases, with the exception of clumpy models that display the largest percentage of Type-1 SEDs, which corresponds to the peak at $\chisquaremin \approx 10$ (see the top clumpy panel in Fig.~\ref{fig:chi2_histograms_models_vs_obs}). The models producing this peak are mostly characterized by values of $\OAtorus=10^{\circ}$ and $\ptorus=0$. Moreover, most synthetic SEDs produced by smooth and two-phase models with $\fcl=0.25$ and $0.5$ with $\chisquaremin<2$, are Type-1 SEDs. Conversely, for clumpy and two-phase models with $\fcl=0.9$ and $0.97$, the majority of synthetic SEDs with $\chisquaremin<2$ are Type-2 SEDs. 

In the right panel of Figure~\ref{fig:bar_plot_percentages_chi2_type1_type2} we observe that, for the SEDs using the ISM dust composition with $\chisquaremin<2$, the percentage of both Type-1 and Type-2 SEDs does not vary significantly with $\fcl$ (except for Type-1 SEDs from clumpy and $\fcl=0.97$) and the percentage of Type-2 SEDs is always larger than the percentage of Type-1 SEDs for a factor of $\sim3$. On the other hand, for the SEDs with $\chisquaremin>10$, the percentage of both Type-1 and Type-2 SEDs decreases as the $\fcl$ increases, being the percentage of Type-1 SEDs always larger than the percentage of Type-2 SEDs.

\begin{figure*}
    \centering
    \includegraphics[trim=0cm 0 0 0, clip,scale=0.6]{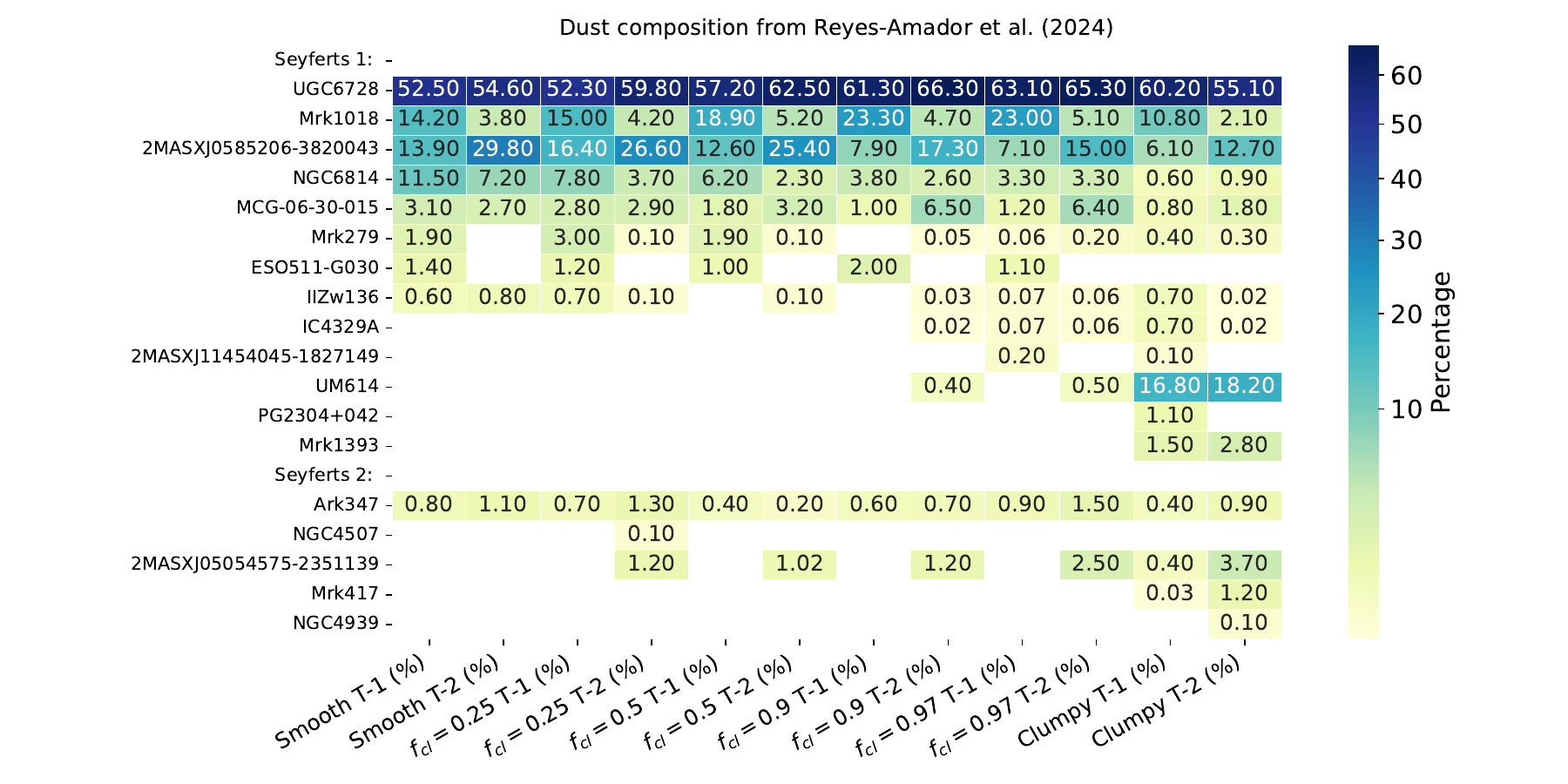}
    \caption{Percentages of SEDs per clumpiness value per type (T-1 and T-2) per object that satisfactorily reproduce each observed spectrum. The synthetic SEDs considered on these calculations are only those using the \citet{Reyes-Amador2024} dust composition and getting $\chisquaremin < 2$.}
\label{fig:percentages_chi2_lowerthan2_obs_my_comp}
\end{figure*}

\begin{figure*}
    \centering
    \includegraphics[trim=0cm 0 0 0, clip,scale=0.6]{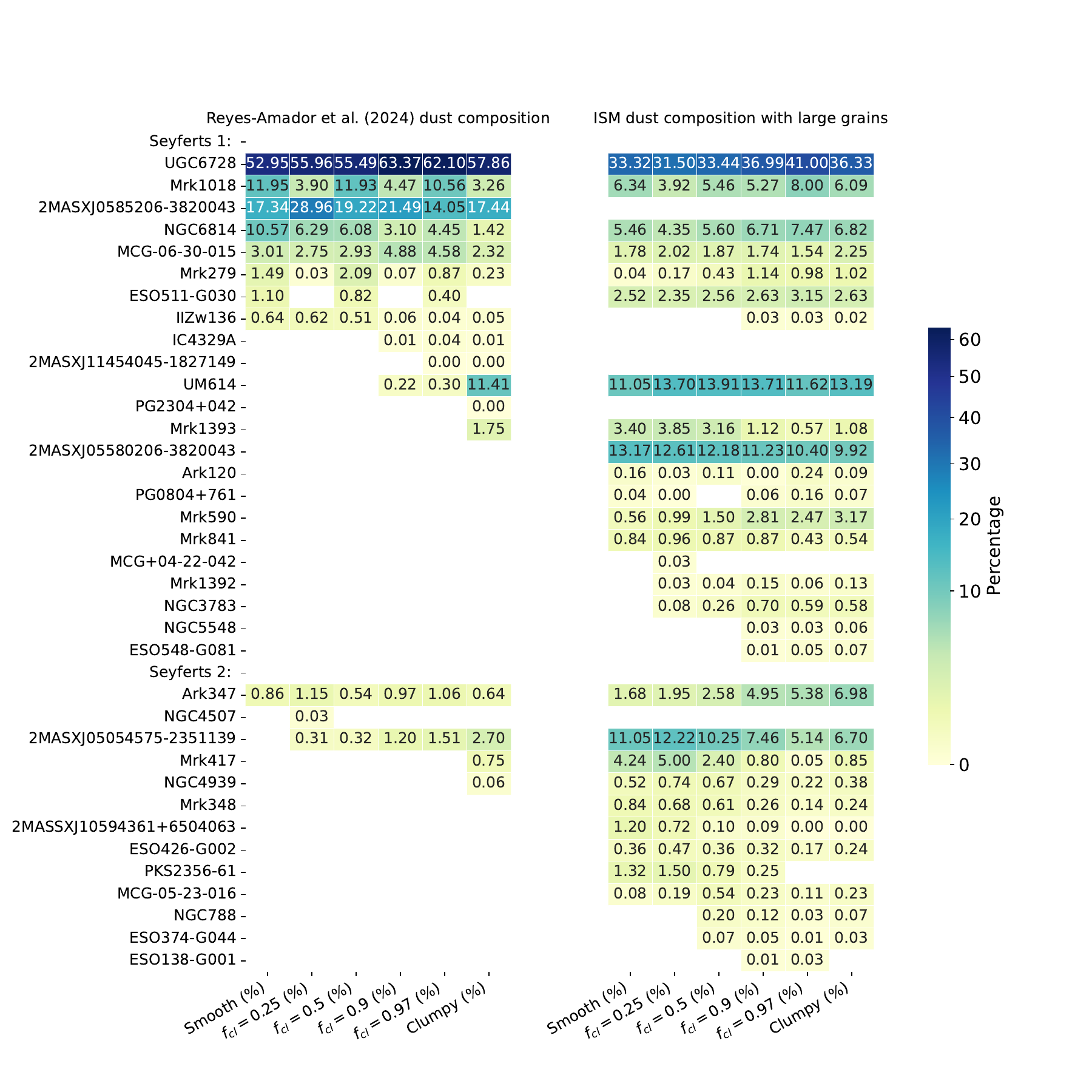}
    \caption{Percentages of SEDs per dust composition per clumpiness value per object that satisfactorily reproduce each observed spectrum. The left panel show the percentage values for the \citet{Reyes-Amador2024} and the right panel for the ISM dust with large grains. The synthetic SEDs considered on these calculations are only those with $\chisquaremin < 2$. \textbf{Note:} values with an order of magnitude $<10^{-2}$ appear as 0.00 but is not zero.}
\label{fig:percentages_chi2_lowerthan2_obs}
\end{figure*}

We then turned our analysis to the synthetic SEDs using the dust composition reported by \citet{Reyes-Amador2024} with $\chisquaremin < 2$. Figure~\ref{fig:percentages_chi2_lowerthan2_obs_my_comp} shows the percentages of SEDs per clumpiness value per type (Type 1 and Type 2) per object that satisfactorily reproduce each observed spectrum from those with a $\chisquaremin < 2$. We found that 18 ($\sim 26 \%$) out of the 68 spectra in the sample (the ones in the column 1 of that figure) are reproduced by the SEDs using the dust composition reported by \citet{Reyes-Amador2024}, from which five are Seyfert 2 and 13 are Seyfert 1 but regardless of their AGN type, most of them are well fitted by both Type-1 and Type-2 SEDs. This figure also illustrates that 10 out of the 18 ($\sim 55\%$) spectra are reproduced by SEDs from more than four clumpiness values, with the spectrum of UGC~6728 being the one with the highest percentage ($> 50\%$) in the models with the six clumpiness values for both Type-1 and Type-2 SEDs. Three spectra ($\sim 16\%$) are reproduced by SEDs from two to four clumpiness values, and five spectra ($\sim 28 \%$) are reproduced by SEDs from only one clumpiness value.

Figure~\ref{fig:percentages_chi2_lowerthan2_obs} shows that the spectra of 36 ($\sim 53\%$) out of the 68 objects in the sample are reproduced properly with at least one of the six clumpiness values and one of the two dust compositions, although with low percentage of SEDs in some cases.
When using the ISM dust composition, a higher number (31 objects) of observed spectra are reproduced compared to the dust composition reported by \citet{Reyes-Amador2024} (18 objects).
However, 13 objects are well fitted by SEDs using either dust composition, of which UGC~6728, Mrk~1018, MCG-06-30-015, Mrk~279, and IIZw~136 are the ones with a total percentage (summing the percentages of the six clumpiness values) of well-fitted SEDs that is larger when using the dust composition reported by \citet{Reyes-Amador2024} compared to the ISM dust composition. The objects with a larger total percentage of well-fitted SEDs using the ISM dust composition are NGC~6814, ESO~511-G030, UM~614, Mrk~1393, Ark~347, 2MASXJ~05054575-2351139, Mrk~417, and NGC~4939.  However, no trend is observed for any spectrum with respect to the dust density distribution or dust composition.

\section{Discussion} \label{sec:discussion}

This work affronts the study of the role of the dust density distribution and the dust chemical composition in RT models of the dusty torus in AGN in a self-consistent way. In this section, we discuss the main results obtained from the analysis of our SED library in the context of previous works that have studied the dusty torus models. 

\subsection{The role of the dust density distribution}  \label{sec:discuss_role_dust_distribution}

We have explored both extremes in the dust distribution: a homogeneous, smooth medium, in which all of the volume of the torus is filled with matter, and a clumpy one, where the distribution is highly discontinuous and a large amount of empty space is present between the dusty clouds. In models with the first distribution, the scattering effects are minimized by the continuous nature of the dust distribution, maximizing the effect of dust absorption \citep[e.g.][]{Fritz2006}. At the other extreme, in clumpy models, a fraction of light from the primary source can not only freely escape by going through the empty space in between the clumps, but dust emission from the clumps themselves, thermal and scattered, can be directly seen by the observer \citep{Nenkova2002,Nenkova2008b}.

In an attempt to understand the role of the dust density distribution in the dusty torus models, \citet{Feltre2012} performed a comparison between smooth and clumpy models presented by \citet{Fritz2006} and \citet{Nenkova2008a,Nenkova2008b}, respectively, trying to roughly match the parameters of the two models. Although the comparison is not straightforward because these two models are not built in the same way, they found that clumpy Type-2 SEDs show less steep continuum at $\rm{\lambda <7 \ \mu m}$ and weaker absorption silicate features compared to the smooth Type-2 SEDs. They noticed that smooth Type-2 SEDs show a $\rm{10 \ \mu m}$ silicate feature in stronger absorption as compared to the $\rm{18 \ \mu m}$ silicate feature. We found the same tendencies for Type-2 SEDs from models with $\fcl \leq 0.5$, as compared to the models with a dust distribution closer to clumpy (see Fig.~\ref{fig:coverage_SEDs} comparing panels of the first row against panels of the second row). For Type-1 SEDs, \citet{Feltre2012} found that clumpy models show the $\rm{18 \ \mu m}$ silicate feature stronger in emission, a characteristic that is also shown in our model SEDs (see Figs.~\ref{fig:example_SEDs} and~\ref{fig:coverage_SEDs}). We find that differences are maximal for edge-on SEDs. This is expected because when the light from the central source crosses through the torus, absorption and scattering processes take different weights on the resulting SED \citep{Pier1992,GranatoDanese1994}, strongly depending on how the dust is distributed within this region. Moreover, these differences are also maximal for edge-on SEDs within the 3-1000\,$\mu$m wavelength range rather than the 5-32\,$\mu$m (see Fig.~\ref{fig:distributions_medians_glob_res}). This is due to the significant difference of flux in edge-on SEDs at $\rm{\lambda<7 \ \mu m}$ (see Figs.~\ref{fig:example_SEDs} and~\ref{fig:coverage_SEDs}). We find that smooth models produce stronger and broader silicate absorption features but weaker and narrower emission features compared to the clumpy distribution. In contrast, clumpy models tend to generate weaker and narrower features in absorption but stronger emission features (see Fig.~\ref{fig:silicate_properties_slopes_all_models}).

If dust is distributed in both phases, the impact of dust absorption and the scattering effect will mix \citep{Stalevski2012}, and the prevalence of one among the other will depend, in principle, on both the fraction of clouds, $\fcl$, and the filling factor, $\ffill$. The advantage of the present analysis compared with that performed by \citet{Feltre2012} is that we make for the first time a comparison of one-to-one SED libraries of models where all parameters are identical, and the RT code is the same, varying these two parameters associated with the clumpiness. The filling factor has a minor effect on the differences between SEDs (see Figs.~\ref{fig:median_residuals_per_wavelength},~\ref{fig:median_residuals_per_wavelength_clumpy_vs_counter}, and~\ref{fig:median_residuals_per_wavelength_my_comp_vs_omaira_comp}). The global differences between smooth models and their two-phase counterparts with $\fcl = 0.25$ and $\fcl=0.5$ are negligible. These differences increase significantly when comparing smooth models with those with $\fcl>0.5$ (see Figures~\ref{fig:distributions_medians_glob_res} and~\ref{fig:median_residuals_per_wavelength}). \citet{Stalevski2012} explored the differences in the SEDs of a two-phase distribution with $\fcl=0.97$ compared to smooth and clumpy distributions. It is important to clarify that they used that value for $\fcl$ because, in combination with $\ffill=0.25$, it aligns with the ratio between high density (clumps) and the low-density (inter-clump dust) phases, which has a value of 100  for typical ISM \citep{Witt1996}. Interestingly, they found that smooth and two-phase models with $\fcl=0.97$ show no significant differences when the clumps are small because, in this case, the number of clumps is so high that each clump occupies one grid cell, resembling a smooth dust distribution. With our model grid, we can study the differences between a two-phase distribution and the other two distributions, not only for $\fcl=0.97$, but also for other lower values and even different $\ffill$. This allowed us to find that the differences between the smooth and their counterparts with a fixed $\fcl$ are larger for lower $\ffill$, as this parameter is proportional to the number of clumps (see Equation~\ref{eq:filling_factor}). A lower $\ffill$ means fewer clumps, which maximizes the differences with the smooth model counterparts. Additionally, \citet{Stalevski2012} found that at edge-on view, compared to smooth models, the $\rm{10 \ \mu m}$ silicate feature in two-phase SEDs shows slightly lower absorption. In contrast, two-phase models exhibit less pronounced NIR emission at a face-on view. This aligns with our results shown in Figure~\ref{fig:median_residuals_per_wavelength}.

Similarly to what has already been noted for smooth models, a purely clumpy dust distribution produces SEDs which are very similar to the respective two-phase counterparts with $\fcl  =0.90$ and $\fcl=0.97$, indicating that above a dust mass fraction in clumps higher than 90\%, the effect produced by clumpy component starts to prevail in the SEDs. Although out of the scope of the present simulations, by increasing the number of grid points for $\fcl$ and $\ffill$, we could determine the precise lower and upper limits for two-phase dust distribution transitions between being distinguishable from smooth and clumpy distributions. \citet{Stalevski2012} found that for a $\ffill=0.25$, the two-phase models result in the biggest difference compared to smooth and clumpy, and that the main differences between clumpy and two-phase model SEDs arise mainly from face-on orientations. Something that, although found in some cases, is not observed systematically in our models since the global differences between clumpy models and their counterparts are similar when compared between the two viewing angles and the two wavelength ranges (see Fig.~\ref{fig:distributions_medians_glob_res}). Additionally, the MIR bump around $\rm{\lambda \sim 30 \ \mu m}$ in smooth model SEDs has a larger flux compared to models with a fraction of dust in clumps (i.e., $\fcl>0$), independently of the viewing angle (see Fig.~\ref{fig:median_residuals_per_wavelength_clumpy_vs_counter}). 

\subsubsection{Is there a preferred dust distribution to reproduce observations?} \label{sec:best_dust_distribution}

\citet{Gonzalez-Martin2019II} found that the spectral slopes ($\alphanir$, $\alphamir$, and $\alphafir$) of the observed spectra of a sample of nearby AGNs are reproduced more accurately for the smooth and two-phase models from \citet{Fritz2006} and \citet{Stalevski2016}, respectively. In the attempt to reproduce the silicate feature details, \citet{Gonzalez-Martin2023} created a new SED library of two-phase models with $\fcl=0.25$ and the maximum dust grain size as an explored parameter, finding that these new models are capable of reproducing all the observed spectral slopes with at least one synthetic SED. Conversely, we found that none of our models --across the six clumpiness values and both dust compositions--are able to reproduce all the observed spectral slopes. This is because of our limited sampling of the parameter space compared to that work. However, the clumpy ($\fcl=1$) dust distribution seems to better represent the spectral properties (used to construct these diagrams) than the smooth and two-phase counterparts. The ranges of values of the three spectral slopes become broader as $\fcl$ decreases (see Figs.~\ref{fig:silicate_properties_slopes_all_models} and~\ref{fig:silicate_properties_slopes_all_models_omaira}). This suggests that models with $\fcl \leq 0.5$ produce a larger fraction of SEDs that cannot describe the spectral slopes of any observed spectrum in the AGN sample we are considering.  

With respect to the properties associated with the shape of the silicate feature ($\peak$, $\strength$, and $\ewidth$),  \citet{Gonzalez-Martin2023} found that the clumpy models from \citet{Nenkova2008b} produce strong $\rm{10 \ \mu m}$ silicate features in emission and the smooth models from \citet{Fritz2006} produce strong $\rm{10 \ \mu m}$ silicate features in absorption, something also found by \citet{Feltre2012}. Our results agree with those findings since we found that models with $\fcl\leq0.5$ produce stronger absorption and weaker emission of the $\rm{10 \ \mu m}$ silicate features, while the opposite happens for models with $\fcl>0.5$. Additionally, \citet{Gonzalez-Martin2019II} and \citet{Gonzalez-Martin2023} found that the ranges of values of the strength of the silicate features of the smooth, two-phase, and clumpy models are broad compared to the data. This is illustrated in Figs.~\ref{fig:silicate_properties_slopes_all_models} and~\ref{fig:silicate_properties_slopes_all_models_omaira}, which show that, in our SED library, there is a large fraction of SEDs that cannot reproduce the silicate properties of the sample. This is an indicator that the parameter space can be selected more realistically or that the full parameter space is not needed to reproduce the observed spectral characteristics, at least for the AGN sample considered in this work. A sample including a larger number of objects, more AGN types, or even high-redshift sources is needed to further explore the adequacy of this model grid. 

\citet{Gonzalez-Martin2019II} found that, when only the AGN model is taken into account (without including circumnuclear contributors), the smooth model from \citet{Fritz2006}, the clumpy model from \citet{Nenkova2008b}, and the two-phase models from \citet{Stalevski2016} are equally good to fit the observations. However, \citet{Gonzalez-Martin2023} reported that their two-phase model with $\fcl=0.25$ and ISM dust composition produced a larger percentage of good fits. Interestingly, they also found that the smooth model from \citet{Fritz2006} and the two-phase model from \citet{Stalevski2016} are equally good for fitting the observations and are much better than the clumpy model from \citet{Nenkova2008b} when circumnuclear contributors are included. Indeed, \citet{Varnava2025} found that the smooth torus model of \citet{Efstathiou1995}, which assumes a tapered disc geometry, provides overall better fits of IR observations of a sample of ULIRGs compared to the smooth model by \citet{Fritz2006} and the two-phase models by \citet{Stalevski2016} and \citet{Siebenmorgen2015}. This is because $\sim 40\%$ of ULIRGs host extremely obscured AGNs with column densities $N_{H}>10^{25}$ \citep{Falstad2021,Garcia-Bernete2022b}, a condition that is well represented by the high optical depths that take into account the tapered disc model proposed by \citep{Efstathiou1995}. Recent observations of the James Webb Space Telescope (JWST) showed that the silicate features of highly obscured AGNs are better explained with models using a smooth distribution \citep{Garcia-Bernete2024,Garcia-Bernete2024b}.

By comparing models homogeneously produced and whose only difference is the dust density distribution, we found that --at least for the specific dataset we analyse-- the largest percentage of good fits is obtained when the dust density distributions are clumpy and two-phase with $\fcl=0.97$, and the ones with the largest percentage of bad fits are the smooth models (see Fig.~\ref{fig:bar_plot_percentages_chi2_type1_type2}).

One of the main results of \citet{Gonzalez-Martin2023} is also that various objects do not show a preference for models with a specific dust distribution. In this work, from the spectra with an acceptable fit (36 out of 68 objects, $\sim 53 \%$ of the sample), we analysed the percentage of best-fitting ($\chisquaremin<2$) synthetic SEDs for each object (see Figures~\ref{fig:percentages_chi2_lowerthan2_obs_my_comp} and~\ref{fig:percentages_chi2_lowerthan2_obs}), finding that only six objects prefer a single clumpiness, a single SED type, and a single dust composition. In contrast, 21 out of these 36 objects prefer the models with the six clumpiness values, two SED types, and the two tested dust compositions. Both SED types are preferred for a single object because the observed spectra are classified based on the characteristics of the optical emission lines, while our synthetic SEDs are classified based on the viewing angle and the opening angle of the torus. Therefore, regardless of whether or not the line of sight passes through the dusty medium, the SEDs can reproduce the MIR spectral shape of both Type-1 and Type-2 observed spectra.

\subsection{The role of the dust chemical composition}

The importance of the dust chemical composition in RT models has been pointed out in previous works. \citet{Sirocky2008} showed that differences in the strength of the $\rm{10 \ \mu m}$ and $\rm{18 \ \mu m}$ silicate features depend on the dust chemistry, something also found by \citet{Feltre2012} who realized that spectral differences between smooth and clumpy models not only are due their dust distribution but also in their different dust chemical compositions. \citet{Martinez-Paredes2020} discussed the different chemical compositions assumed for the dust in the most popular torus models supporting the argument that alternative dust compositions (different to the standard ISM dust composition) could explain the observed features in Type-1 AGNs, something pointed out also by other authors \citep{Sturm2005,Li2008,Smith2010,Garcia-Bernete2017,Garcia-Bernete2022c,Tsuchikawa2022}. 

For the first time, in this work, we discuss the effects purely produced by two different dust chemical compositions. Figure~\ref{fig:silicate_properties_slopes_all_models} shows that the models using the \citet{Reyes-Amador2024} dust composition produce large positive values of $\alphamir$, which increase as the $\fcl$ decreases. This is related to the strong absorption silicate feature at $\rm{10 \ \mu m}$ shown in models with lower $\fcl$ since it affects the $\alphamir$ spectral slope (see Fig.~\ref{fig:coverage_SEDs}). This composition works better with a clumpy dust distribution to reproduce the observed spectral slopes ($\alphanir$, $\alphamir$, and $\alphafir$) and silicate properties ($\strength$, $\peak$, and $\ewidth$), showing that, in average, the observed absorption silicate features are better reproduced than the observed emission silicate features. This leads us to suspect that this chemical composition could perhaps help to explain the intense silicate absorption observed reported by \citet{Garcia-Bernete2024,Garcia-Bernete2022b} in highly obscured AGNs like ULIRGs.  Figure~\ref{fig:silicate_properties_slopes_all_models_omaira} shows that models using the ISM dust composition work better with a smooth dust distribution, reproducing a larger fraction of absorption silicate features rather than emission silicate features. 


However, as pointed out by \citet{Gonzalez-Martin2023}, it should be noted that the diagrams plotted in Figures~\ref{fig:silicate_properties_slopes_all_models} and~\ref{fig:silicate_properties_slopes_all_models_omaira} only provide a quick way to check the reliability of the models, highlighting the inability of a model to reproduce a set of observational features, but they are unable to inform on the success of a given model to properly reproduce all the observed spectral features with one single parameter set. This information is provided by the comparison --e.g., through a $\chisquare$ analysis-- with the full wavelength range of the observed data.

\subsection{Adequacy of the models to reproduce the observations} \label{sec:adequacy_models}

In terms of the $\chisquare$, the models with the ISM dust composition provide a larger fraction of good fits for all the clumpiness values compared with the \citet{Reyes-Amador2024} dust composition (see right panel of Fig.~\ref{fig:bar_plot_percentages_chi2_type1_type2}). The models with the former have a fraction of both Type-1 and Type-2 SEDs producing good fits that do not vary significantly with $\fcl$, except for Type-1 SEDs from clumpy and $\fcl=0.97$ models. In apparent contrast with what we observed in Figure~\ref{fig:silicate_properties_slopes_all_models_omaira}, the results from the $\chisquare$ fitting analysis illustrated in Figures~\ref{fig:chi2_histograms_models_vs_obs} and~\ref{fig:bar_plot_percentages_chi2_type1_type2} suggest that this dust composition works better with a clumpy dust distribution. This is explained by the ability of these models to produce the overall observed spectral features at the same time (see Fig.~\ref{fig:fitting_model_vs_obs} as an example). For models using the \citet{Reyes-Amador2024} dust composition, we found that the clumpier the dust distribution, the greater the fraction of Type-2 SEDs capable of reproducing the observed spectra. In contrast, the fraction of Type-1 SEDs is constant with $\fcl$. Models with $\fcl= \leq 0.5$ produce a larger fraction of Type-1 SEDs, producing good fits compared to Type-2 SEDs, while the opposite is true for models with $\fcl\geq0.9$.

According to our results from the analysis of the well-fitted observed spectra described in the last paragraph of Section~\ref{sec:goodness_of_models}, the ISM dust composition works better for reproducing most of the observed spectra properly rather than the \citet{Reyes-Amador2024} dust composition. However, Figure~\ref{fig:percentages_chi2_lowerthan2_obs} exhibits that some objects (UGC~6728, Mrk~1018, MCG-06-30-015, Mrk~279 Y IIZw~136) prefer synthetic SEDs with the \citet{Reyes-Amador2024} dust composition. We argue that, although a secondary ingredient compared to the dust distribution (and the overall geometry), modifications to dust composition should be tested to obtain a good match with observations. 

Most studies estimating the size of dust structures in AGNs are based on optical/NIR interferometry \citep[e.g.][]{Pfuhl2020,Dexter2020,GravityCollab2023} or reverberation mapping \citep[e.g.][]{Suganuma2006,Mandal2024}, using K-band data which, sampling observations around 2\,$\mu$m, are sensitive to emission coming from the hottest dust, at temperatures close to their sublimation values. These studies generally report dusty torus sizes on the order of sub-pc scales, which aligns with expectations because hotter dust is observed closer to the centre at the inner radius of the torus. On the other hand, interferometric observations targeting cooler dust (temperatures of hundreds of Kelvins) have revealed dusty torus sizes on the order of a few parsecs \citep[e.g.][]{Jaffe2004,Tristram2014,GamezRosas2022,Isbell2022}. We find that a dusty torus with $Y=2$ cannot properly reproduce the observed spectra in our sample (see Appendix~\ref{sec:models_Y2}). For the assumed bolometric luminosity and sublimation temperature (see Section~\ref{sec:the_torus_geometry}), our modelled dusty torus has an inner radius of $\Rin = 0.42$~pc. Therefore, $Y=2$ corresponds to an outer radius of $\Rout = 0.84$~pc. When comparing with observed spectra, the synthetic SEDs are scaled by the bolometric luminosity of the AGN, which affects the $\Rin$ (see Eq.~\ref{eq:sublimation_radius}). The range of bolometric luminosities that our AGN sample comprises is $\log(L_{\text{bol}}) = 41.9$ (for M~106) to $45.6$ \citep[for PG~0804+761, see][]{Gonzalez-Martin2023}, what results in dusty tori with inner radii ranging from $\rm{\Rin=0.02 \ pc}$ to $\rm{1.36 \ pc}$. Such compact dusty tori are not capable of producing the $\rm{5 - 32 \ \mu m}$ emission of the AGN in our sample.

\section{Summary and conclusions} \label{sec:conclussions}

Most of the works discussed in Section~\ref{sec:introduction} aim to explain differences in the SEDs produced by models with varying dust density distributions. However, discrepancies in geometry, dust properties, and RT methods may also influence the results. This paper offers the advantage of isolating the effect of dust density distribution (smooth, clumpy, and two-phase) across a range of parameters, allowing SED differences to be attributed solely to this factor. Moreover, we investigate the effects of two distinct dust chemical compositions (see Sec.~\ref{sec:the_dust_chemical_composition}): the ISM dust composition (49\% of graphites and 51\% of silicates) with large grains (up to $\rm{10 \ \mu m}$) and the oxide/silicate-based (47\% of graphites, $\sim 26.3\%$ of porous alumina, $\sim10.3\%$ of periclase, and $\sim 16.4\%$ of olivine) composition from \citet{Reyes-Amador2024}. We generated an extensive synthetic SED library using RT simulations analysed through model-to-model comparisons at face-on and edge-on viewing angles. These were also compared to 68 mid-infrared AGN spectra from \emph{Spitzer}, focusing on the overall goodness of fit, spectral shape, and silicate feature properties.

Our main results on dust density distribution are:

\begin{itemize}
    \item It significantly influences the SEDs at viewing angles where the line of sight passes through the dusty torus, especially at edge-on orientations. 
    \item Smooth models produce stronger and broader silicate absorption features but weaker and narrower emission features, whereas clumpy models show the opposite. The two-phase models exhibit intermediate characteristics depending on the clumpiness fraction ($\fcl$) and filling factor ($\ffill$).
    \item The global differences between smooth models and their two-phase counterparts with $\fcl = 0.25$ and $\fcl=0.5$, and between clumpy models and their two-phase counterparts with $\fcl=0.9$ and~$0.97$, are minimal.
    \item Models with $\fcl \geq 0.9$ yield more similar $\chisquaremin$ distributions between Type-1 and Type-2 SEDs than those with $\fcl \leq 0.5$.
    \item Clumpy and two-phase models with $\fcl=0.97$ produce most of the good fits to the observed sample.
\end{itemize}

Our main results on dust chemical composition are:

\begin{itemize}
    \item It significantly influences the silicate features at both edge-on and face-on viewing angles.
    \item The ISM dust composition with large grains better reproduces the silicate parameters ($\strength$, $\peak$, and $\ewidth$), mainly for a smooth distribution.
    \item Both dust compositions produce a larger fraction of SEDs exhibiting silicate features in emission rather than in absorption, particularly around $10 < \peak < 12$ and also both yield better $\chisquare$ fits for clumpier dust distributions.
    \item The ISM dust composition with large grains exhibits a higher fraction of good fits, particularly within the Type-2 SEDs, independently of dust density distributions, whereas the \citet{Reyes-Amador2024} dust composition provides more good fits within the Type-1 SEDs for models with $\fcl \leq 0.5$, and within the Type-2 SEDs for models with $\fcl \geq 0.9$.
\end{itemize} 

Our findings emphasize the importance of accounting for both dust density distribution and dust composition in AGN torus models to achieve a better match with observations and pave the way for future studies aiming to constrain the physical properties of the dusty torus in AGNs further, enhancing our understanding of their infrared emission and overall structure.

\section*{Acknowledgements}
This research is mainly funded by the UNAM PAPIIT projects IN109123 and IN110723 and CONAHCyT Ciencia de Frontera project CF-2023-G-100.
OURA thanks CONACyT for the Ph.D. scholarship No. 881295. The authors gratefully acknowledge the computing time granted by LANCAD and
CONACYT on the supercomputer Yoltla/Miztli/Xiuhcoatl at LSVP UAM-Iztapalapa/
DGTIC UNAM/CGSTIC CINVESTAV and the technical staff from IRyA: Daniel Díaz, Alfonso Ginori and Gilberto Zavala. 
We thank the Combined Atlas of Sources with Spitzer IRS Spectra (CASSIS) for providing the data to the community. We acknowledge the developers of the Python packages Numpy, Matplotlib, and Scipy, and the developers of the SKIRT code which proved very crucial to obtain the results presented in this work. IGB is supported by the Programa Atracci\'on de Talento Investigador ``C\'esar Nombela'' via grant 2023-T1/TEC-29030 funded by the Community of Madrid.


\section*{Data Availability}

Our SED library is not available to the public, yet. They will be published in the future, once other aspects are analysed. However, the data generated from the model-to-model ($\globalres$ and $\res$) and the models versus observations ($\alphanir$, $\alphamir$, $\alphafir$, $\strength$, $\peak$, $\ewidth$, and $\chisquare$) analyses could be provided by sending a request to the first author.



\input{mnras_template-arxiv.bbl}




\appendix


\section{Additional results}

\subsection{Examples of SED fitting to observed spectra}

\begin{figure*}
   \centering
    \includegraphics[trim=0cm 0 0 0, clip,scale=0.5]{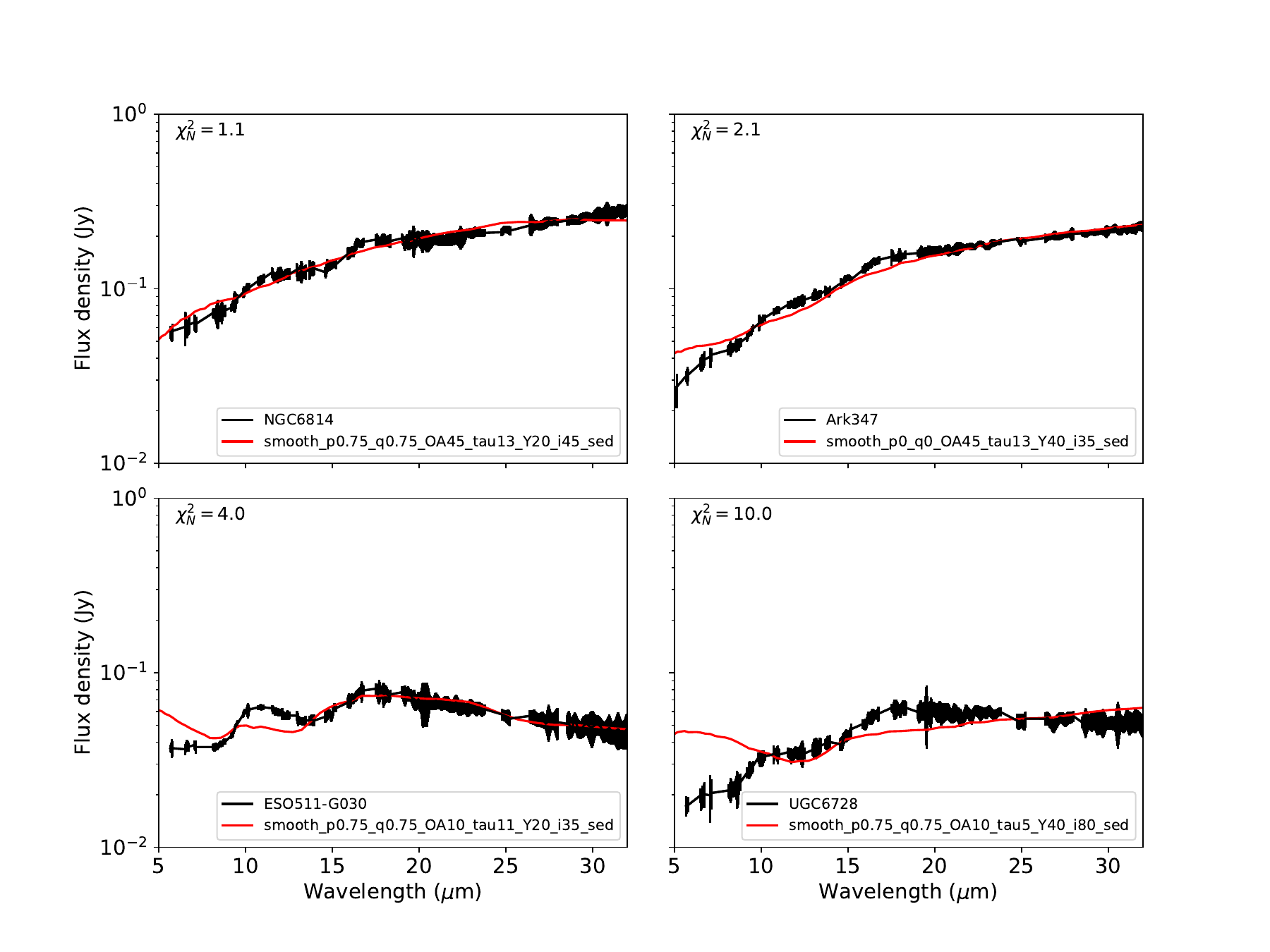}
    \caption{Examples of observed spectra (black) fitted by synthetic SEDs (red) with different $\chisquare$ values. The labels of the synthetic SED indicate the dust density distribution and the parameter values used in the torus models that produced them.}
    \label{fig:fitting_model_vs_obs}
\end{figure*}

As explained in Section~\ref{sec:model_vs_observations}, we implemented $\chisquare$ analysis to our SED library by fitting each synthetic SED to a sample of MIR observed spectra. As examples, Figure~\ref{fig:fitting_model_vs_obs} show the spectra of four objects in our sample: NGC~6814, Ark~347, ESO~511-G030, and UGC~6728 fitted with synthetic SEDs providing fits with $\chisquare = 1.1$, $2.1$, $4.0$, and $10.0$, respectively. In the legends of that figure, we can see that the labels indicate the dust density distribution and the parameter values used in the torus models that produced them. In these cases, all of them were produced by a dusty torus with a smooth dust distribution. 

\subsection{Models with Y=2} \label{sec:models_Y2}

\begin{figure*}
    \centering
    \includegraphics[trim=0cm 0 0 0, clip,scale=0.45]{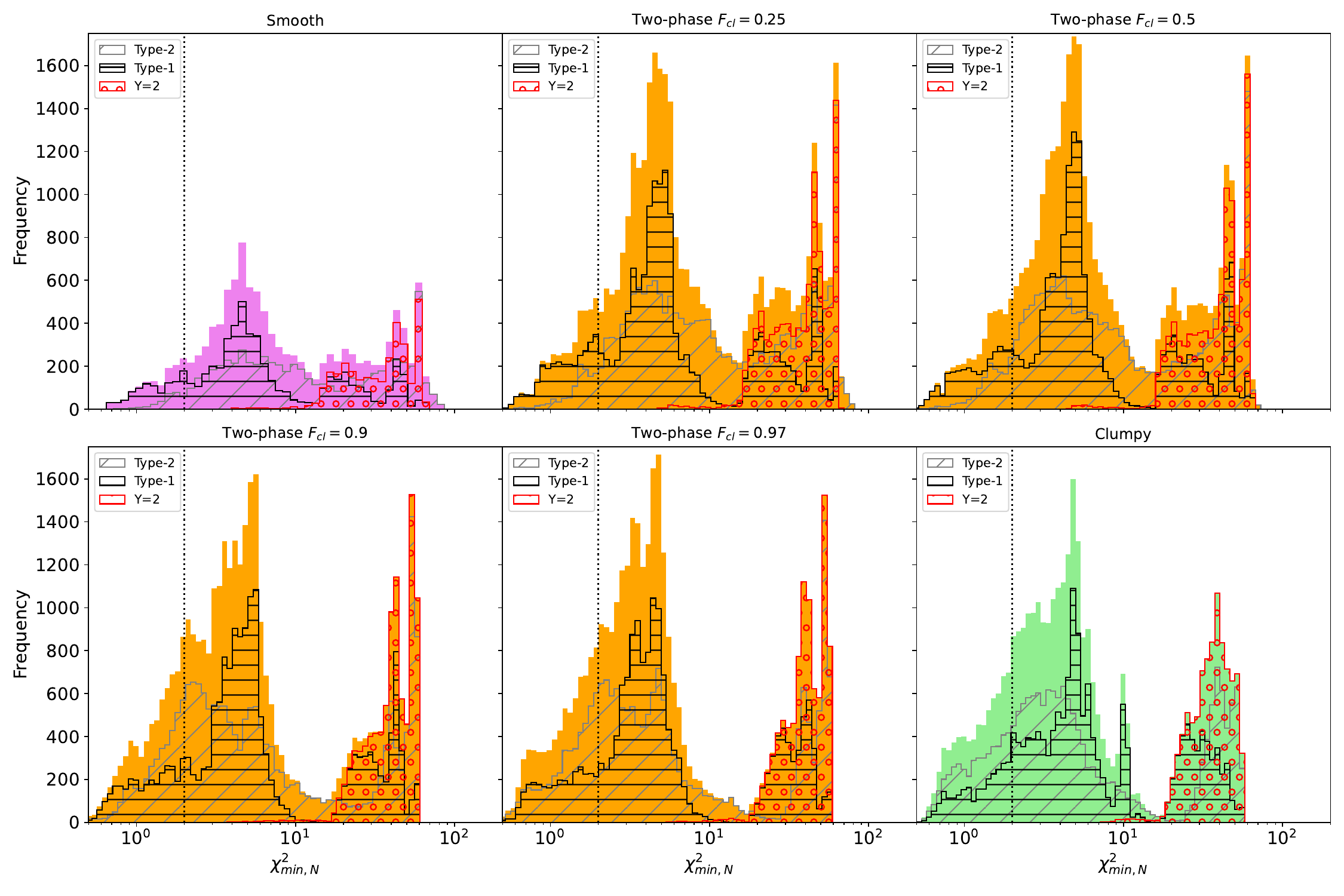}
    \caption{Same as in Fig.~\ref{fig:chi2_histograms_models_vs_obs}, but here, the red histograms filled with circles correspond to the SEDs of models having Y=2.}
    \label{fig:chi2_histograms_models_vs_obs_Y2}
\end{figure*}

We explored $Y=2$ in the parameter space of models using the \citet{Reyes-Amador2024} dust composition. We analysed the resulting synthetic SEDs through the calculation of $\chisquare$ comparing with the observed spectra, finding that, regardless of the dust density distribution (smooth, two-phase, or clumpy), this parameter value fails to reproduce the observations. Figure~\ref{fig:chi2_histograms_models_vs_obs_Y2} shows the distributions of the $\chisquaremin$ for the models with the six clumpiness values using the \citet{Reyes-Amador2024} dust composition, which include the $\chisquare$ obtained from the SEDs of models with $Y=2$. Notably, most of them produce $\chisquaremin > 10$. Only a negligible fraction produces lower $\chisquaremin$, being around $\chisquare \gtrsim 5$. This implies that a dusty torus with a $\Rout$ twice as large as the given $\rm{\Rin = 0.42 \ pc}$ can not reproduce the MIR observed spectra of our sample. 

\section{Technical information of the RT simulations}

\subsection{The spatial grid}
\label{sec:the_spatial_grid}

Before performing the Monte Carlo calculation, the spatial domain must be discretized into cells where the properties of the medium are assumed to be uniform. These cells should be carefully defined to accurately represent the dust density at each point within the grid. Therefore, it is crucial to subdivide the simulation space using a grid that effectively captures the characteristics of the dust distribution, with particular attention to the variations in dust density and optical depth. To ensure this, a convergence test can be conducted by briefly analysing the grid structure and its ability to represent the medium accurately (see Section~\ref{sec:analysis_model_convergence}). For each simulation, we used $\rm{5 \times 10^6}$ photon packets, 200 density samples, and a maximum number of iterations of 30 to ensure that the simulation ends when the convergence is reached. For the smooth models, we used a 2D cylindrical grid (\textsc{Cylinder2DSpatialGrid}) with 80,000 cells, which is based on cylindrical coordinates and perfect for axisymmetric geometries. For the clumpy and two-phase models, we used an octtree spatial grid (\textsc{PolicyTreeSpatialGrid}) with a number of cells varying between $\rm{\sim 3 \times 10^6}$ and $\rm{\sim 15 \times 10^6}$ depending on the values in the parameters of both the geometry and the spatial grid itself. This grid recursively subdivides the cuboidal domain into 8 cuboidal subcells until each cell satisfies certain criteria, or until a maximum level of subdivision has been reached.

\subsection{Analysis of the model convergence} \label{sec:analysis_model_convergence}

SKIRT provides several methods to evaluate the convergence of calculations in a simulation. The most efficient is the \textsc{ConvergenceInfoProbe}, which generates information related to the theoretical and discretized dust distribution. This consists of the dust mass expected for given values of the parameters, as compared to the one calculated after the space domain discretization, and in the expected optical depth, along the coordinates axes at a specific wavelength, as compared to the value calculated after discretization. Both quantities are presented as the input model values (defined analytically by the medium system) and the grid-discretized values (calculated from the finite-resolution spatial grid used in the simulation). Comparing both sets of values as following

\begin{equation} \label{eq:convergence}
    \text{Convergence} = \frac{\text{Input}-\text{Gridded}}{\text{Input}},
\end{equation}
we can determine whether the configured spatial grid adequately captures the material distribution in the simulation. In an ideal scenario, the input and gridded values would be equal, resulting in $\text{Convergence}=0$.

\begin{figure*}
    \centering
    \includegraphics[trim=0cm 0 0 0, clip,scale=0.45]{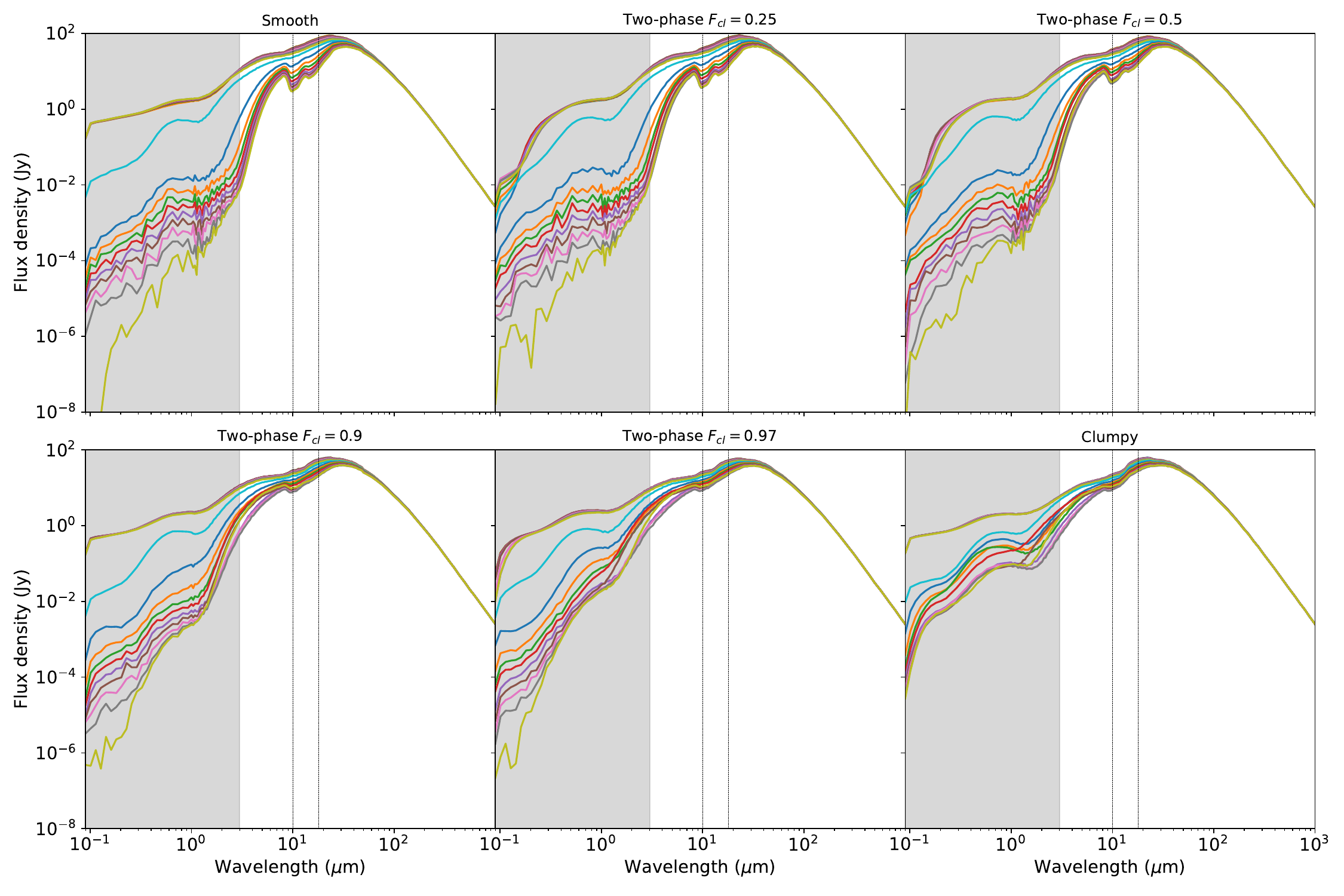}
    \caption{Synthetic SEDs with comparable parameter values viewed at the 19 values of $i$ obtained from the models with the six clumpiness values using the \citet{Reyes-Amador2024} dust composition. For all the models, the parameter values are $\ptorus=0.75$, $\qtorus=0.75$, $\OAtorus=45^\circ$, $Y=20$, $\opttau=9$. For the two-phase and clumpy models, $\ffill=0.15$. The gray-shaded region illustrates the wavelengths affected by non-zero convergence and Monte Carlo noise.}
    \label{fig:example_SEDs_convergence}
\end{figure*}

For the smooth models, we evaluated convergence in both the total dust mass and the optical depth at $\rm{0.55 \ \mu m}$ ($\convtau$) along the three coordinate axes ($x$, $y$, and $z$), applying a threshold $\text{Convergence}<10\%$ for each quantity, since we found it is flexible enough not to produce significant impacts in the wavelength region of interest in synthetic SEDs.  For the clumpy and two-phase models, we omitted the evaluation of convergence in the optical depth along the $x$ and $y$ axes. This is because the input values correspond to a smooth dust density distribution, while the gridded values is calculated for a clumpy distribution. The difference in this analysis comes from how SKIRT handles the simulation: the input values are determined before the \textsc{ClumpyGeometryDecorator} operates to the dust density, whereas the gridded values are calculated afterwards. Consequently, these values will differ by construction. However, for the optical depth at $\rm{0.55 \ \mu m}$ along the $z$ axis ($\convtauz$), the value must ideally be zero regardless of the dust density distribution, as there is no material along that direction due to the geometry of the torus. Although perfect convergence ($\text{Convergence}=0$) is theoretically expected, this is not achievable in practice due to minor imperfections in the construction of the spatial grid. Therefore, for clumpy and two-phase models, we evaluated only the convergence of $\convtauz$ and the total dust mass. To account for the added complexity of clumpy and two-phase dust density distributions, we relaxed the threshold to $\text{Convergence}<15\%$ for $\convtauz$, which still produces unaffected synthetic SEDs in the wavelength region of interest. From the total clumpy and two-phase RT simulations, only three failed to meet this condition.
After analysing the convergence in the RT simulations, we found that all smooth models satisfied the condition $\text{Convergence}<10\%$ with $\text{Convergence}=0\%$ for $\convtauz$, indicating no difference between the input (theoretical) and the gridded (simulated) optical depth at $\rm{0.55 \ \mu m}$ along the $z$ axis. In contrast, most two-phase and clumpy models exhibited non-zero convergence $0 < \text{Convergence}<15\%$ for $\convtauz$. The impact of the non-zero convergence for $\convtauz$ is a reduction in flux at wavelengths $\rm{\lambda < 3 \ \mu m}$ in Type-1 synthetic SEDs (i.e., those where the line-of-sight does not pass through the dusty torus). This effect is evident in the synthetic SEDs of two-phase models with $\fcl = 0.25, 0.5, 0.97$ shown in Figure~\ref{fig:example_SEDs_convergence}. Fortunately, the dust emission becomes significant at $\rm{\lambda > 3 \ \mu m }$, a spectral region that is not affected by the flux reduction caused by non-zero convergence, combined with the Monte Carlo noise present in Type-2 synthetic SEDs of smooth and two-phase models (also shown in Figure~\ref{fig:example_SEDs_convergence}).

\bsp	
\label{lastpage}
\end{document}